\documentclass[11pt]{article}
\pdfoutput=1
\usepackage{soul}
\usepackage{jheppub}
\usepackage{amsfonts}
\usepackage{amsmath}
\allowdisplaybreaks[4]         
\usepackage{amssymb}   
\usepackage{euscript}       
\usepackage{color}          
\usepackage{tensor}     
\usepackage{caption}
\usepackage{subcaption}   

\newcommand{\req}[1]{(\ref{#1})} 

\newcommand{\bea}{\begin{eqnarray}}
\newcommand{\eea}{\end{eqnarray}}
\newcommand{\ba}{\begin{eqnarray}}
\usepackage{braket}
\newcommand{\ea}{\end{eqnarray}}

\newcommand{\beq}{\begin{equation}}
\newcommand{\eeq}{\end{equation} }
\newcommand{\beqa}{\begin{eqnarray}}
\DeclareMathOperator{\Tr}{Tr}
\newcommand{\eeqa}{\end{eqnarray}}
\newcommand{\beqar}{\begin{eqnarray*}}
\newcommand{\eeqar}{\end{eqnarray*}}

\newcommand{\be}{\begin{equation}}
\newcommand{\ee}{\end{equation}}

\renewcommand{\req}[1]{(\ref{#1})}

\newcommand{\ssc}{\scriptscriptstyle}

\newcommand{\W}{\mathcal{W}}

\newcommand{\E}{\mathcal{E}}

\newcommand{\cO}{\mathcal{O}}

\newcommand{\Lag}{\mathcal{L}}
\newcommand{\Hsq}{\left( H^2 \right)}

\definecolor{shadecolor}{rgb}{.25,.25,.25}

\title{ \boldmath Higher-derivative holography with a chemical potential}

\author[a]{Pablo A. Cano,}
\author[b]{\'Angel J. Murcia,}
\author[a,c,d]{Alberto Rivadulla S\'anchez}
\author[a]{and Xuao Zhang}

\affiliation[a]{Instituut voor Theoretische Fysica, KU Leuven.
Celestijnenlaan 200D, B-3001 Leuven, Belgium \vspace{0.1cm}}
\affiliation[b]{Instituto de F\'isica Te\'orica UAM/CSIC. C/Nicol\'as Cabrera, 13-15, C.U. Cantoblanco, 28049 Madrid, Spain}
\affiliation[c]{Departamento de F\'isica de Part\'iculas, Universidade de Santiago de Compostela, E-15782 Santiago de Compostela, Spain\vspace{0.1cm}}
\affiliation[d]{Instituto Galego de F\'isica de Altas Enerx\'ias (IGFAE), Universidade de Santiago de Compostela, E-15782 Santiago de Compostela, Spain\vspace{0.1cm}}

\emailAdd{pabloantonio.cano@kuleuven.be}
\emailAdd{angel.murcia@csic.es}
\emailAdd{alberto.rivadulla.sanchez@usc.es}
\emailAdd{xuao.zhang@kuleuven.be}

\date{\today}
\abstract{We carry out an extensive study of the holographic aspects of any-dimensional higher-derivative Einstein-Maxwell theories in a fully analytic and non-perturbative fashion. 
We achieve this by introducing the $d$-dimensional version of Electromagnetic Quasitopological gravities: higher-derivative theories of gravity and electromagnetism that propagate no additional degrees of freedom and that allow one to study charged black hole solutions analytically. These theories contain non-minimal couplings, that in the holographic context give rise to a modified $\braket{JJ}$ correlator as well as to a general $\braket{TJJ}$ structure whose coefficients we compute. We constrain the couplings of the theory by imposing CFT unitarity and positivity of energy (which we show to be equivalent to causality in the bulk) as well as positive-entropy bounds from the weak gravity conjecture. 
The thermodynamic properties of the dual plasma at finite chemical potential are studied in detail, and we find that exotic zeroth-order phase transitions may appear, but that many of them are ruled out by the physical constraints. We further compute the shear viscosity to entropy density ratio, and we show that it can be taken to zero while respecting all the constraints, providing that the chemical potential is large enough. We also obtain the charged R\'enyi entropies and we observe that the chemical potential always increases the amount of entanglement and that the usual properties of R\'enyi entropies are preserved if the physical constraints are met. Finally, we compute the scaling dimension and magnetic response of twist operators and we provide a holographic derivation of the universal relations between the expansion of these quantities and the coefficients of $\braket{JJ}$ and $\braket{TJJ}$.
 }

\begin{document} 
\maketitle
\flushbottom

\section{Introduction}
\label{sec:Introduction}
Higher-derivative theories of gravity play a relevant role in the context of the AdS/CFT correspondence \cite{Maldacena,Witten,Gubser}, as they can lead to new insights on the physics of conformal field theories. 
On the one hand, certain higher-derivative terms capture finite $N$ and finite coupling effects in the boundary CFT, as is the case, for instance, for the corrections that appear explicitly in type IIB string theory \cite{Gross:1986iv,Grisaru:1986px,Gubser:1998nz,Buchel:2004di,Myers:2008yi}. In this situation, one is typically interested in a perturbative treatment of the corrections, as the $1/N$ and $1/\lambda$ effects are supposed to be small. 
On the other hand, higher-derivative gravities can be used to probe more general universality classes of CFTs than those covered by Einstein gravity \cite{Nojiri:1999mh,Blau:1999vz,Buchel:2008vz,Hofman:2008ar,Hofman:2009ug}. In other words, they allow one to explore a larger region in the space of CFTs via holography. A paradigmatic example of this is provided by the three-point function of the stress-energy tensor $\braket{TTT}$, which for a general $d$-dimensional CFT depends on three parameters. As is well known, for holographic CFTs dual to Einstein gravity only one of these parameters is non-vanishing, but one can achieve a general $\braket{TTT}$ structure by considering a higher-curvature theory in the bulk \cite{deBoer:2009pn,Buchel:2009sk,Myers:2010jv}. It is also worth noting that, since for a given CFT all the parameters of this correlator could be of order one, from this point of view it even makes sense to study the higher-derivative theory in a non-perturbative fashion. 

The program of studying the holographic aspects of higher-derivative theories as models for more general classes of CFTs has provided many insights into the dynamics of highly-interacting quantum field theories. One of the most impressive applications of this approach consists in unveiling universal properties valid for arbitrary CFTs, whose determination from first principles is sometimes obscure. In this line we can mention the holographic $c$-theorem established by Refs.~\cite{Myers:2010tj,Myers:2010xs}, the universal behavior of corner contributions to the entanglement entropy found in Refs.~\cite{Bueno1,Bueno2}, and more recently, the universal relationship between the free energy of a CFT in a squashed sphere and the coefficients of $\braket{TTT}$ observed in \cite{Bueno:2018yzo,Bueno:2020odt} --- see also \cite{Perlmutter:2013gua,Mezei:2014zla,Chu:2016tps,Li:2018drw} for other interesting examples.
On broader terms, higher-order gravities allow one to inspect which features of holographic CFTs dual to Einstein gravity are general and which ones can be changed. In this way, it is natural to wonder about the possible effects of higher-derivative terms on the holographic predictions regarding, for example, hydrodynamics, entanglement structure, superconductors, etc. --- see \textit{e.g.} \cite{Buchel:2004di,Kats:2007mq,Brigante:2007nu,Myers:2008yi,Cai:2008ph,Ge:2008ni,Gregory:2009fj,Jing:2010zp,Cai:2010cv,Hung:2011xb,Hung:2011nu,deBoer:2011wk,Galante:2013wta,Hung:2014npa,Bianchi:2016xvf,Dey:2016pei,Edelstein:2022xlb} for a necessarily incomplete list of references on these topics. 

In this paper we are interested in higher-derivative bulk theories that contain not only the metric, but also a vector field, which according to the holographic duality couples to a current operator $J^a$ in the boundary.  Similarly to the case of pure gravity, the higher-derivative terms permit us to study more general classes of dual CFTs.  An important quantity in this regard is the mixed correlator $\braket{TJJ}$, which has a fixed form for holographic Einstein-Maxwell theory, but which for a general CFT may contain an additional structure. The presence of this extra structure can be encoded in the energy-flux parameter $a_2$ of Ref.~\cite{Hofman:2008ar}, which is zero for EM theory, but which can get a non-vanishing value for higher-derivative theories --- in particular, it requires non-minimal couplings.  It is interesting to note that, for QCD, one has $a_2\approx -3/2$, so if one wanted to provide a holographic approximation to this theory one would need to consider higher-derivative operators with order one couplings. 

The presence of a vector field also allows us to explore the effect of a chemical potential in the CFT.  It is then interesting to study how the holographic predictions for certain properties of the CFT, such as, \textit{e.g.}, the hydrodynamics of charged plasmas \cite{Mas:2006dy,Son:2006em,Maeda:2006by} or entanglement and R\'enyi entropies \cite{Belin:2013uta}, change when we vary the couplings of the higher-order terms.
Although some of these questions have already been explored, most of the analyses so far have followed a perturbative approach \cite{Liu:2008kt,Cremonini:2008tw,Cremonini:2009sy,Myers:2009ij,Myers:2010pk,Cai:2011uh}, or have either stick to particular models, \textit{e.g.}, \cite{Cai:2008ph,Ge:2008ni,Ge:2009ac}. On the other hand, our goal is to perform a non-perturbative analysis of this type of theories taking into account all kinds of interactions between gravity and electromagnetism. This includes, in particular, non-minimal couplings of the form $RFF$, which, to the best of our knowledge, have not been studied in a non-perturbative fashion in the holographic context yet.  As we show, these are actually the most interesting terms to be added to the Einstein-Maxwell action due to their effects on the dual CFT.  

A key question in order to carry out an exact exploration rather than a perturbative one is to have a bulk theory which is amenable to analytic computations, which is typically not the case when there are higher derivatives involved. In the case of pure gravity, this is the reason why most of the literature has focused on a subset of theories with special properties, including Lovelock \cite{Lovelock1,Lovelock2,Wheeler:1985nh,Boulware:1985wk,Cai:2001dz,Padmanabhan:2013xyr}, Quasitopological \cite{Oliva:2010eb,Myers:2010ru,Dehghani:2011vu,Cisterna:2017umf} and Generalized Quasitopological gravities (GQG) \cite{PabloPablo,Hennigar:2016gkm,PabloPablo2,Hennigar:2017ego,PabloPablo3,Ahmed:2017jod,Bueno:2019ycr}, which are the only non-trivial ones in $D=4$. All of these theories actually belong to the GQG class \cite{Hennigar:2017ego,PabloPablo3}, and all of them share the following properties: they allow for the analytic study of static black hole solutions and they only propagate  a massless graviton on maximally symmetric backgrounds. Furthermore, this family of theories forms a basis for an effective-field-theory (EFT) extension of GR \cite{Bueno:2019ltp}, so they provide a general enough set of higher-order gravities. 
These theories can be minimally coupled to a Maxwell field while keeping all of their properties, but this is not a sufficiently general theory as it misses higher-derivative terms involving the vector field. For this reason, Ref.~\cite{Cano:2020qhy} introduced the family of Electromagnetic Quasitopological gravities (EQGs) in $D=4$, as extensions of the GQG theories that contain a vector field that can be coupled to gravity in many different forms, including non-minimal couplings. These theories were also recently studied in $D=3$ by Ref.~\cite{Bueno:2021krl}.

In this paper, we generalize this construction to arbitrary dimensions, and we will use the corresponding Electromagnetic Quasitopological theories to learn many aspects about holography in the presence of a chemical potential beyond the Einstein-Maxwell model.  Let us summarize our different contributions in each section. 

\begin{itemize}
\item In Section~\ref{sec:EQG} we introduce the family of EQG theories. These are most naturally written in terms of a $(d-2)$-form $B$ (in $D=d+1$ dimensions), and are characterized by possessing static black hole solutions  magnetically charged under $B$ whose metric depends on a single function. We explain how the $B$-field can be dualized into a vector field, in terms of which the solutions become electrically charged.  We provide a four-derivative EQG with four different operators, in which we focus for the rest of the paper. We additionally obtain EQGs at arbitrary order in the field strength and in the curvature. 

\item In Section~\ref{sec:BHs} we study the asymptotically AdS black hole solutions with spherical, planar and hyperbolic horizons of the four-derivative EQG we introduced in the previous section. 

\item In Section~\ref{sec:dictionary} we establish several basic entries of the holographic dictionary of these theories. We review the coefficients of the $\braket{TT}$ and $\braket{TTT}$ correlators and we carry out a detailed computation of the central charge $C_{J}$ of $\braket{JJ}$ as well as of the parameter $a_2$ that controls the angular distribution of energy radiated after a local insertion of $J$  is performed \cite{Hofman:2008ar}. Using these results we obtain explicitly the coefficients of the three-point function $\braket{TJJ}$.  

\item In Section~\ref{sec:constraints} we constrain the couplings of the bulk higher-derivative theory by imposing several physical conditions. We analyze unitarity and positivity-of-energy bounds on the boundary, and we show that the latter are exactly equivalent to avoidance of superluminal propagation of electromagnetic waves in the bulk. We further study the constraints coming from the mild form of the weak gravity conjecture (WGC) \cite{Cheung:2018cwt,Hamada:2018dde}, which demands that the corrections to the entropy of stable black holes at fixed mass and charge be positive, and that has recently been explored in the case of AdS in Ref.~\cite{Cremonini:2019wdk}. 

\item In Section~\ref{sec:thermo} we study the thermodynamic properties of the dual CFT by computing the grand-canonical potential of black holes of arbitrary topology. Focusing on the case of planar black holes, we carry out a detailed analysis of the phase structure of charged plasmas as a function of the chemical potential $\mu$. We find that new phases appear with respect to the Einstein-Maxwell prediction, and sometimes a zeroth-order phase transition can take place. However, we find that the physical constraints disfavor the values of the couplings giving rise to this situation. 

\item In Section~\ref{sec:hydro} we compute the shear viscosity at finite chemical potential and we study the ratio between this quantity and the entropy density. Taking into account all the physical constraints, we show that the behavior of $\eta/s$ is quite different for our holographic models depending on the sign of $a_2$. For $a_2<0$, this ratio is always a growing function of the chemical potential and we establish absolute bounds on it valid for any dimension and any value of the chemical potential. On the other hand, we show that for $a_2>0$ one can get $\eta/s=0$ for a sufficiently high chemical potential even if all the physical constraints are satisfied. 

\item In Section~\ref{sec:renyi} we study the charged R\'enyi entropies $S_n$ \cite{Belin:2013uta} and the generalized twist operators for the holographic Electromagnetic Quasitopological theories. We prove that, as long as the unitarity constraints are met, a small chemical potential always increases the amount of entanglement. Furthermore, we show that, if the WGC bounds are also satisfied, then the R\'enyi entropies satisfy a series of standard inequalities as a function of the index $n$, but we observe that these can be violated if the WGC does not hold. We also compute the scaling dimension and magnetic response of generalized twist operators. The expansion of these quantities around $n=1$ and  $\mu=0$ is known to be dictated by the coefficients of $\braket{TT}$, $\braket{JJ}$ and $\braket{TJJ}$ in a specific form and we provide a holographic derivation of these relationships that match exactly the results of \cite{Belin:2013uta} obtained from first principles.

\item We conclude in Sec.~\ref{sec:conclusions}, where we discuss the significance of our findings and comment on future directions. 
\end{itemize}


\section{Electromagnetic quasitopological gravities in arbitrary dimension}
\label{sec:EQG}

\subsection{Gravity, $(d-2)$-forms and their electromagnetic dual}\label{subsec:dual}
In this paper, we consider $(d+1)$-dimensional theories of gravity and of a $(d-2)$-form $B$ of the form\footnote{Gravity actions like this one need to be supplemented with York-Gibbons-Hawking boundary terms to make the variational problem well posed \cite{York:1972sj,Gibbons:1976ue}. Although on general grounds we expect the existence of these terms, determining them is a difficult problem and they are only known explicitly for certain higher-order theories such as Lovelock gravity \cite{Myers:1987yn,Teitelboim:1987zz}. However, it is possible to obtain simple effective boundary terms as long as we restrict to asymptotically AdS spacetimes with the boundary placed at infinity \cite{Bueno:2018xqc} --- we address this in Section \ref{sec:euclideanaction}.}
\begin{equation}\label{eq:LHtheory}
	I = \int d^{d+1}\! x \sqrt{|g|} \, \mathcal{L} (g_{\mu\nu}, R_{\mu\nu\rho\sigma}, H_{\mu_1 \cdots \mu_{d-1}})\, ,
\end{equation}
where $R_{\mu\nu\rho\sigma}$ is the Riemann tensor of the metric $g_{\mu\nu}$, and the $(d-1)$-form $H$ is the field strength of $B$, $H = d B$. The Lagrangian is supposed to be a scalar function built out of these tensors, and we implicitly assume that it has a polynomial form or that it can be expanded as such. In particular, we are interested in theories that reduce to the standard Einstein-$(d-2)$-form Lagrangian for small curvatures and field strengths,
\begin{equation}\label{eq:twodertheory}
\mathcal{L}=\frac{1}{16 \pi G}\bigg[ R +\frac{d(d-1)}{L^2}-\frac{2}{(d-1)!}H_{\mu_1\ldots \mu_{d-1}}H^{\mu_1\ldots \mu_{d-1}}+\ldots\bigg]\, .
\end{equation}

These theories are invariant under diffeomorphisms and under gauge transformations $B\rightarrow B+d \Lambda$, where $\Lambda$ is a $(d-3)$-form, and their equations of motion obtained from the variation of the action read

\begin{align}\label{eq:EinsteinEq}
\mathcal{E}_{\mu\nu}=\tensor{P}{_{(\mu}^{\rho \sigma \gamma}} \tensor{R}{_{\nu) \rho \sigma \gamma}} -\frac{1}{2}g_{\mu \nu} \mathcal{L}+2 \nabla^\sigma \nabla^\rho P_{(\mu| \sigma|\nu)\rho}-(d-1)\tensor{\mathcal{M}}{_{(\mu}^{\alpha_{1} \ldots \alpha_{d-2}}} \tensor{H}{_{\nu) \alpha_{1} \ldots \alpha_{d-2}}}&=0\, ,\\
\label{eq:MaxwelEq}
\mathcal{E}^{\nu_1\ldots \nu_{d-2}}=\nabla_{\mu}\mathcal{M}^{\mu\nu_1\ldots \nu_{d-2}}&=0\,,
\end{align}
where 

\begin{equation}
P^{\alpha \beta \rho \gamma}=\frac{ \partial \mathcal{L}}{ \partial R_{\alpha \beta \rho \gamma}}\, ,\quad \mathcal{M}^{\alpha_{1} \ldots \alpha_{d-1}}=-\frac{1}{2}\frac{ \partial \mathcal{L}}{ \partial H_{\alpha_{1} \ldots \alpha_{d-1}}}\, .
\label{eq:defpandm}
\end{equation}
Our interest in these theories lies on the fact that they allow for black hole solutions magnetically charged under the form $B$, as we explain below. Furthermore, the $(d-2)$-form can be related to a $1$-form (a vector field) by means of a duality transformation, and therefore we can map any of these theories to a higher-derivative extension of Einstein-Maxwell theory, which is the interpretation in which we are most interested. 

Let us quickly review the process of dualization.  Starting from the theory \req{eq:LHtheory}, we can dualize the $(d-2)$-form $B$ into a 1-form by introducing the Bianchi identity $dH=0$ in the action as follows\footnote{We introduce the factor $1/(4\pi G)$ bearing in mind that  the Lagrangian $\mathcal{L}$ will contain an overall $1/(16\pi G)$ normalization.}

\begin{equation}
\tilde I=\int d^{d+1}{x}\sqrt{|g|}\left\{\mathcal{L} (g_{\mu\nu}, R_{\mu\nu\rho\sigma}, H_{\mu_1 \cdots \mu_{d-1}})+\frac{1}{4\pi G (d-1)!} A_{\alpha_{1}}\partial_{\alpha_{2}}H_{\alpha_{3}\ldots\alpha_{D}}\epsilon^{\alpha_{1}\ldots\alpha_{D}}\right\}\, .
\end{equation}
At this point, $A_{\mu}$ is a Lagrange multiplier whose variation yields the Bianchi identity of $H$, which is now considered as a fundamental variable instead of $B$.  We can integrate by parts to express the action as

\begin{equation}
\begin{aligned}
\tilde I&=\int_{\mathcal{M}} d^{d+1}{x}\sqrt{|g|}\left\{\mathcal{L} (g_{\mu\nu}, R_{\mu\nu\rho\sigma}, H_{\mu_1 \cdots \mu_{d-1}})+\frac{1}{4\pi G (d-1)!}(\star F)_{\alpha_{1}\ldots\alpha_{D-2}}H^{\alpha_{1}\ldots\alpha_{D-2}} \right\}\\
&-\frac{1}{4\pi G}\int_{\partial\mathcal{M}} d^{D-1}x\sqrt{|h|}n^{\mu}A^{\nu}(\star H)_{\mu\nu}\, ,
\end{aligned}
\label{eq:generaldual}
\end{equation}
where we have defined $F=dA$. The variation with respect to $A_{\mu}$ still yields the Bianchi identity of $H$, but now it becomes clear that the variation with respect to $H$ yields an algebraic relation between this field and $F$, namely

\begin{equation}\label{eq:FstarH}
F=4\pi G (d-1)! \star\frac{\partial \mathcal{L}}{\partial H}\, .
\end{equation}
Then, one should invert this relation in order to get $H(F)$, and inserting this back into the action one would get the dual theory for the vector $A_{\mu}$. Note that the dualization process also generates a boundary term, which is precisely the term that makes the variational principle for the vector well-posed, and that, when computing the Euclidean action,  corresponds to working in the canonical ensemble (fixed electric charge).  

It is important to notice that the dual Lagrangian $\tilde{\mathcal{L}}$ is the Legendre transform of $\mathcal{L}$ with respect to $H$. Then, by the properties of the Legendre transform one can write the inverse relation between $H$ and $F$ as follows

\begin{equation}\label{eq:HstarF}
H=-8\pi G \star\frac{\partial \tilde{\mathcal{L}}}{\partial F}\, .
\end{equation}
This relation is useful because it allows us to identify the electric and magnetic charges in both frames. In fact, in the frame of the $(d-2)$-form we will have solutions with magnetic charge, which in the frame of the vector field correspond to electrically charged solutions. We define this charge in either frame as
\begin{equation}\label{eq:chargedef0}
q=\frac{1}{4\pi G}\int_{\mathcal{S}_{d-1}}H=-2\int_{\mathcal{S}_{d-1}}\star\frac{\partial \tilde{\mathcal{L}}}{\partial F}\, ,
\end{equation}
where the integral is performed over any spacelike co-dimension two hypersurface $\mathcal{S}_{d-1}$ that encloses the charge source. In the case of black hole solutions, $\mathcal{S}_{d-1}$ can be any surface that encloses the black hole horizon. 

Inverting \req{eq:FstarH} explicitly in order to obtain the dual Lagrangian is in general not possible. However, an important type of theories that we will consider in this paper are those quadratic in $H$, and all of them can be written as
\begin{equation}
I=\frac{1}{16\pi G}\int d^{d+1}x\sqrt{|g|}\Big\{\mathcal{L}_{\rm grav}-\frac{2}{(d-1)!}\tensor{(H^2)}{_{\mu\nu}^{\rho\sigma}}\tensor{Q}{^{\mu\nu}_{\rho\sigma}}\Big\}\, ,
\end{equation}
where $\mathcal{L}_{\rm grav}=R+\ldots$ only depends on the curvature, and where we are introducing the notation\footnote{That the most general quadratic Lagrangian can be written using only the object $\tensor{(H^2)}{_{\mu\nu}^{\rho\sigma}}$ (\textit{i.e}, with only four free indices) can be proven by writing the Lagrangian in terms of $\star H$ first.} 

\begin{equation}
	{\Hsq^{\mu_1 \cdots \mu_n}}_{\nu_1 \cdots \nu_n} \equiv H^{\mu_1 \cdots \mu_n \mu_{n+1} \cdots \mu_{d-1}} H_{\nu_1 \cdots \nu_n \mu_{n+1} \cdots \mu_{d-1}}\, .
	\label{eq:defH2}
\end{equation}

In this case, it is possible to find the dual theory explicitly. The relation \req{eq:FstarH} can be written in this case as

\begin{equation}\label{eq:FstarHlinear}
(\star F)_{\alpha_{1}\ldots\alpha_{d-1}}=\tensor{Q}{^{\mu\nu}_{[\alpha_{1}\alpha_{2}}}H_{\alpha_{3}\ldots\alpha_{d-1}]\mu\nu}\, .
\end{equation}
This can be inverted in the following way. Let us first introduce the following tensor,
\begin{equation}\label{eq:Qtilde}
\tensor{\tilde Q}{^{\mu\nu}_{\rho\sigma}}=\frac{12}{(d-1)(d-2)}\tensor{Q}{^{[\alpha\beta}_{\alpha\beta}}\tensor{\delta}{^{\mu}_{\rho}}\tensor{\delta}{^{\nu]}_{\sigma}}\, ,
\end{equation}
and its inverse, that we denote by $\tensor{(\tilde Q^{-1})}{^{\mu\nu}_{\rho\sigma}}$, and which by definition is determined from the equation

\begin{equation}\label{eq:inverseQ}
\tensor{(\tilde Q^{-1})}{^{\mu\nu}_{\alpha\beta}}\tensor{\tilde Q}{^{\alpha\beta}_{\rho\sigma}}=\tensor{\delta}{^{[\mu}_{[\rho}}\tensor{\delta}{^{\nu]}_{\sigma]}}\, .
\end{equation}
Then, one can check that \req{eq:FstarHlinear} is inverted by 

\begin{equation}
H_{\alpha_{1}\ldots\alpha_{d-1}}=\frac{1}{2}\epsilon_{\alpha_{1}\ldots\alpha_{d-1}\rho\sigma}\tensor{(\tilde Q^{-1})}{^{\rho\sigma}_{\alpha\beta}}F^{\alpha\beta}\, .
\end{equation}
and the dual action reads simply

\begin{equation}
\begin{aligned}
\tilde I=&\frac{1}{16\pi G}\int d^{d+1}x\sqrt{|g|}\Big\{\mathcal{L}_{\rm grav}-F_{\mu\nu}F^{\rho\sigma}\tensor{(\tilde Q^{-1})}{^{\mu\nu}_{\rho\sigma}}\Big\}\\
+&\frac{1}{4\pi G}\int_{\partial\mathcal{M}} d^{D-1}x\sqrt{|h|}n^{\mu}A^{\nu}\tensor{(\tilde Q^{-1})}{^{\alpha\beta}_{\mu\nu}}F_{\alpha\beta}\, .
\end{aligned}
\end{equation}

When the Lagrangian contains terms beyond quadratic order in $H$, such as $(H^2)^2$, the equation \req{eq:FstarH} becomes a tensorial polynomial equation, whose resolution is more involved. One could nevertheless solve it by assuming a series expansion in $F$.

\subsection{Electromagnetic quasitopological gravities: general definition}\label{sec:EQGdef}

We are interested in studying the charged static solutions with spherical, planar or hyperbolic sections of the theories \req{eq:LHtheory}.  A general metric ansatz for these configurations reads
\begin{equation}
	ds_{N, f}^2 = - N^2(r) f(r) dt^2 + \frac{dr^2}{f(r)} + r^2 d \Sigma^2_{k, (d-1)} ~,
	\label{eq:Nfmetric}
\end{equation}
where the metric $d \Sigma^2_{k, (d-1)}$ is given by

\begin{equation}
	d \Sigma_{k, (d-1)}^2 = \left\lbrace 
	\begin{array}{ll}
		d\Omega_{(d-1)}^2 & \text{for } k = 1 \text{ (spherical),} \\[0.3em]
		\displaystyle \frac{1}{L^2} dx_{(d-1)}^2 & \text{for } k = 0 \text{ (flat),} \\[0.6em]
		d\Xi_{(d-1)}^2 & \text{for } k = -1 \text{ (hyperbolic).}
	\end{array} \right. \,
\end{equation}

\noindent
In addition, we assume the following magnetic ansatz for the $H$ field,
\begin{equation}\label{eq:magneticH}
H_{Q}=Q\, \omega_{k, (d-1)}\, ,
\end{equation}
where $Q$ is a constant related to the magnetic charge and $\omega_{k, (d-1)}$ is the volume form of $d \Sigma^2_{k, (d-1)}$, whose integral yields the volume of this space, that we denote by $V_{k, (d-1)}=\int \omega_{k, (d-1)}$.
It is obvious that this $H$ satisfies the Bianchi identity $dH=0$, but one can also check that, for any theory of the form \req{eq:LHtheory}, it also solves its equation of motion \req{eq:MaxwelEq} when we use the metric \req{eq:Nfmetric}. 
Since we do not have to worry about the ``Maxwell equation'' anymore, the problem of finding the solutions becomes simpler:
 one only has to solve the equations for the metric functions $N$ and $f$, that, as shown in \cite{Cano:2020qhy}, can be obtained by means of the reduced Lagrangian, 
\begin{equation}
	L_{N, f} = \sqrt{|g|} \Lag |_{ds^2_{N, f}, H_Q}\, .
\end{equation}

\noindent
The equations of motion are obtained simply by varying this Lagrangian with respect to the functions $f$ and $N$,

\begin{equation}
	\E_N = \frac{\delta L_{N, f}}{\delta N} , \qquad \E_f = \frac{\delta L_{N, f}}{\delta f} \,.
\end{equation}
One can then prove that $\E_N=\E_f =0$ imply that the Einstein equations \req{eq:EinsteinEq} are satisfied, taking into account that $H_{Q}$ solved its own equation \req{eq:MaxwelEq},

So far the analysis is completely general, but typically one would not be able to solve these equations for a generic Lagrangian. For this reason, it is interesting to restrict to a subset of theories, introduced as Electromagnetic Quasitopological gravities (EQG) in \cite{Cano:2020qhy} (in $d+1=4$), that make possible to perform analytic computations. These theories are simply characterized by the condition that
\begin{equation}\label{eq:fcondition}
\frac{\delta L_{N, f}}{\delta f}\Big|_{N=\text{const.}}\equiv 0 \quad \forall\,\, f(r)\, .
\end{equation}
In other words, for these theories the reduced Lagrangian $L_{N, f}$ is a total derivative when $N(r)$ takes a constant value. In the purely gravitational case, this definition gives rise to the Generalized Quasitopological gravities \cite{PabloPablo,Hennigar:2017ego,PabloPablo3,Ahmed:2017jod,Bueno:2019ycr}, which include Quasitopological \cite{Oliva:2010eb,Myers:2010ru,Dehghani:2011vu,Cisterna:2017umf} and Lovelock gravities \cite{Lovelock1,Lovelock2,Padmanabhan:2013xyr} as particular cases. Our construction extends the definition of those theories to include a $(d-2)$-form (or equivalently, a vector field upon dualization), allowing one to study charged black hole solutions. Let us note that the standard two-derivative theory \req{eq:twodertheory} satisfies \req{eq:fcondition} and therefore belongs to the EQG class. In general, all of the theories in this family satisfy a number of properties, which are the same as for their four-dimensional counterparts studied in \cite{Cano:2020qhy}, and that we summarize here. 

\begin{enumerate}
\item The degrees of freedom that propagate in maximally symmetric backgrounds are the same as in the two-derivative theory. This is particularly relevant for the gravitational sector of the theory, since general higher-order gravities typically propagate a massive ghost-like graviton and a scalar mode along with the massless graviton. The condition \req{eq:fcondition} guarantees that these modes are absent on the vacuum. 
\item The theory allows for charged solutions of the form \req{eq:Nfmetric}, \req{eq:magneticH} with $N(r)=N_k=\text{const.}$, \textit{i.e}, characterized by a single function $f(r)$. 
\item The equation for the function $f(r)$, which is obtained from $\E_N\big|_{N=N_k}=0$, can be integrated once, and the integration constant is proportional to the total mass of the spacetime. 
\item For some theories the integrated equation for $f(r)$ is algebraic and hence it can be solved trivially: if this happens, the theory is of the ``Quasitopological'' subclass. Other times the integrated equation is a second order ODE for $f(r)$, and that type of theories is of the ``Generalized Quasitopological'' subclass.
\item In all cases, the thermodynamic properties of charged black holes can be accessed analytically. 
\end{enumerate}

In this paper we will only deal with the Quasitopological class of Lagrangians, which already constitute a quite extensive set, as we show below.

\subsection{Four-derivative EQGs}

Let us begin by classifying the theories belonging to the EQG family at the four-derivative level. There are four types of terms one could include in the Lagrangian at that order, namely, those of the types $R^2$, $R H^2$, $H^4$ and $(\nabla H)^2$, although our interest lies mostly on the first two. 
In the case of quadratic curvature Lagrangians, we know there are three independent densities,

\begin{equation}
\mathcal{L}_{R^2}=\lambda_{1}R^2+\lambda_{2}R_{\mu\nu}R^{\mu\nu}+\lambda_{3}R_{\mu\nu\rho\sigma}R^{\mu\nu\rho\sigma}\, ,
\end{equation}
but there is only one combination of these that satisfies the ``single-function" condition \req{eq:fcondition}: the Gauss-Bonnet density (\textit{i.e.}, the quadratic Lovelock Lagrangian),
\begin{equation}
\mathcal{X}_{4}=R^2-4R_{\mu\nu}R^{\mu\nu}+R_{\mu\nu\rho\sigma}R^{\mu\nu\rho\sigma}\, .
\end{equation}

\noindent
That Lovelock gravity satisfies \req{eq:fcondition} and possesses single-function solutions of the form \req{eq:Nfmetric} is well known \cite{Wheeler:1985nh,Boulware:1985wk,Cai:2001dz,Camanho:2011rj}, so let us turn our attention to the next case. 

Regarding the operators of the form $R H^2$, there are again three of them, that can be written as\footnote{There is a fourth contraction of the form $\tensor{\Hsq}{^{\mu\nu}_{\rho\sigma}} R\indices{_\mu^\rho_\nu^\sigma}$, but it can be checked that this is related to the term multiplied by $\alpha_3$ by means of the Bianchi identity of the Riemann tensor.}

\begin{equation}
\mathcal{L}_{R H^2}=\alpha_1 H^2 R + \alpha_2 \tensor{\Hsq}{^{\mu}_{\nu}} \tensor{R}{^{\nu}_{\mu}} + \alpha_3 \tensor{\Hsq}{^{\mu\nu}_{\rho\sigma}} \tensor{R}{^{\rho\sigma}_{\mu\nu}}\, ,
\end{equation}
where we recall that we are using the notation introduced in Eq.~\req{eq:defH2}.  Evaluating this Lagrangian on \req{eq:Nfmetric} and \req{eq:magneticH} with $N(r)$ equal to a constant value $N_{k}$, we obtain

\begin{equation}
\begin{aligned}
r^{d-1}\mathcal{L}_{R H^2} \Big|_{N_k, f}&=\frac{Q^2 (d-1)!}{ r^{d+1}}\Big[\left(-2 \alpha _3+\alpha _1 (1-d) (d-2)-\alpha _2 (d-2)\right) (f-k)\\
&+f' \left(2 \alpha _1 (1-d) r-\alpha _2 r\right)-\alpha _1 r^2 f''\Big]\, ,
\end{aligned}
\end{equation}
where we included the factor $r^{d-1}$ from the volume element $\sqrt{|g|}$. In order for this Lagrangian to belong to the EQG family we apply the condition \req{eq:fcondition} that tells us that the quantity above should be a total derivative. It is straightforward to compute the functional derivative of this Lagrangian with respect to $f$ and we find that there is a single condition in order for it to vanish identically, 

\begin{equation}
	\alpha_3 = - (2d-1) (d-1) \alpha_1 - (d-1) \alpha_2\, .
	\label{eq:alpha3singlefunction}
\end{equation}

\noindent
Therefore, there are two linearly independent contractions of the form $H^2 R$ that we can add to the two-derivative Lagrangian and maintain single-function solutions.
Moving to the next case, in general dimensions there are two independent operators of the form $H^4$ that do not violate parity, which can be chosen as\footnote{In order to see that there are only two independent terms, it is clearer to work in terms of the two-form $G = \star H$. There are only two inequivalent quartic contractions: $\left(G_{\mu\nu}G^{\mu\nu}\right)^2$ and  $\tensor{G}{_{\mu}^{\nu}}\tensor{G}{_{\nu}^{\alpha}}\tensor{G}{_{\alpha}^{\beta}}\tensor{G}{_{\beta}^{\mu}}$.} 

\begin{equation}
\mathcal{L}_{H^4}=\beta_{1}\left(H^2\right)^2+\beta_2\tensor{\Hsq}{^{\mu}_{\nu}}\tensor{\Hsq}{^{\nu}_{\mu}}.
\end{equation}
When evaluated on \req{eq:Nfmetric} and \req{eq:magneticH} we see that both on-shell densities are independent of $f(r)$ and therefore they both belong to the EQG class straightforwardly. However, it will be enough for our purposes to only keep one of them, as both terms contribute to spherical/planar/hyperbolic black hole solutions in the exact same way. Thus, we will take for simplicity the $(H^2)^2$ operator. 
Finally, we find that there are no terms of the form $(\nabla H)^2$ belonging to the EQG class. 

Therefore, introducing appropriate normalization factors, we have the following four-derivative EQG theory
\begin{equation}
\begin{aligned}
I_{\rm EQG,4} = & \frac{1}{16 \pi G}\int d^{d+1}x\sqrt{|g|} \bigg[ R +\frac{d(d-1)}{L^2}-\frac{2}{(d-1)!}H^2+\frac{\lambda}{(d-2)(d-3)}L^2\mathcal{X}_{4}\\
&+ \frac{2 \alpha_1 L^2}{(d-1)!} \left( H^2R-(d-1)(2d-1)\tensor{R}{^{\mu\nu}_{\rho\sigma}}\tensor{\Hsq}{^{\rho\sigma}_{\mu\nu}}\right) + 
\\
&+ \frac{2 \alpha_2 L^2}{(d-1)!}  \left(\tensor{R}{^{\mu}_{\nu}} \tensor{\Hsq}{^{\nu}_{\mu}}-(d-1)\tensor{R}{^{\mu\nu}_{\rho\sigma}}\tensor{\Hsq}{^{\rho\sigma}_{\mu\nu}}\right)+\frac{\beta L^2}{(d-1)!^2}\Hsq^2\Bigg] \,.
\end{aligned}
\label{eq:EQTfour}
\end{equation}
This is the theory in which we are going to focus in the rest of the paper. Certainly, the most interesting part of it are the non-minimally coupled terms $R H^2$, which have not been considered before in the literature. 

Interestingly, having four independent parameters, this theory is general enough from the point of view of Effective Field Theory. 
As shown by Refs.~\cite{Liu:2008kt,Myers:2009ij}, an EFT extension of Einstein-Maxwell theory (or in our case, Einstein-$(d-2)$-form theory) only requires four independent parity-preserving terms, as the rest of higher-derivative operators can be removed via field redefinitions. We have checked that our Lagrangian above indeed spans this basis of four independent operators, which means that we can capture any parity-preserving four-derivative correction to Einstein-Maxwell theory. It could be particularly interesting to use it to capture the corrections arising from supersymmetric theories in $d=4$ \cite{Hanaki:2006pj,Bobev:2021qxx,Liu:2022sew}. Although five dimensional supergravity theories with higher-derivative corrections also have parity-breaking Chern-Simons terms, that we are not including, it turns out those terms do not affect most (or none) of the results we are going to discuss in this paper. 

There is a crucial difference between our approach and the EFT one, though, which is the fact that we are going to carry out a fully non-perturbative analysis of our theory \req{eq:EQTfour}, while in EFT one is usually restricted to the linear perturbative regime. Of course, one can always recover this perturbative regime from our analytic and exact results by expanding linearly in the couplings.  However, the exact result is clearly more interesting and it could serve as an educated guess for the behavior of these theories and their holographic duals beyond the limited perturbative approach.

Let us close this section by taking note of the electromagnetic dual theory of \eqref{eq:EQTfour}. The fact that we have an $H^4$ term makes it difficult to invert \eqref{eq:FstarH} explicitly, so obtaining a closed expression for the dual action is involved (although perhaps not impossible). However, it is easy to obtain the dual action if we perform a derivative expansion. In that case we can write $H(F)=H_0(F)+H_2(F) L^2+\mathcal{O}(L^4)$, and the inversion of \eqref{eq:FstarH} at each order in $L$ is straightforward. We find that the dual theory, to fourth order in derivatives, reads

\begin{equation}\label{eq:dualEQGfour}
\begin{aligned}
\tilde{I}_{\mathrm{EQT},4}&=\frac{1}{16 \pi G } \int d^{d+1} x \sqrt{\vert g \vert} \Bigg [ R+ \frac{d(d-1)}{L^2}- F^2+ \frac{\lambda }{(d-2)(d-3) } L^2 \mathcal{X}_4 \\&+ \frac{L^2}{d-2} R F^2\left(3d\alpha_{1}+\frac{d\alpha_{2}}{(d-1)}\right) -\frac{2L^2}{d-2}F_{\mu\alpha}\tensor{F}{_{\nu}^{\alpha}}R^{\mu\nu}\left(4(2d-1)\alpha_{1}+\frac{(3d-2)\alpha_{2}}{(d-1)}\right)\\
&+\frac{2L^2}{d-2}F_{\mu\nu}F_{\rho\sigma}\tensor{R}{^{\mu\nu\rho\sigma}}((2d-1)\alpha_{1}+\alpha_{2})+ \frac{\beta}{4} L^2 (F^2)^2 +\mathcal{O}(L^4)\Bigg]\, ,
\end{aligned}
\end{equation}
and it contains an infinite tower of higher-order terms that we could also compute.

We study the black hole solutions of \req{eq:EQTfour} in Section~\ref{sec:BHs} below, but let us first show how analogous versions of this theory exist at higher orders.

\subsection{EQGs at all orders}
\label{subs:eqgsallorders}

It is possible to construct EQGs in any spacetime dimension $D=d+1$ at arbitrary order in the curvature tensor and the field strength. In the case of pure gravity theories, Quasitopological and Generalized Quasitopological at all orders were obtained in Ref.~\cite{Bueno:2019ycr}, so let us focus here in the case of non-minimally coupled theories.  In analogy with the four-dimensional theories identified in \cite{Cano:2020qhy}, we have been able to find the following infinite families of EQGs:
\begin{align}
\label{eq:lagegqa}
\mathcal{L}^{(a)}_{d,n,m}&=\left(2n \tensor{R}{_{\mu}^{\alpha}}\tensor{\delta}{_{\nu}^{\beta}}+g_{d,n,m}\tensor{R}{^{\alpha\beta}_{\mu\nu}}\right)\tensor{\left(R^{n-1}\right)}{^{\mu\nu}_{\rho\sigma}}(H^2)^{\rho\sigma}{}_{\alpha\beta} (H^2)^{m-1}\, ,\\[0.5em]
\mathcal{L}^{(b)}_{d,n,m}&=\left(n R\tensor{\left(R^{n-1}\right)}{^{\mu\nu}_{\rho\sigma}}+\kappa  \tensor{\left(R^{n}\right)}{^{\mu\nu}_{\rho\sigma}}+2n(n-1) \tensor{R}{_{\gamma}^{\mu}} \tensor{R}{^{\beta}_{\rho}}\tensor{\left(R^{n-2}\right)}{^{\gamma\nu}_{\beta\sigma}} \right) (H^2)_{\mu\nu}{}^{\rho\sigma}(H^2)^{m-1}\, ,
\label{eq:lagegqb}
\end{align} 
where
\begin{equation}
g_{d,n,m}=-d(n-1)-2(d-1)m\, , \quad \kappa_{d,n,m}=\frac{g_{d,n,m}}{2}(1-g_{d,n,m}) \, .
\end{equation}
If we evaluate the previous Lagrangians on the ansatz given by \req{eq:Nfmetric} and \req{eq:magneticH} we find:
\begin{align}
\mathcal{L}^{(a)}_{d,n,m}&=\frac{2^{n} Q^{2m} ((d-1)!)^{m}}{r^{2(d-1)m}}\psi_k^{n-1} \left (- n \mathcal{H}_{k,d}+g_{d,n,m} \left ( \frac{k-f}{r^2} \right)\right) \, , \\[0.5em]
\mathcal{L}^{(b)}_{d,n,m} &= \frac{2^{n-2} Q^{2m} ((d-1)!)^{m}}{r^{2(d-1)m}} \psi_{k}^{n-2} \left ( 4 \kappa_{d,n,m} \psi_k^2-2n \left ((d-1) \mathcal{H}_{k,d}+ \mathcal{G}_d\right) \psi+2n(n-1) \mathcal{H}_{k,d}^2   \right) \,,
\end{align}
where
\begin{align}
\psi_{k}&=\frac{k-f}{r^2}\, ,\\[0.3em]
\mathcal{H}_{k,d}&=\frac{-(d-2) k+(d-2)f+r f'}{r^2}+\frac{f N'}{r N}\, , \\[0.3em]
\mathcal{G}_d&=\frac{2(d-1) f N'+4 r f N''+6 r N'f'+N((2(d-1))f'+2r f'')}{2 Nr}\,.
\end{align}

That these theories define truly EQGs can be seen from the fact that the reduced Lagrangians (taking into account the volume element) become a total derivative when evaluated on $N(r)=$ constant. Explicitly, we have
\begin{equation}
\begin{split}
\left. r^{d-1} \mathcal{L}^{(a)}_{d,n,m} \right \vert_{ ds^2_{1,f} , H_Q} &= \frac{d}{dr} \mathcal{I}^{(a)}_{k,d,n,m}\, , \\[0.5em]
\left. r^{d-1} \mathcal{L}^{(a)}_{d,n,m} \right \vert_{ ds^2_{1,f} , H_Q} &= \frac{d}{dr} \mathcal{I}^{(b)}_{k,d,n,m}\, ,
\end{split}
\end{equation}
where
\begin{equation}
\begin{split}
\mathcal{I}^{(a)}_{d,n,m}&=2^n Q^{2m} ((d-1)!)^m r^{d+2m(1-d)} \psi_k^n  \, , \\[0.3em]
\mathcal{I}^{(b)}_{d,n,m}&=2^{n-1}Q^{2m}((d-1)!)^m  r^{d+2m(1-d)} \psi_k^{n-1} \bigg( (1-2m+d(2m+2n-1)) \psi_k+n r \psi_k' \bigg ) \,.
\end{split}
\end{equation}
As a result, we can write infinite examples of EQGs at any order in the curvature and the field strength by considering linear combinations of $\mathcal{L}^{(a)}_{d,n,m}$ and $\mathcal{L}^{(b)}_{d,n,m}$ with arbitrary coefficients. For instance,  the four-derivative theory \eqref{eq:EQTfour} can be expressed as
\begin{equation}
\begin{aligned}
I_{\rm EQG,4} = & \frac{1}{16 \pi G}\int d^{d+1}x\sqrt{|g|} \bigg[ R +\frac{d(d-1)}{L^2}-\frac{2}{(d-1)!}H^2+\frac{\lambda}{(d-2)(d-3)}L^2\mathcal{X}_{4}\\
&+ \frac{2 L^2\alpha_1}{(d-1)!}\mathcal{L}^{(b)}_{d,1,1}+\frac{L^2 \alpha_2}{(d-1)!}\mathcal{L}^{(a)}_{d,1,1}+\frac{\beta L^2}{(-3d+4)(d-1)!^2}\mathcal{L}^{(a)}_{d,0,2}\bigg] \,.
\end{aligned}
\end{equation}

At any order in derivatives, the most general EQG one can write from the theories \eqref{eq:lagegqa} and \eqref{eq:lagegqb} is
\begin{equation}
I_{\mathrm{EQG}, \mathrm{gen}}=\frac{1}{16 \pi G}\int d^{d+1} x \sqrt{\vert g \vert}\left \lbrace  R+\frac{d(d-1)}{L^2}-\frac{2}{(d-1)!}H^2 +\mathcal{L}^{\mathrm{EQG}} \right \rbrace \, ,
\end{equation}
where 
\begin{equation}
\mathcal{L}^{\mathrm{EQG}} =\sum_{n=1}^\infty \sum_{m=1}^\infty L^{2(n+m-1)}\left (\lambda_{n,m} \mathcal{L}_{n,m}^{(a)}+\gamma_{n,m} \mathcal{L}^{(b)}_{n,m} \right )\,.
\end{equation}
One could of course add to these theories the pure (Generalized) Quasitopological theories of Ref.~\cite{Bueno:2019ycr} that we mentioned before. 

Since, by construction, this theory is an EQG, it has solutions of the form \req{eq:Nfmetric} and \req{eq:magneticH} with $N(r)=N_k=$ const. The equation of motion for $f(r)$, after integration on the radial variable, reads
\begin{equation}
\begin{split}
k-f-&\frac{m}{(d-1) r^{d-2}} + \frac{2 Q^2}{(d-1)(d-2) r^{2(d-2)}}+ \frac{r^2}{L^2}+ \\&+\sum_{n,p=1}^\infty \frac{2^{n-1} L^{2(p+n-1)} Q^{2p} ((d-1)!)^p \psi_k^{n-1}}{(d-1)r^{2(d-1)p}}(\alpha_{n,p}k+\beta_{n,p} f)=0\,,
\end{split}
\end{equation}
where $m$ is an integration constant proportional to the mass and we have implicitly defined
\begin{equation}
\begin{split}
\alpha_{n,p}&=  2\lambda_{n,p}+(d-1)(-1+2p+2n)\gamma_{n,p} \, , \\[0.3em]
\beta_{n,p}&=2(n-1)\lambda_{n,p}+(d-1-(4d-2)n+2 d n^2+2p(d-1)(2n-1))\gamma_{n,p}  \,.
\end{split}
\end{equation}
Notice that this equation is algebraic in $f$ and therefore these theories belong to the Quasitopological subclass. However, theories of the Generalized Quasitopological type (with a second order equation for $f(r)$) must exist as well. 
The study of these theories and their black hole solutions as well as their holographic properties is intended to be carried out elsewhere.

\section{AdS vacua and black hole solutions}\label{sec:BHs}
In this section we focus on the solutions of the theory \req{eq:EQTfour}, and we start by determining its AdS vacua. 
As is well-known, the higher-derivative terms modify the length scale of AdS, $\tilde L$, which no longer coincides with the cosmological-constant scale $L$. It is customary to denote 
\begin{equation}
\tilde L=\frac{L}{\sqrt{f_{\infty}}}
\end{equation}
for a dimensionless constant $f_{\infty}$, so that for pure AdS space the Riemann tensor takes the form 
\begin{equation}\label{eq:RiemAdS}
\tensor{R}{^{\mu\nu}_{\rho\sigma}}=-\frac{2 f_{\infty}}{L^2}\tensor{\delta}{^{[\mu}_{[\rho}}\tensor{\delta}{^{\nu]}_{\sigma]}}\, .
\end{equation} 
Taking this into the Einstein equations \req{eq:EinsteinEq}, one finds that $f_{\infty}$ must satisfy

\begin{equation}\label{eq:AdSvacuum}
1-f_{\infty}+\lambda f_{\infty}^2=0\, ,
\end{equation}
which is the well-known result for Gauss-Bonnet gravity \cite{Myers:2010jv}. This polynomial equation has two real roots if $\lambda\le 1/4$, but only one is continuously connected to the Einstein gravity vacuum when $\lambda=0$, and this is
\begin{equation}\label{eq:finfty}
f_{\infty}=\frac{1}{2\lambda}\left[1-\sqrt{1-4\lambda}\right]\, .
\end{equation}
When $\lambda>1/4$ there is no AdS solution, so this is the maximum value $\lambda$ can take.  As corresponding to Lovelock gravity, but also to the complete family of Generalized Quasitopological gravities, the linearized gravitational equations around this vacuum are identical to the linearized Einstein equations, up to the identification of an effective Newton's constant that determines the coupling to matter \cite{Aspects}. In the case of GB gravity, the effective Newton's constant reads 

\begin{equation}\label{eq:Geff}
G_{\rm eff}=\frac{G}{1-2\lambda f_{\infty}}\, .
\end{equation}
Observe that the denominator in this expression is the slope of the AdS vacuum equation \req{eq:AdSvacuum}. This is in fact no accident and the same property holds for all theories with an Einstein-like spectrum \cite{Bueno:2018yzo,CanoMolina-Ninirola:2019uzm}. We also note that $G_{\rm eff}$ is divergent in the limit $\lambda\rightarrow 1/4$, which is known as the critical theory \cite{Crisostomo:2000bb,Fan:2016zfs}.

Let us now obtain the static spherically/plane/hyperbolic-symmetric solutions of \req{eq:EQTfour}. By construction, this theory belongs to the EQG class, and therefore it allows for solutions of the form  \req{eq:Nfmetric} and \req{eq:magneticH}with $N(r)=N_{k}=$ constant. 
As a matter of fact, the equation  $\delta L_{N,f} / \delta f=0$ computed from the reduced Lagrangian implies that $N'(r)=0$, so that these are the only solutions.  Then, we only have to find the function $f(r)$ by solving the equation $\delta L_{N,f} / \delta N |_{N = N_k} = 0$. This equation takes the form of a total derivative --- as it should happen according to the results in  \cite{PabloPablo3} --- and explicitly it reads

\begin{equation}
	\begin{aligned}
	\frac{\delta L_{N,f}}{ \delta N} = & \frac{d}{dr} \Bigg[ (d-1) \frac{r^{d}}{L^2} \left(1- \frac{L^2}{r^2}(f(r)-k)  +\lambda \frac{L^4}{r^4}(f(r)-k)^2 \right)\\
	& + \frac{2 Q^2}{d-2} \frac{1}{r^{d}} \Big(  r^2 + (d-1) (d-2) L^2 \alpha_1 f(r)  + (d-2) k L^2 \big( 3 (d-1) \alpha_1 + \alpha_2 \big) \Big) \Bigg] = 0\, .
	\end{aligned}
	\label{eq:EOMforf}
\end{equation}
We note that the integrated equation is algebraic in $f(r)$, not differential, which characterizes this theory as belonging to the proper Quasitopological subclass.  Let us also remark that in this equation one should take $\lambda=0$ in $d=3$, as in that case the GB invariant does not really contribute to the equations of motion (note that the normalization factor of the GB term in \req{eq:EQTfour} diverges for $d=3$, so the limit $d\rightarrow 3$ would seem to give a finite contribution\footnote{This and similar observations were noted by Ref.~\cite{Glavan:2019inb} to propose a non-trivial $D\rightarrow 4$ limit for GB gravity, but the validity of this approach has been contested \cite{Lu:2020iav,Gurses:2020ofy,Hennigar:2020lsl}.}).
Equating the argument of the derivative to a constant $m$, which will be related to the physical mass of the black hole, and introducing
\begin{equation}
X:=\frac{L^2}{r^2}(f(r)-k)\, ,
\end{equation}
we can rewrite the equation as follows,

\begin{equation}\label{eq:poleqX}
\lambda X^2-\Gamma(r) X+1+\Upsilon(r)=0\, ,
\end{equation}
where

\begin{align}
\Gamma(r)=&1-\frac{2\alpha_1 L^2 Q^2}{r^{2(d-1)}}\, ,\\\notag
\Upsilon(r)=&-\frac{m L^2}{(d-1)r^d}+\frac{2L^2Q^2}{(d-1)(d-2)r^{2(d-1)}}\left(1+k(d-2)\frac{L^2}{r^2}(4(d-1)\alpha_1+\alpha_2)\right)\\
&-\frac{\beta L^4 Q^4}{(3d-4)(d-1)r^{4(d-1)}}\, .
\end{align}
This is simply a quadratic polynomial in $X$ and thus we can solve it straightforwardly obtaining

\begin{equation}\label{eq:fsolEQG}
f(r)=k+\frac{r^2}{2\lambda L^2}\left[\Gamma(r)\pm \sqrt{\Gamma^2(r)-4\lambda(1+\Upsilon(r))}\right]\, .
\end{equation}
We have two roots, that correspond to two solutions connected to different AdS vacua at $r\rightarrow \infty$. We should choose the one that reduces to the Einstein gravity result in the limit $\lambda=0$, and this is the one with the ``$-$'' sign. It is worth noting that, when $\lambda=0$ (which is always the case for $d\le 3$), this solution simply becomes

\begin{equation}
f(r)=k+\frac{r^2(1+\Upsilon(r))}{L^2\Gamma(r)}\, .
\end{equation}
Let us then identify the physical properties of this solution. For $r \rightarrow \infty$, $f(r)$ behaves as

\begin{equation}
	f(r) = f_{\infty}\frac{r^2}{L^2} +k - \frac{m}{(d-1)(1-2\lambda f_{\infty})r^{d-2}} + \cO \left( \frac{1}{r^{2(d-2)}} \right) + \cdots ~,
	\label{eq:Asymptotic for of f(r)}
\end{equation}
where $f_{\infty}$ is given by \req{eq:finfty}. Therefore, it asymptotes to the AdS vacuum that we have determined above. 
On the other hand, the mass $M$ is identified by looking at the following term in the asymptotic expansion of $f$ \cite{Abbott:1981ff,Deser:2002jk,Senturk:2012yi,Adami:2017phg,Altas:2018pkl},

\begin{equation}
-\frac{16\pi G_{\rm eff} M}{(d-1)N_k V_{k,d-1}}\frac{1}{r^{d-2}}\in f(r)\, .
\end{equation}
where $G_{\rm eff}$ is the effective Newton's constant  and the factor $N_k$ takes into account the normalization of the time coordinate at infinity, which is equivalent to a change of units. 
Also note that, in the cases in which the volume of the transverse sections $V_{k,d-1}$ is infinite, one would instead define an energy density $\rho=M/V_{k,d-1}$.

Using the value of $G_{\rm eff}$ given by \req{eq:Geff}, we get that the physical mass of the black hole is

\begin{equation}
	M = \frac{N_k V_{k, d-1}}{16 \pi G} m\, ,
	\label{eq:Physicalmass}
\end{equation}
which is proportional to $m$, as mentioned before. On the other hand, we define the magnetic charge of the $(d-2)$-form $B$ by

\begin{equation}\label{eq:chargedef}
q=\frac{1}{4\pi G}\int_{\mathcal{S}_{d-1}}H\, ,
\end{equation}
where the integral is performed over any spacelike co-dimension two hypersurface $\mathcal{S}_{d-1}$ that encloses $r=0$. Note that, as we discussed around \req{eq:chargedef0}, this quantity is also the electric charge of the dual theory. It is straightforward to see that

\begin{equation}
q=\frac{V_{k,d-1}}{4\pi G} Q\, ,
\label{eq:physcharge}
\end{equation}
and again in the cases $k=0,-1$ one could define instead a charge density $q/V_{k,d-1}$. 

It will also be important for later purposes to determine the electrostatic potential of the dual theory. The field strength of the dual vector field $A_{\mu}$ is obtained according to \req{eq:FstarH}. Evaluating that expression on the metric \req{eq:Nfmetric} and on the $H$-field \req{eq:magneticH}, we find that it corresponds to a pure electric field,

\begin{equation}
	\begin{aligned}
		F = & dt \wedge dr N_k Q\Bigg[ - \frac{1}{r^{d-1}} - \frac{L^2 \alpha_1}{r^{d+1}} \left( 3d(d-1) k -3 d (d-1) f(r) + 2 (d-1) r f'(r) + r^2 f''(r) \right) \\
		& \hspace{1.5cm} - \frac{L^2 \alpha_2}{r^{d+1}} \left( d k - d f(r) + r f'(r) \right) + \frac{L^2 Q^2 \beta}{r^{3(d-1)}} \Bigg] \, .
	\end{aligned}
\end{equation}
Surprisingly, this can be written explicitly as a total derivative, $F_{tr} = -\Phi'(r)$, where

\begin{equation}
	\begin{aligned}
		\Phi(r) = &- N_k Q \Bigg[ \frac{1}{(d-2) r^{d-2}} + \frac{L^2 \alpha_1}{r^d} \left( 3 (d-1) k - 3 (d-1) f(r) - r f'(r) \right) \\
		& \hspace{1cm} + \frac{L^2 \alpha_2}{r^d} \left( k - f(r) \right) - \frac{L^2 Q^2 \beta}{(3d-4) r^{3d-4}} \Bigg]+\Phi_{\infty} \, 
	\end{aligned}
	\label{eq:Electricpotentialgeneral}
\end{equation}
is the electrostatic potential. We are adding an integration constant $\Phi_{\infty}$ that represents the value of the potential at infinity. 

The solution given by \req{eq:fsolEQG} represents a black hole as long as the function $f(r)$ has a zero $f(r_{+})=0$ (which would correspond to a horizon) which is smoothly connected to infinity (this is, there should be no singularities between $r=r_{+}$ and $r\rightarrow \infty$). It is easier to look at the position of the horizon directly from \req{eq:poleqX}. In fact, at the horizon we have $X(r_{+})=-k L^2/r_{+}^2$, and hence we get
\begin{equation}
\lambda \frac{k^2 L^4}{r_{+}^4}+\Gamma(r_{+})\frac{k L^2}{r_{+}^2}+1+\Upsilon(r_{+})=0\, .
\end{equation}
We cannot obtain the value of $r_{+}$ explicitly from this equation, but it is useful to express instead the mass as a function of $r_{+}$ and the charge,

\begin{equation}
\begin{aligned}\label{eq:massrplus}
M=&\frac{N_k V_{k, d-1}}{16 \pi G} \Bigg[(d-1)\Big(kr_{+}^{d-2}+\frac{r_{+}^d}{L^2}+\lambda k^2 L^2 r_{+}^{d-4}\Big)\\
&+\frac{2Q^2}{(d-2)r_+^{d-2}}\left(1+k(d-2)\frac{L^2}{r_+^2}(3(d-1)\alpha_1+\alpha_2)\right)-\frac{\beta L^2 Q^4}{(3d-4)r_+^{3d-4}}\Bigg]\, .
\end{aligned}
\end{equation}

The Hawking temperature of the black hole is given by $T = N_k f'(r_{+}) / 4 \pi$. This can be easily evaluated by differentiating the equation \eqref{eq:poleqX} with respect to $r$ and evaluating at $r_{+}$, which yields
\begin{equation}
	\begin{aligned}
		T = & \frac{N_k}{4 \pi r_+ \left(1 - 2 L^2 Q^2 \alpha_1 r_+^{-2(d-1)} + 2 k L^2 \lambda r_+^{-2} \right) } \Bigg[ \left( (d-2) k + d \frac{r_+^2}{L^2} + (d-4) k^2 \lambda \frac{L^2}{r_+^2} \right) \\
		& - \frac{2 Q^2}{(d-1) r_+^{2(d-1)}} \left(r_+^2 + d k L^2 (3(d-1)\alpha_1 + \alpha_2) \right) + \frac{\beta L^2 Q^4}{(d-1) r_+^{2(2d-3)}} \Bigg] \, .
	\end{aligned}
 	\label{eq:Temperature}
\end{equation}

On the other hand, we must impose the electrostatic potential \eqref{eq:Electricpotentialgeneral} to vanish at the horizon.\footnote{The reason for this is clearer if one works in Euclidean signature, $t=i\tau$: the vector $A=A_{\tau}d\tau$ would be singular at $r=r_{+}$ unless $A_{\tau}(r_{+})=0$.} In this way, the asymptotic value of the potential reads

\begin{equation}
	\Phi_{\infty} = N_k Q \Bigg[ \frac{1}{(d-2) r_+^{d-2}} + \frac{L^2 \alpha_1}{r_+^d} \left( 3 (d-1)k - r_+ \frac{4 \pi T}{N_k} \right) + \frac{L^2 \alpha_2 k}{r_+^d} - \frac{L^2 Q^2 \beta}{(3d - 4) r_+^{3d-4}} \Bigg] \, .
	\label{eq:phiinf}
\end{equation}

Finally, let us compute the entropy of the black hole. This is given by the Iyer-Wald's formula \cite{Wald:1993nt,Iyer:1994ys}

\begin{equation}
	S = -2\pi \int_{\Sigma} d^{d-1} x \sqrt{h} \frac{\partial \Lag}{\partial R_{\mu\nu\rho\sigma}} \epsilon_{\mu\nu} \epsilon_{\rho\sigma}\ ,
\end{equation}
where $h$ is the determinant of the induced metric at the horizon, and $\epsilon_{\mu\nu}$ is the binormal, normalized as $\epsilon_{\mu\nu} \epsilon^{\mu\nu} = -2$. Evaluating this expression, one finds the value of the entropy

\begin{equation}
	S = \frac{r_+^{d-1} V_{k, d-1}}{4 G} \left( 1 + \frac{2 L^2 Q^2 \alpha_1}{r_+^{2d-2}} + \frac{2 L^2 k (d-1) \lambda}{(d-3) r_+^2} \right)\, .
	\label{eq:entropy1}
\end{equation}
We will further discuss the thermodynamic properties of these black holes in Section~\ref{sec:thermo}.

\section{Holographic dictionary}
\label{sec:dictionary}
The family of Electromagnetic Quasitopological gravities introduced in Sec.~\ref{sec:EQGdef} is most naturally written in terms of a $(d-2)$-form field. However, as we saw in Sec.~\ref{subsec:dual}, this $(d-2)$-form can be dualized into a vector field, and hence these theories are actually equivalent to higher-derivative extensions of Einstein-Maxwell theory. While we will perform many computations in the frame of the $(d-2)$-form, their holographic aspects are better understood in terms of the vector field in the ``Maxwell frame".

Vector fields in the bulk of AdS couple to currents in the boundary theory. In our case, we are working with a dimensionless gauge field $A_{\mu}$, but the holographic dictionary actually requires that the vector has dimensions of energy. Thus, the field that couples to the dual current, $J^a$, is not $A_{\mu}$ but rather 
\begin{equation}
\tilde A_{\mu}=\ell_{*}^{-1} A_{\mu}\, ,
\end{equation}
where $\ell_{*}$ is a length scale that should be fixed by the particular duality in each case. Here we do not know what the dual theory is, so we keep $\ell_{*}$ general. This implies that, for instance, the chemical potential in the dual theory is identified as 
\begin{equation}
\mu=\lim_{r\rightarrow\infty}\tilde A_{t}=\lim_{r\rightarrow\infty} \ell_{*}^{-1} A_{t} \, .
\end{equation}

In this section we compute other entries of the holographic dictionary of these theories: the two-point function $\langle JJ \rangle$ and the energy flux after an insertion of $J^a$, which is equivalent to the 3-point function $\langle TJJ \rangle$. We also review the $\langle TT \rangle$ and $\langle TTT \rangle$ correlators. 

Our goal is to study the electromagnetic dual of the four-derivative Electromagnetic Quasitopological theory given by \req{eq:EQTfour}. Observe however that the term $H^4$ will not play any role in this section, since in order to compute $\langle JJ \rangle$  and $\langle TJJ \rangle$ we only need the quadratic terms. Thus, we can ignore the $H^4$ term for all practical purposes. In addition, in this section we do not really need to stick to the EQG family, so out of generality we can consider the action
\begin{equation}
	\begin{aligned}
		I & = \frac{1}{16 \pi G}\int d^{d+1}x\sqrt{|g|} \bigg[ R +\frac{d(d-1)}{L^2}+\frac{\lambda}{(d-2)(d-3)}L^2\mathcal{X}_{4}-\frac{2}{(d-1)!}\tensor{(H^2)}{_{\mu\nu}^{\rho\sigma}}\tensor{Q}{^{\mu\nu}_{\rho\sigma}}\Bigg] \, ,
	\end{aligned}
	\label{eq:EQTfour2}
\end{equation}
where $\tensor{Q}{^{\mu\nu}_{\rho\sigma}}$ contains the three possible couplings at linear order in the curvature,

\begin{equation}\label{QEQT}
\tensor{Q}{^{\mu\nu}_{\rho\sigma}}=\tensor{\delta}{^{[\mu}_{[\rho}}\tensor{\delta}{^{\nu]}_{\sigma]}}\left(1-\alpha_{1}L^2 R\right)-\alpha_{2} L^2 \tensor{R}{^{[\mu}_{[\rho}}\tensor{\delta}{^{\nu]}_{\sigma]}}-\alpha_{3}L^2\tensor{R}{^{\mu\nu}_{\rho\sigma}}\, .
\end{equation}

\noindent
Then, the tensor $\tilde Q$ defined in \req{eq:Qtilde} reads
\begin{equation}\label{QtildeEQT}
\begin{aligned}
\tensor{\tilde Q}{^{\mu\nu}_{\rho\sigma}}=&\tensor{\delta}{^{[\mu}_{[\rho}}\tensor{\delta}{^{\nu]}_{\sigma]}}\left[1 - L^2 R\left(\alpha_{1}+\frac{\alpha_2}{d-1}+\frac{2\alpha_3}{(d-1)(d-2)}\right)\right]\\
&+2L^2\left(\frac{\alpha_2}{d-1}+\frac{4\alpha_3}{(d-1)(d-2)}\right)\tensor{R}{^{[\mu}_{[\rho}}\tensor{\delta}{^{\nu]}_{\sigma]}}-\frac{2\alpha_3}{(d-1)(d-2)}L^2\tensor{R}{^{\mu\nu}_{\rho\sigma}}\, ,
\end{aligned}
\end{equation}
and we can write the dual theory using the inverse of this tensor as
\begin{equation}
	\begin{aligned}
		\tilde{I} & = \frac{1}{16 \pi G}\int d^{d+1}x\sqrt{|g|} \bigg[ R +\frac{d(d-1)}{L^2}+\frac{\lambda}{(d-2)(d-3)}L^2\mathcal{X}_{4} -(\tilde Q^{-1})^{\mu\nu\rho\sigma}F_{\mu\nu}F_{\rho\sigma}\Bigg] \, .
	\end{aligned}
	\label{eq:EQTfourdual}
\end{equation}
The EQG case \req{eq:EQTfour} is then recovered by setting

\begin{equation}\label{eq:alpha3EQG}
\alpha_3 = - (2d-1) (d-1) \alpha_1 - (d-1) \alpha_2\, .
\end{equation}

\subsection{Stress tensor 2- and 3-point functions}
It is a well-known fact that holographic higher-order gravities give rise to a different correlator structure of the dual stress-energy tensor. For the Gauss-Bonnet correction in  \req{eq:EQTfourdual} this effect is well-known \cite{deBoer:2009pn,Camanho:2009vw,Buchel:2009sk}, and thus we only need to quote the results from the literature. 

The 2-point function of the stress-energy tensor in any CFT has the form

\begin{equation}
\langle T_{ab}(x)T_{cd}(0)\rangle=\frac{C_{T}}{|x|^{2d}}\mathcal{I}_{ab,cd}(x)\, ,
\end{equation}
where $\mathcal{I}_{ab,cd}(x)$ is a fixed tensorial structure and $C_{T}$ is the central charge. Holographically, this correlator is determined by studying linearized gravitational fluctuations around the AdS vacuum and evaluating the action on this solution. Now, since the linearized equations of GB gravity are identical to those of Einstein gravity upon a renormalization of Newton's constant,  the value of $C_{T}$ is essentially obtained from the one in GR by replacing $G$ by $G_{\rm eff}$ in  Eq.~\req{eq:Geff}, this is

\begin{equation}\label{eq:CT}
C_{T}=\frac{(1-2\lambda f_{\infty})\Gamma(d+2)}{8(d-1)\Gamma(d/2)\pi^{(d+2)/2}}\frac{\tilde L^{d-1}}{G}\, .
\end{equation}
We recall that $\tilde L=L/\sqrt{f_{\infty}}$ is the AdS radius, where $f_{\infty}$ is given by \req{eq:finfty}.

On the other hand, the 3-point function $\langle TTT \rangle$ in theories that preserve parity is only characterized by three constants \cite{Osborn:1993cr}. The Ward identity of the stress tensor provides a relation between these constants and the central charge $C_{T}$, so only two additional parameters are necessary to determine the 3-point function. These parameters can be chosen to be the coefficients $t_{2}$ and $t_{4}$ that measure the energy fluxes at infinity after an insertion of the stress tensor \cite{Hofman:2008ar}. In fact, the explicit relation between the coefficients $\mathcal{A}$, $\mathcal{B}$, $\mathcal{C}$  of the 3-point function and the parameters $t_{2}$ and $t_{4}$ was found in Ref.~\cite{Buchel:2009sk}.

In holographic Einstein gravity one finds $t_2=t_4=0$, and thus higher-order gravities allow one to explore more general universality classes of dual CFTs. In particular, in Gauss-Bonnet gravity the coefficient $t_2$ is non-vanishing for $d>3$ and it reads \cite{Buchel:2009sk}
\begin{equation}\label{eq:t2GB}
t_{2}=\frac{4 \lambda f_{\infty}}{1-2\lambda f_{\infty}}\frac{d(d-1)}{(d-2)(d-3)}\, .
\end{equation}
On the other hand, $t_4=0$ for the theory \req{eq:EQTfourdual}. A non-vanishing $t_4$ can be achieved by introducing other higher-derivative terms such as Quasitopological \cite{Myers:2010jv} and Generalized Quasitopological gravity \cite{Bueno:2018xqc,Bueno:2020odt}, or more general theories with an Einstein-like spectrum \cite{Li:2019auk}. However, since our focus in this paper is the presence of non-minimally coupled gauge fields, it will be enough to stick to the case of the Gauss-Bonnet correction.

\subsection{Current 2-point function}
In a CFT, the two-point function of any pair of operators is constrained by conformal symmetry up to a proportionality constant. In the case of a current $J^a$, we have

\begin{equation}
\langle J_a(x) J_b(y) \rangle=\frac{C_{J}}{|x-y|^{2(d-1)}}I_{ab}(x-y)\, ,
\end{equation}
where the quantity $I_{ab}(x)$ is defined as
\begin{equation}\label{eq:Iab}
I^{ab}(x)=g^{ab}-2\frac{x^ax^b}{x^2} \, ,
\end{equation}
and the constant $C_{J}$ is the central charge of the current $J$. As a first example, let us compute this constant for a CFT dual to the following theory,

\begin{equation}\label{eq:Fexample}
\begin{aligned}
I_{\rm example}=\frac{1}{16\pi G}\int d^{d+1}x\sqrt{|g|}\bigg\{&R+\frac{d(d-1)}{L^2}-F^2+\epsilon_1 L^2 R F^2+\epsilon_2 L^2R_{\mu\nu}F^{\mu\alpha}\tensor{F}{^{\nu}_{\alpha}}\\
&+\epsilon_3L^2R_{\mu\nu\rho\sigma}F^{\mu\nu}F^{\rho\sigma}\bigg\}\, .
\end{aligned}
\end{equation}
Notice that, in terms of $\tilde A_{\mu}=\ell_{*}^{-1} A_{\mu}$, the Maxwell term in the action can be written as $-\frac{1}{4 g^2}\tilde F^2$, from where we identify the gauge coupling constant $g$,
\begin{equation}
g^{-2}=\frac{\ell_{*}^2}{4\pi G} \, .
\end{equation}
Now, in order to compute $C_J$, we have to consider a small perturbation of $A_{\mu}$ around pure AdS space and to evaluate the action in the corresponding solution with appropriate boundary conditions. Since in this example we do not have a GB term in the action, the AdS curvature is simply
\begin{equation}
\tensor{R}{^{\mu\nu}_{\rho\sigma}}=-\frac{2}{L^2}\tensor{\delta}{^{[\mu}_{[\rho}}\tensor{\delta}{^{\nu]}_{\sigma]}}\, ,
\end{equation}  
and we have the following 
\begin{equation}
F^2-\epsilon_1 L^2 R F^2-\epsilon_2L^2R_{\mu\nu}F^{\mu\alpha}\tensor{F}{^{\nu}_{\alpha}}-\epsilon_3L^2R_{\mu\nu\rho\sigma}F^{\mu\rho}F^{\nu\sigma}\big|_{\text{AdS}}=\left(1+d(d+1)\epsilon_1+d \epsilon_2+2\epsilon_3\right)F^2\, .
\end{equation}
Thus, around pure AdS spacetime, the only effect of the non-minimal couplings is to rescale the gauge coupling constant, so that we get an effective constant that reads

\begin{equation}
g_{\rm eff}^{-2}=g^{-2}\left(1+d(d+1)\epsilon_1+d \epsilon_2+2\epsilon_3\right)\, .
\end{equation}
Therefore, it is already clear that the central charge $C_{J}$ in the theory \req{eq:Fexample} is the same one as in Einstein-Maxwell theory, but replacing $g$ by $g_{\rm eff}$. This yields

\begin{equation}\label{eq:CJ}
C_{J}^{\rm example}=\left(1+d(d+1)\epsilon_1+d \epsilon_2+2\epsilon_3\right)C_{J}^{\rm \ssc EM}\, ,
\end{equation}
where the Einstein-Maxwell central charge $C_{J}^{\rm\ssc EM}$ reads\footnote{This charge is four times that of \cite{Belin:2013uta} to account for the different normalization of the vector field.}

\begin{equation}
C_{J}^{\rm\ssc EM} = \frac{\Gamma(d)}{\Gamma(d/2 - 1)} \frac{\ell_{*}^2\tilde L^{d-3}}{4 \pi^{d/2 + 1} G}\, ,
\end{equation}
and in this case $\tilde L=L$. 
Note that unitarity requires that $C_{J}>0$, which sets a bound on the couplings $\epsilon_i$. 

Let us now turn to the case of interest for this paper, corresponding to the theory for the $(d-2)$-form \req{eq:EQTfour2}, which we expressed in the Maxwell frame in \req{eq:EQTfourdual}. The most difficult aspect of this theory is that it involves computing the inverse of a tensor, $\tensor{\tilde Q}{^{\mu\nu}_{\rho\sigma}}$. However, this can be trivially inverted on an AdS background. On account of the GB term, the AdS radius is in this case is $\tilde L=L/\sqrt{f_{\infty}}$, and when evaluated on the curvature tensor \req{eq:RiemAdS}, both tensors \req{QEQT} and \req{QtildeEQT} take the following value 

 \begin{equation}
\begin{aligned}
\tensor{\tilde Q}{^{\mu\nu}_{\rho\sigma}}=\alpha_{\rm eff}\tensor{\delta}{^{[\mu}_{[\rho}}\tensor{\delta}{^{\nu]}_{\sigma]}}\, ,
\end{aligned}
\end{equation}
where
\begin{equation}
\begin{aligned}\label{eq:Alphaeffectivedef}
\alpha_{\rm eff}=1+f_{\infty}\alpha_{1}d(d+1)+f_{\infty}\alpha_{2}d+2f_{\infty} \alpha_{3}\, .
\end{aligned}
\end{equation}
Thus, the inverse of this tensor is simply

 \begin{equation}
\begin{aligned}
\tensor{(\tilde Q^{-1})}{^{\mu\nu}_{\rho\sigma}}=\frac{1}{\alpha_{\rm eff}}\tensor{\delta}{^{[\mu}_{[\rho}}\tensor{\delta}{^{\nu]}_{\sigma]}}\, .
\end{aligned}
\end{equation}
Therefore, around an AdS vacuum, the quadratic term of the field $\tilde A_{\mu}=\ell_{*}^{-1} A_{\mu}$ in the action \req{eq:EQTfourdual} is given by 

\begin{equation}
\mathcal{L}_{\tilde F^2}=-\frac{1}{4g_{\rm eff}^2}\tilde F^2\, ,\quad g_{\rm eff}^2=\frac{4\pi G}{\ell_{*}^2}\alpha_{\rm eff}\, .
\end{equation}

Following the same logic as in the previous example, we conclude that the central charge $C_{J}$ is the same as for Einstein-Maxwell theory, but rescaled by the constant $\alpha_{\rm eff}$, 

\begin{equation}\label{eq:CJgen}
C_{J}=\frac{C_{J}^{\rm \ssc EM}}{\alpha_{\rm eff}}\, .
\end{equation}
Interestingly, since the duality transformation has the effect of inverting the effective gauge coupling, the combination $\alpha_{\rm eff}$ appears in the denominator rather than in the numerator of $C_{J}$. Thus, the 2-point function can now diverge for finite values of the couplings $\alpha_{i}$ while it vanishes if we take any of these couplings to infinity. In any case, due to unitarity we have to impose the constraint
\begin{equation}
\alpha_{\rm eff}>0\, ,
\end{equation}
which sets a bound on the $\alpha_{i}$ parameters.  For the Electromagnetic Quasitopological gravity \req{eq:EQTfour}, this reduces to the result
\begin{equation}
\begin{aligned}\label{eq:alphaeffEQG}
\alpha^{\rm EQG}_{\rm eff}=1-f_{\infty}\alpha_{1}(3d^2-7d+2)-f_{\infty}\alpha_{2}(d-2)\, .
\end{aligned}
\end{equation}

\subsection{Energy fluxes}\label{sec:energyflux}

We wish now to perform a conformal collider thought experiment as introduced in Ref.~\cite{Hofman:2008ar}. 
Consider a CFT$_{d}$ in flat space $ds^2=-dt^2+ \delta_{ij}dx^i dx^j$ in its vacuum state, that we denote by $|0\rangle$. For future reference, we note that the bulk geometry dual to this CFT in this state is pure AdS in the Poincar\'e patch, expressed as

\begin{equation}\label{eq:AdSxcoord}
	\begin{aligned}
		ds^2=&\frac{\tilde L^2}{z^2}\left[-(dx^{0})^2+\delta_{ij}dx^i dx^j+dz^2\right]\, ,
	\end{aligned}
\end{equation}
with $x^{0}=t$.  We then want to perform an insertion of a current operator of the form $\epsilon_{i}J^i$, where $\epsilon_i$ is a constant polarization tensor, and we wish to obtain the energy flux measured at infinity.  More precisely, we consider an operator of the form
\begin{equation}
 \mathcal{O}_{E}=\int d^dx \epsilon_{i}J^i e^{-i E x^0} \psi(x/\sigma)\, ,
\end{equation}
where  $\psi(x/\sigma)$ is a distribution function that localizes the insertion at $x^{a}=0$ for $\sigma\rightarrow 0$, and $E$ is the energy. In terms of the cartesian coordinates $x^{a}$, the operator for the energy flux in the direction $\vec n$ is given by

\begin{equation}\label{eq:flux-operator}
	\mathcal E\left(\vec n\right)=\lim_{r\to \infty}r^{d-2}\int_{-\infty}^\infty dx^{0}\, \tensor{T}{^{0}_{i}} \left(x^{0},r \vec{n}\right)n^i\, ,
\end{equation}
where $r^{2}\equiv \delta_{ij} x^i x^j$. We are interested in the expectation value for the energy flux after the insertion of the operator  $\mathcal{O}_{E}$,

\begin{equation}\label{eq:flux1}
	\langle\mathcal E\left(\vec n\right)\rangle=\frac{\langle 0|\mathcal{O}_{E}^{\dagger}\mathcal E\left(\vec n\right)\mathcal{O}_{E}|0\rangle}{\langle 0|\mathcal{O}_{E}^{\dagger}\mathcal{O}_{E}|0\rangle} ~.
\end{equation}

By making use of the $O(d-1)$ 
symmetry of the problem, one can then see that the expectation value of this energy flux takes the form \cite{Hofman:2008ar}

\begin{equation}\label{eq:flux}
	\langle\mathcal E\left(\vec n\right)\rangle_{J}=\frac{E}{\Omega_{(d-2)}}\left[1+a_2\left(\frac{|\epsilon\cdot n|^2}{|\epsilon|^2}-\frac{1}{d-1}\right)\right]\ ,
\end{equation}
where  $\Omega_{(d-2)}$ is the volume of the $(d-2)$-sphere of unit radius and $a_{2}$ is a theory-dependent constant. By the construction of $\langle\mathcal E\left(\vec n\right)\rangle$, it is clear that it involves an integrated $\langle TJJ \rangle$ correlator over an integrated two-point function $\langle JJ \rangle$.  As it turns out,  the three-point function $\langle TJJ \rangle$ is constrained by conformal symmetry up to two constants.  The parameter $a_{2}$ is clearly a function of these constants, and the Ward symmetry of the stress-energy tensor provides an additional relation between these and $C_{J}$. Therefore, the 3-point function $\langle TJJ \rangle$ is fully determined by the central charge $C_{J}$ together with the parameter $a_2$. We show the explicit relation in the next section.

Holographically, the energy fluxes can be obtained by evaluating the gravitational action on the background of a shock wave, given by the metric
\begin{equation}\label{eq:shockwave}
	\begin{aligned}
		ds^2=&\frac{\tilde L^2}{u^2}\left[\delta(y^+)\W\left(y^i,u\right)\left(dy^+\right)^2-dy^+dy^-+\sum_{i=1}^{d-2}\left(dy^i\right)^2+du^2\right]\, .
	\end{aligned}
\end{equation}
It is important that the coordinates $(y^a,u)$ are not the same as the original cartesian coordinates $(x^a,z)$ of \req{eq:AdSxcoord}, but related to them according to 

\begin{equation}\label{eq:Coordinatesyintermsofx}
	y^{+}=-\frac{1}{x^{+}}\ , \quad y^{-}\equiv x^{-}-\frac{\sum_{i=1}^{d-2}(x^i)^2}{x^{+}}-\frac{z^2}{x^{+}}\ , \quad y^i\equiv \frac{x^i}{x^{+}}\,, \quad u=\frac{z}{x^{+}}\, ,
\end{equation}
for $i=1, 2, \dots, d-2$, and where $x^{\pm}=x^{0}\pm x^{d-1}$.  We refer to the Refs.~\cite{Hofman:2008ar,Myers:2010jv} for additional details on this construction. 
This metric is a solution of the gravitational field equations if $\W$ satisfies the equation
\begin{equation}\label{eq:WEOM}
	\partial_u^2\W -\frac{d-1}{u}\partial_u \W+\sum_{i=1}^{d-2}\partial_i^2\W=0\, ,
\end{equation}
which holds for Einstein gravity and for general higher-derivative extensions of it \cite{Horowitz:1999gf}.
We are interested in the following solution of the previous equation,

\begin{equation}\label{eq:solW}
	\W(y^i,u)=\frac{\W_0\,u^d}{\left(u^2+\sum_{i=1}^{d-2}(y^{i}-y^{i}_0)^2\right)^{d-1}}\, ,
\end{equation}
where $\W_0$ is a normalization constant and $y^i_0=n^i/(1+n^{d-1})$, where $n^i$ are the components of the vector $\vec{n}$ in the frame described by the coordinates $x^i$, related to $y^i$, $y^+$ and $y^-$ as given in \eqref{eq:Coordinatesyintermsofx}. 

Now, since we want to measure energy fluxes of an excited state, we must consider a perturbation of the vector field $A_{\mu}$ on top of this background. 
In particular, an insertion with the operator \req{eq:flux-operator} is dual to a non-normalizable perturbation of the vector field. Choosing for instance a constant polarization in the $x^1$ direction, this means that we must consider a vector with boundary condition $A_{x^1}\propto z^0 e^{-i E x^0}$ when $z\rightarrow 0$. When extended to the bulk and expressed in the $(y^a,u)$ coordinate system, it is known \cite{Hofman:2008ar} that this kind of perturbation behaves near $y^+=0$ as

\begin{equation}\label{eq:Adelta}
A_{y^1}(y^{+}\approx 0, y^{-},y^i,u)\sim e^{i E y^{-}/2}\delta(y^1)\ldots \delta(y^{d-2})\delta(u-1)\, .
\end{equation}
This will be important later, as the shock wave is localized at $y^{+}=0$ and hence we will eventually have to evaluate $A_{\mu}$ at $y^{+}=0$.

Working directly in terms of the $(y^a,u)$ coordinates, we may simply consider a perturbation of the form
\begin{equation}\label{eq:Ashock}
	A=dy^{1} A_{y^1}+dy^{+}A_{y^{+}}\, .
\end{equation}
The non-vanishing components of its field strength tensor $F_{\mu\nu} = \partial_\mu A_\nu - \partial_\nu A_\mu$ are simply

\begin{equation} 
F_{\mu \nu} = 2\partial_{[\mu|} A_{y^1} \delta^{1}_{|\nu]}+2\partial_{[\mu|} A_{y^+} \delta^{+}_{|\nu]}\, .
\end{equation}
In principle, the dynamics of the field $A$ is determined by the action with higher-order corrections, in the background \eqref{eq:shockwave}. However, if we ignore contact terms (this is, terms of the form $A \W$) in its equations of motion, they reduce simply to Maxwell's equations, 

\begin{equation}
	\nabla_\mu F^{\mu \nu} = 0 ~,
\end{equation}
in the same way that the dual Lagrangian on vacuum AdS is equal to the Maxwell Lagrangian with a modified coupling constant. By imposing the following condition,
\begin{equation}
\partial_{-}A_{y^{+}}=\frac{1}{2}\partial_{y^1}A_{y^1}\, ,
\end{equation}
which ensures that the perturbation is transverse, $\nabla_{\mu}A^{\mu}=0$, the Maxwell equations are reduced to the following equation for $A_{y^1}$
\begin{equation}
	-4\partial_{+}\partial_{-}A_{y^1}+\partial_{u}^2A_{y^1}-\frac{d-3}{u}\partial_{u}A_{y^1}+\sum_{i=1}^{d-2}\partial_{i}^2A_{y^1}=0\, .
\end{equation}
The solution to this equation with the boundary conditions discussed above (note that they are expressed in terms of the $x$ coordinates) then develops the behavior in \req{eq:Adelta}.

In order to compute the energy flux we have to evaluate the on-shell action and extract the piece proportional to $\W A^2$ (since this is the piece in the action that couples to $TJJ$). For our theory \req{eq:EQTfourdual},
this requires us to evaluate first the tensor $\tensor{\tilde Q}{^{\mu\nu}_{\rho\sigma}}$, and then compute the components of its inverse $\tensor{(\tilde Q^{-1})}{^{\mu\nu}_{\rho\sigma}}$ using the relation \req{eq:inverseQ}. The tensor $\tensor{\tilde Q}{^{\mu\nu}_{\rho\sigma}}$ is given by \req{QtildeEQT}, and taking into account that the shockwave \req{eq:shockwave} is an Einstein space satisfying 
\begin{equation}
R_{\mu\nu}=- \frac{d f_{\infty}}{L^2}g_{\mu\nu}\, ,
\end{equation}
we find that

\begin{equation}\label{QtildeEQTOS}
\begin{aligned}
\tensor{\tilde Q}{^{\mu\nu}_{\rho\sigma}}=&\alpha_{\rm eff}\tensor{\delta}{^{[\mu}_{[\rho}}\tensor{\delta}{^{\nu]}_{\sigma]}}-\frac{2\alpha_3}{(d-1)(d-2)}L^2\tensor{W}{^{\mu\nu}_{\rho\sigma}}\, .
\end{aligned}
\end{equation}
Here the constant $\alpha_\text{eff}$ is given by \eqref{eq:Alphaeffectivedef} and $\tensor{W}{^{\mu\nu}_{\rho\sigma}}$ is the Weyl tensor, whose non-vanishing components read

\begin{equation}\label{eq:Components Qtilde shockwave background}
	\begin{aligned}
    \tensor{W}{^{-i}_{+j}} &= \delta(y^+)\frac{f_{\infty}  u}{L^2} \left[ u \partial_i \partial_j \W - \delta^i_j \partial_u \W \right]\, , \\
    \tensor{W}{^{-i}_{u+}} & = \tensor{W}{^{u-}_{+i}} = -\delta(y^+)\frac{f_{\infty}}{L^2} u^2 \partial_i \partial_u \W\, , \\
	\tensor{W}{^{u-}_{u+}} &= \delta(y^+)\frac{f_{\infty} u}{L^2} \left[ u \partial_i \partial_i \W - (d-2) \partial_u \W \right]\, ,
	\end{aligned}
\end{equation}
plus those obtained interchanging indices. These expressions have been simplified by using the equation of motion \eqref{eq:WEOM}, since we will use them to evaluate the on-shell action. We note that, as corresponding to a wave, the Weyl tensor satisfies

\begin{equation}
\tensor{W}{^{\mu\nu}_{\rho\sigma}}\tensor{W}{^{\rho\sigma}_{\alpha\beta}}=0\, ,
\end{equation}
and therefore, the inverse of $\tilde Q$ simply reads

\begin{equation}\label{QtildeOsinvs}
\begin{aligned}
\tensor{(\tilde Q^{-1})}{^{\mu\nu}_{\rho\sigma}}=&\frac{1}{\alpha_{\rm eff}}\tensor{\delta}{^{[\mu}_{[\rho}}\tensor{\delta}{^{\nu]}_{\sigma]}}+\frac{2\alpha_3}{\alpha_{\rm eff}^2(d-1)(d-2)}L^2\tensor{W}{^{\mu\nu}_{\rho\sigma}}\, .
\end{aligned}
\end{equation}
We are then ready to evaluate the on-shell action \req{eq:EQTfourdual}. Since we are only interested in the piece of the form  $\W A^2$, we only need to compute the following term,
\begin{equation}
\begin{aligned}
\tilde{I}_{W A^2} & = -\frac{1}{16 \pi G}\int d^{d+1}x\sqrt{|g|}\tensor{(\tilde Q^{-1})}{^{\mu\nu\rho\sigma}}F_{\mu\nu}F_{\rho\sigma}\\
&=-\frac{1}{16 \pi G}\int d^{d+1}x\sqrt{|g|} \bigg[\frac{1}{\alpha_{\rm eff}}F^2+\frac{2\alpha_3 L^2}{\alpha_{\rm eff}^2(d-1)(d-2)}W^{\mu\nu\rho\sigma}F_{\mu\nu}F_{\rho\sigma}\Bigg] \, .
\end{aligned}
\end{equation}
Since the only component of the inverse metric that depends on $\W$ is $g^{--}$, we have
\begin{equation}
F^2=2(F_{-1})^2g^{--}g^{11}+\ldots=-\frac{8f_{\infty}^2 u^4 \delta(y^{+})\W}{L^4}(\partial_{-}A_{y^1})^2+\ldots\, ,
\end{equation}
where the ellipsis denote terms that do not depend on $\W$ and therefore are irrelevant for this computation. On the other hand, we have

\begin{equation}
W^{\mu\nu\rho\sigma}F_{\mu\nu}F_{\rho\sigma}=4W^{- 1 - 1}(F_{- 1})^2=-\delta(y^+)\frac{8 f_{\infty}^3  u^6}{L^6} \left[\partial_1^2 \W - \frac{1}{u}\partial_u \W \right](\partial_{-}A_{y^1})^2\, .
\end{equation}

\noindent
Then, putting these two contributions together and integrating by parts, we find

\begin{equation}\label{eq:Wphi2}
\begin{aligned}
\tilde{I}_{W A^2} & = -\frac{1}{4\pi G\alpha_{\rm eff}}\int du d^{d}y\frac{\tilde L^{d-3}}{u^{d-3}}\delta(y^{+})\W A_{y^1}\partial_{-}^2 A_{y^1}\left[1+\frac{2f_{\infty}\alpha_{3}}{\alpha_{\rm eff}(d-1)(d-2)}T_2\right]\, ,
\end{aligned}
\end{equation}
where we defined 

\begin{equation}
T_2 = \frac{u(u \partial_1 \partial_1 \W - \partial_u \W)}{\W}\, .
\end{equation}

\noindent
Since the shock wave localizes the integral to $y^+=0$, and since $A_{y^1}$ behaves as in \req{eq:Adelta}, we have to evaluate the integrand at $u = 1$ and $y^i = 0$, which can be done in a straightforward manner by plugging in the solution for $\W$ \eqref{eq:solW}. Taking into account that the perturbation in \req{eq:Ashock} has a polarization $\epsilon=(\epsilon_{1},0,\ldots,0)$, we have the following value of $T_2$,
\begin{equation}
T_2 \Big|_{u = 1, y^i = 0} = d (d-1) \left( n_1^2 - \frac{1}{d-1} \right) =  d (d-1) \left( \frac{|\epsilon\cdot n|^2}{|\epsilon|^2} - \frac{1}{d-1} \right)\, .
\end{equation}
Therefore, comparing the expressions of the energy flux \eqref{eq:flux} and the on-shell action \eqref{eq:Wphi2}, we immediately read off the coefficient $a_2$,

\begin{equation}\label{eq:valueofa2}
a_2 = \frac{2 d \alpha_3 f_\infty }{(d-2)\alpha_{\rm eff}} = \frac{2 d \alpha_3 f_\infty}{(d-2)(1+f_{\infty}\alpha_{1}d(d+1)+f_{\infty}\alpha_{2}d+2f_{\infty} \alpha_{3})}\, ,
\end{equation}
where we have made use of \req{eq:Alphaeffectivedef}. 
In the case of EQG, given by the action \req{eq:EQTfour}, this result reduces to

\begin{equation}\label{eq:a2EQG}
	a_2^{\rm EQG} =  - \frac{2 d (d-1) \left( (2d-1) \alpha_1 + \alpha_2 \right) f_\infty}{(d-2) ( 1 - (3d^2 - 7d + 2) f_\infty \alpha_1 -(d-2) f_\infty \alpha_2)}\, .
\end{equation}

\subsection{Three-point function $\langle TJJ \rangle$}
The three point correlator $\langle TJJ\rangle$ in position space in a CFT is constrained by conformal symmetry to have the form \cite{Osborn:1993cr,Erdmenger:1996yc}
\be
    \langle T_{ab}(x_1)J_c(x_2)J_d(x_3)\rangle = \frac{t_{abef}(X_{23})g^{eg}g^{fh}I_{cg}(x_{21})I_{dh}(x_{31})}{|x_{12}|^d |x_{13}|^d |x_{23}|^{d-2} }\ ,
\ee
where $I_{ab}(x)$ is the structure introduced in Eq.~\req{eq:Iab} and

\be
   \begin{aligned}
    t_{abcd}(X^a) &= \hat a h^{(1)}_{ab}(\hat{X}^a)g_{cd} + \hat b h^{(1)}_{ab}(\hat{X}^a)h^{(1)}_{cd}(\hat{X}^a) + \hat c h^{(2)}_{abcd}(\hat{X}^a) + \hat e h^{(3)}_{abcd} \, , \\
    h^{(1)}_{ab}(\hat{X}^a) &= \hat{X}_a\hat{X}_b - \frac{1}{d}g_{ab} \, , \\
    h^{(2)}_{abcd}(\hat{X}^a) &= 4\tensor{\hat{X}}{_{(a}}\tensor{g}{_{b)(d}}\tensor{\hat{X}}{_{c)}}-\frac{4}{d}\hat{X}_a\hat{X}_b g_{cd} - \frac{4}{d}\hat{X}_c\hat{X}_dg_{ab} + \frac{4}{d^2}g_{ab}g_{cd} \, , \\
        h^{(3)}_{abcd} &= g_{ac}g_{bd} + g_{ad}g_{bc} - \frac{2}{d} g_{ab}g_{cd} \, , \\
   \end{aligned}
   \label{quantitiesTJJ}
\ee
where we also have 
\be
    x_{12}^a = x_1^a - x_2^a \, ,\quad X_{23}^a = \frac{x_{13}^a}{|x_{13}|^2} - \frac{x_{23}^a}{|x_{23}|^2} \, ,\quad \hat{X}_{12}^a = \frac{X_{12}^a}{|X_{12}|}\, ,
\ee
and so on with their corresponding permutations. This expression depends on four theory-dependent constants $\hat a$, $\hat b$, $\hat c$, and $\hat e$. However, only two of them are free parameters because of the following constraints coming from current conservation:
\be
    d\hat a - 2\hat b + 2(d-2)\hat c = 0,\qquad \hat b - d(d-2)\hat e = 0\, .
\ee
Following Ref.~\cite{Belin:2013uta}, we will work in terms of $\hat c$ and $\hat e$. In addition, there is one Ward identity that relates the central charge $C_{J}$ to these coefficients, namely, 

\be\label{eq:CJed}
    C_J = \frac{2\pi^{d/2}}{\Gamma\left(\frac{d+2}{2}\right)}(\hat c+\hat e)\, .
\ee
This reduces the number of independent parameters to just one, and this one can be related to the coefficient $a_2$ entering into the expectation value of the energy flux \req{eq:flux}. As is clear from Eq.~\req{eq:flux1}, this flux involves an integrated $\langle TJJ\rangle$ correlator, and therefore it is a somewhat straightforward (but tedious) field theory computation to obtain the desired relationship. This was performed in general dimensions by Ref.~\cite{Chowdhury:2012km}, finding the result\footnote{The final result offered by Ref.~\cite{Chowdhury:2012km} (their formula (6.14)) has minus sign with respect to the value we show here. However, we have reviewed their computations and we believe that this sign is a typo. Also, our formula here coincides with the value of $a_2$ for the $d=4$ case provided by Ref.~\cite{Hofman:2008ar}. }

\be \label{eq:a2ed}
    a_2 = \frac{(d-1)((d-2)d\hat e-\hat c)}{(d-2)(\hat c+\hat e)}\, .
\ee
In this way, we can fully determine the 3-point function $\langle TJJ\rangle$ from $C_{J}$ and $a_2$. Inverting the two equations above we can indeed write 
\begin{align}
\hat c&=\frac{C_{J}(d-2)\Gamma\left(\frac{d+2}{2}\right)}{2\pi^{d/2}(d-1)^3}\left(d(d-1)-a_2\right)\, ,\\
\hat e&=\frac{C_{J}\Gamma\left(\frac{d+2}{2}\right)}{2\pi^{d/2}(d-1)^3}\left(d-1+(d-2)a_2\right)\, .
\end{align}
Finally, using the values of $a_2$ and $C_J$ found for our EQGs, given by Eqs.~\req{eq:a2EQG} and \req{eq:CJgen} respectively, we have
\begin{equation}\label{eq:TJJEQG}
\begin{aligned}
\hat{c}^{\rm EQG} & = \frac{d (d-2) d! L^{d-3} \ell_{*}^2 \left[ d - 2 - (d-1) (3d^2 - 10d + 2) f_\infty \alpha_1 - (d^2 - 4d + 2) f_\infty \alpha_2 \right]}{2^5 (d-1)^2 \pi^{d+1} f_\infty^{(d-3)/2} G \left[ 1 - (d-2) f_\infty \left( (3d-1) \alpha_1 + \alpha_2 \right) \right]^2}\, ,\\[0.5em]
\hat{e}^{\rm EQG} & = \frac{d (d-2) (d-2)! L^{d-3} \ell_*^2 \left[ 1 - (d-1) (7d - 2) f_\infty \alpha_1 - (3d - 2) f_\infty \alpha_2 \right]}{2^5 (d+1) \pi^{d+1} f_\infty^{(d-3)/2} G \left[ 1 - (d-2) f_\infty \left( (3d-1) \alpha_1 + \alpha_2 \right) \right]^2}\, .
\end{aligned}
\end{equation}
This result will be important for us in Section~\ref{sec:renyi}.

\section{Causality, unitarity and weak-gravity-conjecture constraints}\label{sec:constraints}
The theory \req{eq:EQTfour}, which is going to be the focus of our holographic explorations for the rest of the paper, depends on four free parameters. As we have seen in the previous section, these parameters modify several entries of the holographic dictionary allowing us to probe more general universality classes of holographic CFTs than those covered by Einstein-Maxwell theory. 
However, they are not completely free, as one must demand that the hypothetical dual theory satisfies reasonable physical properties, such as unitarity. Thus, we must determine the allowed values of these parameters if we want to obtain any sensible answers from holography.  

\subsection{Unitarity in the boundary}
In the boundary theory, several constraints are found by demanding that the different correlators and energy fluxes defined in the previous section respect unitarity.

There is an even more fundamental condition that our theory must satisfy: the existence of an AdS vacuum. From Eq.~\req{eq:finfty}, which determines the AdS scale $\tilde L=L/\sqrt{f_{\infty}}$ we see that this happens if
\begin{equation}
\lambda\le \frac{1}{4}\, ,\quad (d>3)
\end{equation}
which we take into account from the start. 

\subsubsection*{Constraints from $\langle TT\rangle$ and $\langle TTT\rangle$}
One first condition comes from demanding that the central charge of the stress-tensor two-point function be positive, $C_{T}>0$. This is also directly interpreted as a unitarity condition in the bulk, as it is equivalent to imposing $G_{\rm eff}>0$, hence preventing the graviton from having a negative energy. In the presence of the Gauss-Bonnet term, the central charge is given by \req{eq:CT}, and therefore we must impose

\begin{equation}
1-2\lambda f_{\infty}>0\, .
\end{equation}
One can see this is always satisfied for all the allowed values of $\lambda$, $\lambda<1/4$, and therefore this condition does not provide additional constraints. 

On the other hand, a stronger bound is achieved by demanding positivity of the energy 1-point function. Analogously to what we saw in Sec.~\ref{sec:energyflux}, the expectation value of the energy flux produced after an insertion of the stress-energy tensor $\epsilon_{ij}T^{ij}$ in general reads \cite{Hofman:2008ar}

\begin{equation}\label{eq:fluxT}
\langle\mathcal E\left(\vec n\right)\rangle_{T}=\frac{E}{\Omega_{(d-2)}}\left[1+t_{2}\left(\frac{\epsilon^{*}_{ij}\epsilon_{il}n^jn^l}{\epsilon^{*}_{ij}\epsilon_{ij}}-\frac{1}{d-1}\right)+t_{4}\left(\frac{|\epsilon_{ij}n^{i}n^{j}|}{\epsilon_{ij}^{*}\epsilon_{ij}}-\frac{2}{d^2-1}\right)\right]\ .
\end{equation}
For holographic CFTs dual to \req{eq:EQTfour}, we have $t_4=0$ while $t_2$ is given by \req{eq:t2GB}.
The energy flux must be positive in any direction $n^i$ and for any choice of polarization $\epsilon_{ij}$. These conditions were analyzed by Ref.~\cite{Buchel:2009sk} in general dimensions, finding that $\lambda$ is bound to the following interval,

\begin{equation}\label{eq:lambdaconstr}
-\frac{(3d + 2) (d - 2)}{4 (d + 2)^2} \leq \lambda \leq \frac{(d - 2) (d - 3) (d^2 - d + 6)}{4 (d^2 - 3d + 6)^2}\, .
\end{equation}
We note that $\lambda=1/4$ is not allowed by the upper bound in any dimension, while the lower bound prevents $\lambda$ to become too negative.

\subsubsection*{Constraints from $\langle JJ\rangle$ and $\langle TJJ\rangle$}
\label{sec:TJJconstr}
The unitarity constraints on the Gauss-Bonnet coupling were known since Refs.~ \cite{deBoer:2009pn,Camanho:2009vw,Buchel:2009sk}. Let us now discuss the novel constraints on the parameters $\alpha_1$ and $\alpha_2$ of the non-minimally coupled terms. These work very similarly to the gravitational case and follow from the unitarity of $\langle JJ\rangle$ and the energy one-point function. 

The central charge of the current two-point function is given by Eq.~\req{eq:CJgen}, and, as we already discussed there, its positivity implies that 
\begin{equation}\label{eq:CJconstr}
\alpha_{\rm eff}^{\rm EQG}=1-f_{\infty}\alpha_{1}(3d^2-7d+2)-f_{\infty}\alpha_{2}(d-2)>0\, .
\end{equation}
Again, since this quantity is, up to a constant, the coupling constant of the Maxwell field, its positivity is equivalent to demanding that photons carry positive energy in the bulk. 

We can obtain more interesting bounds from the energy flux created after an insertion of the current operator, given by \req{eq:flux}. Demanding that the energy flux is positive in any direction, we find that the parameter $a_2$ must satisfy
\begin{equation}
-\frac{d-1}{d-2}\le a_{2}\le d-1\, ,
\end{equation}
where the upper bound comes from $\vec{n}\perp \vec{\epsilon}$ and the lower bound from $\vec{n} \propto \vec{\epsilon}$. Using the value of $a_2$ for our Electromagnetic Quasitopological gravities, given by \req{eq:a2EQG}, this translates into 

\begin{equation}\label{eq:a2EQGv2}
	-1\le  - \frac{2 d  \left( (2d-1) \alpha_1 + \alpha_2 \right) f_\infty}{ 1 - (3d^2 - 7d + 2) f_\infty \alpha_1 -(d-2) f_\infty \alpha_2} \le d-2\, .
\end{equation}
Now, since the denominator of this expression is precisely $\alpha_{\rm eff}^{\rm EQG}$, which is assumed to be positive, by multiplying the whole inequality by $\alpha_{\rm eff}^{\rm EQG}$ we can express the two constraints as follows

\begin{align}\label{eq:a2c1}
1-(7d^2-9d+2) f_{\infty}\alpha_1-(3d-2) f_{\infty}\alpha_2&\ge 0\, ,\\\label{eq:a2c2}
1-\frac{(d-1)(3d^2-14d+4)}{(d-2)} f_{\infty}\alpha_1-\frac{d^2-6d+4}{(d-2)} f_{\infty}\alpha_2&\ge 0\, .
\end{align}
One should not forget to impose \req{eq:CJconstr} together with these constraints. We note that the last inequality has a different character depending on the dimension: the coefficient of $\alpha_1$ is positive for $d=3,4$ and negative for $d\ge 5$, while that of $\alpha_2$ is positive for $d=3,4,5$ and changes sign for $d\ge 6$. For instance, if $\alpha_2=0$ we find that $\alpha_1$ must lie within the interval
\begin{equation}\label{eq:a1alone}
\begin{aligned}
-\frac{(d-1)(-3d^2+14d-4)}{(d-2)}&\le f_{\infty}\alpha_{1}\le \frac{1}{7d^2-9d+2}\, , \quad d=3,4\, ,\,\, \alpha_2=0\, ,\\
\end{aligned}
\end{equation}
but the lower bound disappears for $d\ge 5$. 

\begin{figure}[t!]
	\centering
	\begin{subfigure}{0.49\linewidth}
		\includegraphics[width=\linewidth]{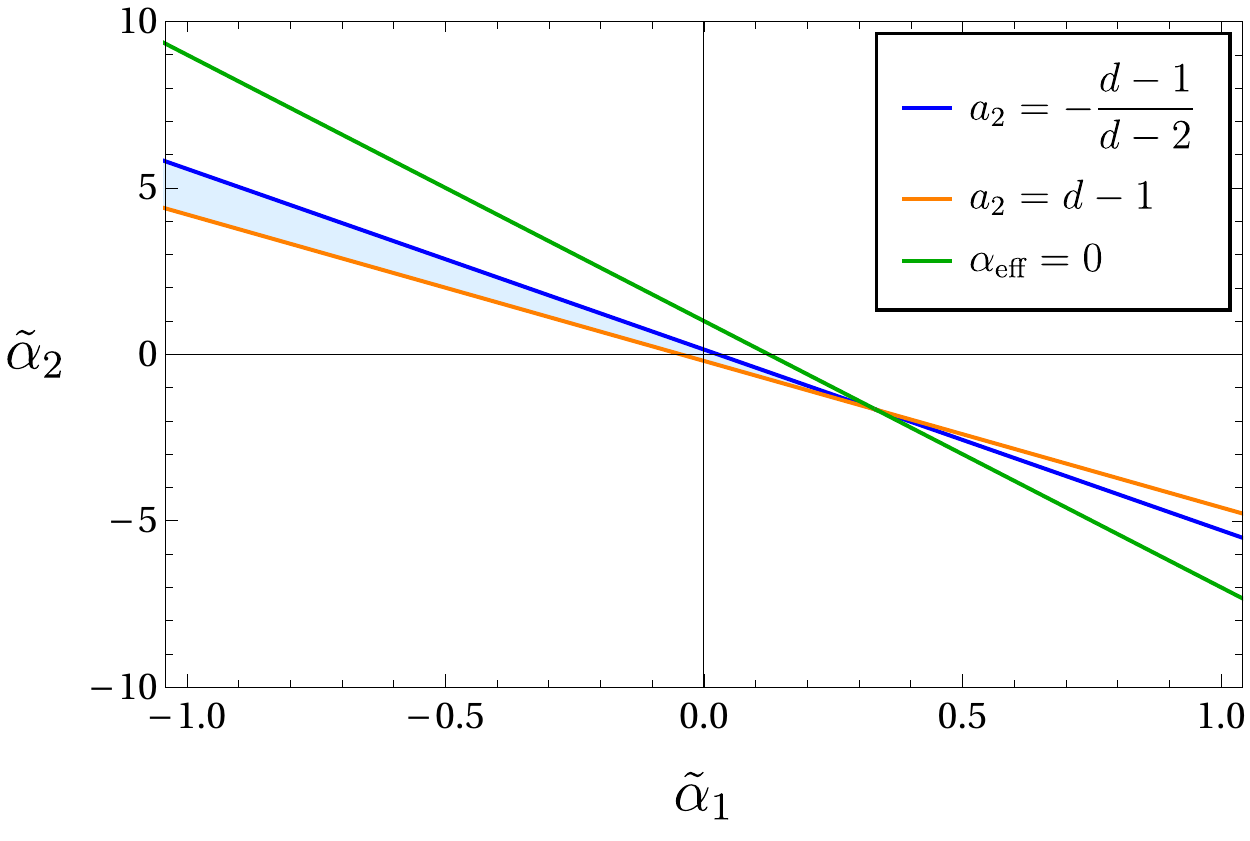}
		\caption{$d = 3$}
		
	\end{subfigure}
	\centering
	\begin{subfigure}{0.49\linewidth}
		\includegraphics[width=\linewidth]{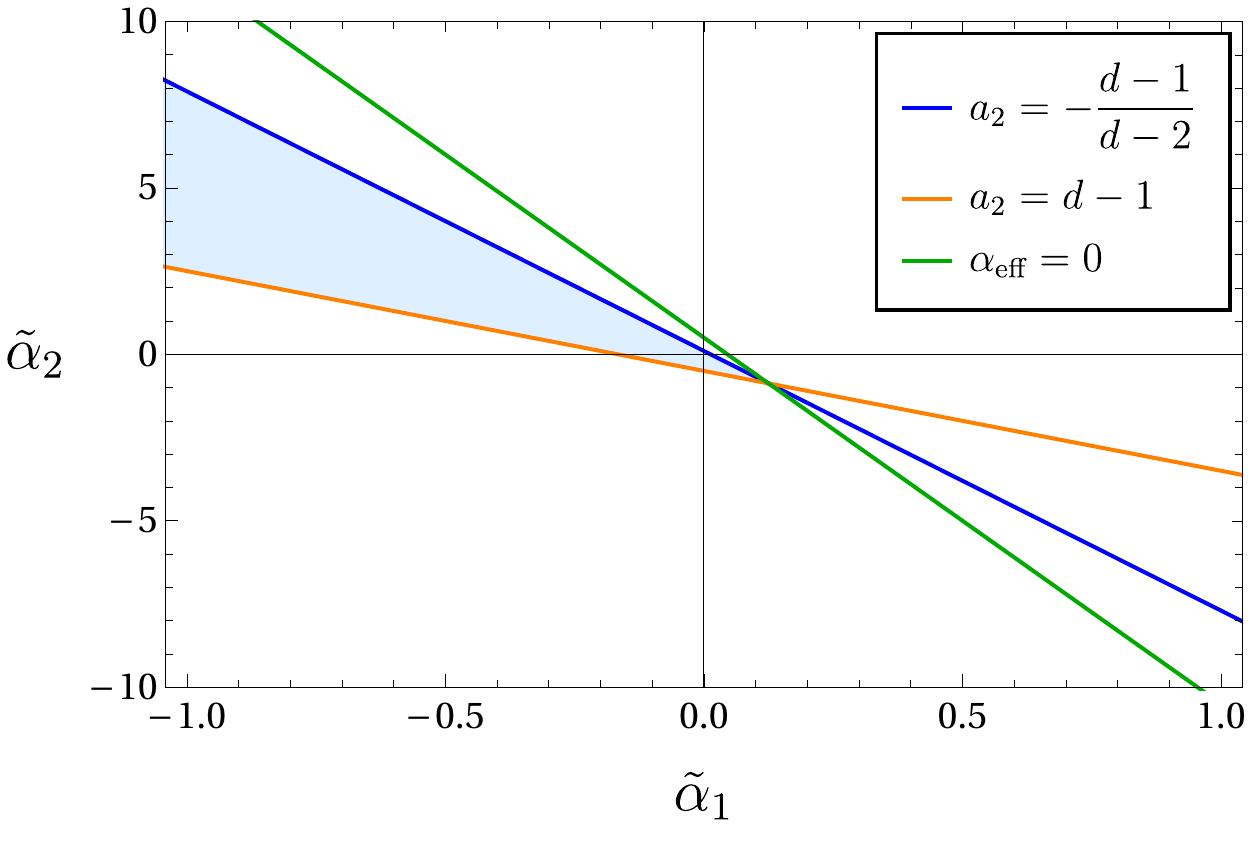}
		\caption{$d = 4$}
	\end{subfigure}
	\centering
	\begin{subfigure}{0.49\linewidth}
		\includegraphics[width=\linewidth]{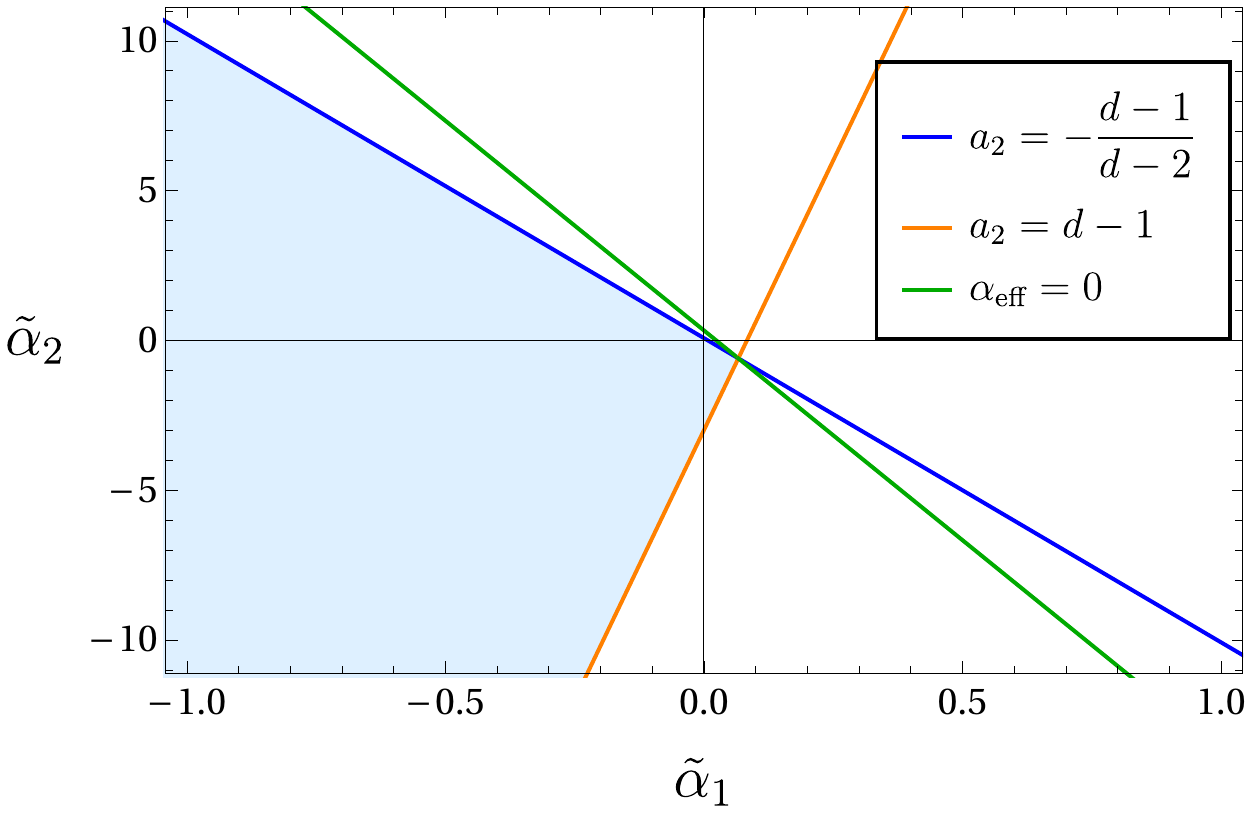}
		\caption{$d = 5$}
	\end{subfigure}
	\centering
	\begin{subfigure}{0.49\linewidth}
		\includegraphics[width=\linewidth]{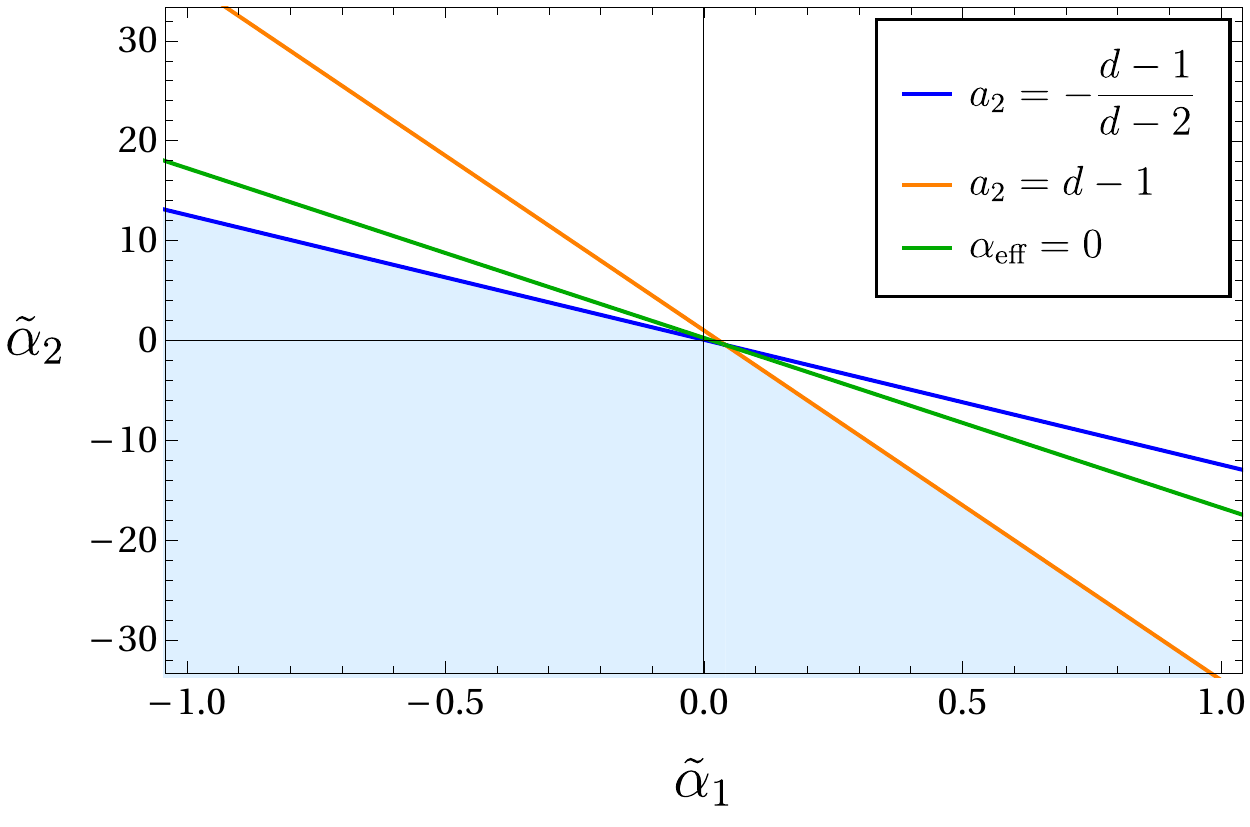}
		\caption{$d = 6$}
	\end{subfigure}
	\caption{Bounds in the constants $\tilde{\alpha}_1 = f_\infty \alpha_1$ and $\tilde{\alpha}_2 = f_\infty \alpha_2$ obtained from unitarity and positivity of energy fluxes, given by \req{eq:CJconstr}, \req{eq:a2c1} and \req{eq:a2c2}, in different dimensions. The allowed region in each case is shaded in blue, and is infinite. For any $d > 6$ the allowing region looks qualitatively similar to that obtained for $d = 6$.}
	\label{fig:Boundsalpha12}
\end{figure}

We note that the bounds are imposed directly on the renormalized couplings $f_{\infty} \alpha_{1,2}$ rather than on the original couplings. However, observe that the value of $f_{\infty}$ is always close to one for the allowed values of $\lambda$ in \req{eq:lambdaconstr} (and it is one in $d=3$). In Fig.~\ref{fig:Boundsalpha12} we show the different constraints and the allowed region in the $(f_{\infty}\alpha_1,f_{\infty}\alpha_2)$ plane. We see that the permitted region grows bigger with the dimension. A very interesting property is that, in $d=3,4,5$, there is an absolute upper bound for $\alpha_1$, regardless of the value of $\alpha_2$. This value is found at the intersection of the three constraints and it reads

\begin{equation}\label{eq:a1abs}
f_{\infty}\alpha_1\le \frac{1}{d(d-2)}\, ,\quad (d=3,4,5)\, .
\end{equation}
Likewise, there is an absolute lower bound for $\alpha_2$ in $d=3,4$:

\begin{equation}\label{eq:a2abs}
f_{\infty}\alpha_2\ge -\frac{2d-1}{d(d-2)}\, ,\quad (d=3,4)\, .
\end{equation}
For higher dimensions, these parameters can take values in the full real line, but interestingly they cannot both be too positive. In fact, only very small values are allowed in that case, as follows from the graph (d) in Fig.~\ref{fig:Boundsalpha12}. 

\subsection{Causality in the bulk}
On general grounds, it is to be expected that physically consistent bulk theories give rise to consistent dual CFTs, and vice-versa. Hence, the unitarity constraints we have discussed must also have a meaning in the bulk. In the case of the constraints coming from the two-point functions $\langle TT\rangle$ and $\langle JJ\rangle$, the interpretation is direct, as the positivity of the central charges is related to that of the energy of gravitational and electromagnetic waves in the bulk. 
However, the bulk interpretation of the constraints coming from the positivity of the energy one-point function is more subtle. At least in the case of Lovelock gravity, it is known that demanding $\langle\mathcal E\left(\vec n\right)\rangle_{T}\ge 0$ is equivalent to enforcing the bulk theory to respect causality  \cite{Camanho:2009vw,Buchel:2009tt,Camanho:2009hu,Camanho:2013pda}, in the sense that one avoids superluminal propagation of gravitational waves \cite{Brigante:2007nu,Brigante:2008gz,Ge:2008ni,Buchel:2009tt}.\footnote{This connection is less clear in other theories outside the Lovelock family \cite{Hofman:2009ug,Myers:2010jv}.}
Here we investigate the analogous connection between causality of electromagnetic waves and positivity of $\langle\mathcal E\left(\vec n\right)\rangle_{J}$, given by \req{eq:flux}.

Our starting point is a neutral planar black hole solution of the theory \req{eq:EQTfour}, with a metric

\begin{equation}\label{eq:fmetricplanar}
	ds^2 = - \frac{f(r)}{f_{\infty}} dt^2 + \frac{dr^2}{f(r)} + \frac{r^2}{L^2} dx_{(d-1)}^2 \,,
\end{equation}
where the function $f(r)$ is given by

\begin{equation}
	f(r) = \frac{r^2}{2 \lambda L^2} \left( 1 - \sqrt{1 - 4 \lambda + \frac{4 \lambda L^2 m}{(d-1) r^d}} \right)\, .
\end{equation}
Note that this is the metric \req{eq:Nfmetric} in which we have set $N_{0}^2=1/f_{\infty}$, so that the speed of light at the boundary is one. 
In order to study the speed of electromagnetic waves in the theory \req{eq:EQTfour}, we can either use its formulation in terms of the $(d-2)$-form $B$ or in terms of the dual vector $A$ --- the result will be independent of the frame employed. Let us consider then a perturbation of the $(d-2)$-form in this black hole background. At linear order, the equation for $B$ can be written as 
%
%
\begin{equation}
	\nabla_{\alpha_1} \left( \tensor{\tilde{Q}}{^{[\mu\nu}_{\mu\nu}}H^{\alpha_1 \ldots \alpha_{d-1}]} \right) = 0\, ,
	\label{eq:EOMH}
\end{equation}
where $\tilde{Q}\indices{^{\mu\nu}_{\rho\sigma}}$ is the tensor introduced in \eqref{QtildeEQT}. 
Particularized to the EQG case, this tensor reads

\begin{equation}\label{QtildeEQT2}
\begin{aligned}
\tensor{\tilde Q}{^{\mu\nu}_{\rho\sigma}}=&\tensor{\delta}{^{[\mu}_{[\rho}}\tensor{\delta}{^{\nu]}_{\sigma]}}\left[1+ \left(\frac{3d\alpha_{1}}{d-2}+\frac{d\alpha_{2}}{(d-1)(d-2)}\right) L^2 R \right]\\
&-2\left(\frac{4(2d-1)\alpha_{1}}{d-2}+\frac{(3d-2)\alpha_{2}}{(d-1)(d-2)}\right) L^2 \tensor{R}{^{[\mu}_{[\rho}}\tensor{\delta}{^{\nu]}_{\sigma]}}+\frac{2}{d-2}((2d-1)\alpha_{1}+\alpha_{2})L^2\tensor{R}{^{\mu\nu}_{\rho\sigma}}\, .
\end{aligned}
\end{equation}
When evaluated on the metric \req{eq:fmetricplanar}, it takes the form

\begin{equation}
\tensor{\tilde Q}{_{\mu\nu}^{\rho\sigma}}=\gamma_{1} \tensor{\rho}{_{[\mu}^{[\rho}}\tensor{\rho}{_{\nu]}^{\sigma]}}+2 \gamma_{2} \tensor{\rho}{_{[\mu}^{[\rho}}\tensor{\sigma}{_{\nu]}^{\sigma]}}+\gamma_{3} \tensor{\sigma}{_{[\mu}^{[\rho}}\tensor{\sigma}{_{\nu]}^{\sigma]}}\, ,
\label{eq:Qspheric}
\end{equation}
where 

\begin{equation}
\tensor{\rho}{^{\alpha}_{\beta}}=\tensor{\delta}{^{\alpha}_{t}}\tensor{\delta}{^{t}_{\beta}}+\tensor{\delta}{^{\alpha}_{r}}\tensor{\delta}{^{r}_{\beta}}\, ,\quad 
\tensor{\sigma}{^{\alpha}_{\beta}}=\sum_{i=1}^{d-1}\tensor{\delta}{^{\alpha}_{i}}\tensor{\delta}{^{i}_{\beta}}\, ,
\end{equation}
are the projectors in the $(t,r)$ and transverse directions, and $\gamma_{1,2,3}$ are the following functions,
\begin{align}
	\begin{split}
		\gamma_1 = 1 & - \frac{\alpha_1 L^2}{r^2} \left[ 3 d (d-1) f - 2 (d-1) r f' - r^2 f'' \right] - \frac{\alpha_2 L^2}{r^2} \left[ d f - r f' \right]\, ,
		\label{eq:gamma1}
	\end{split} \\
	\begin{split}
		\gamma_2 = 1 & - \frac{\alpha_1 L^2}{r^2} \left[ (3d^2 - 11d + 4) f + 2 d r f'(r) - r^2 f'' \right]  \\
		& - \frac{\alpha_2 L^2}{2 (d-1) r^2} \left[ 2 (d^2 - 4d + 2) f + (d + 1) r f' - r^2 f'' \right]\, ,
		\label{eq:gamma2}
	\end{split}\\
	\begin{split}
		\gamma_3 = 1 & - \frac{\alpha_1 L^2}{(d-2) r^2} \left[ (d-5) (3d^2 - 10d + 4) f + 2 (3d^2 - 11d + 4) r f' + 3 d r^2 f'' \right] \\
		& - \frac{\alpha_2 L^2}{(d-1) (d-2) r^2} \left[ (d-3) (d^2 - 6d + 4) f + 2 (d^2 - 4d + 2) r f' + d r^2 f'' \right]\, .
		\label{eq:gamma3}
	\end{split} 
\end{align}

Now, let us consider the following fluctuation of $B$, with a polarization orthogonal to $x^{1}$, 

\begin{equation}\label{eq:Bansatz}
	B = \psi(r) e^{-i \omega t + i k x^1} dx^2 \wedge \cdots \wedge dx^{d-1}\, .
\end{equation}
Its field strength $H = d B$ is given by

\begin{equation}
	\begin{aligned}
		H = e^{-i \omega t + i k x^1} \Big( & \psi'(r) dr \wedge dx^2 \wedge \cdots \wedge dx^{d-1} - i \omega dt \wedge dx^2 \wedge \cdots \wedge dx^{d-1} \\
		& + i k dx^1 \wedge dx^2 \wedge \cdots \wedge dx^{d-1} \Big)\, ,
	\end{aligned}\label{eq:Hformwave}
\end{equation}
and one can see that, with this ansatz, the equations of motion \eqref{eq:EOMH} are reduced to a single component (corresponding to the indices $\alpha_2 \ldots \alpha_{d-1} = x^{2}\ldots x^{d-1}$), so that we do not need to activate other components of $B$. Since we want to study the small wavelength limit $\omega , k \rightarrow \infty$ we only need to keep the derivatives with respect to $t$ and $x^{1}$. Under this approximation, we get

\begin{equation}
\nabla_{\alpha} \left( \tensor{\tilde{Q}}{^{[\mu\nu}_{\mu\nu}}H^{\alpha x^{2} \ldots x^{d-1}]} \right)\propto \frac{L^{2(d-2)}}{r^{2(d-2)}}\left(-\frac{f_{\infty}}{f(r)}(i\omega)^2\gamma_{2}+\frac{L^2}{r^{2}}(ik)^2\gamma_{1}\right)B^{x^{2}\ldots x^{d-1}}\, ,
\end{equation}

\noindent
and hence we get the following dispersion relation

\begin{equation}\label{disp1}
\frac{\omega^2}{k^2}=\frac{\gamma_{1}L^2f(r)}{\gamma_{2} f_{\infty} r^2}\, .
\end{equation}
If we expand this near infinity, we obtain 

\begin{equation}
	\frac{\omega^2}{k^2} = 1 - \frac{L^2 m \left[ 1 - (7d^2 - 9d + 2)f_{\infty}\alpha_1 - (3d - 2) f_{\infty}\alpha_2 \right]}{(d-1) (2 - f_\infty) \alpha_\text{eff}^\text{EQG} r^d} + \mathcal{O} \left( \frac{1}{r^{2d}} \right)\, .
\end{equation}
Now, this is the phase velocity (squared) of the wave front, and consistency with causality requires that it be smaller than the speed of light, $\omega / k \leq 1$.
Since  $f_{\infty}<2$ and we take $\alpha_\text{eff}^\text{EQG} > 0$, the condition $\omega / k \leq 1$ implies

\begin{equation}\label{eq:causality1}
	1 - (7d^2 - 9d + 2)f_{\infty}\alpha_1 - (3d - 2) f_{\infty}\alpha_2 \geq 0\, ,
\end{equation}
which matches precisely the constraint \eqref{eq:a2c1} computed from the lower bound in the allowed range of values of $a_2$. Now, playing with several values of the parameters that respect this bound, it appears that no other constraints are necessary: once \req{eq:causality1} is satisfied, then $\omega^2/k^2\le 1$ everywhere inside the bulk.  A more thorough of these causality constraints deeper in the bulk interior would be convenient, though. 

We can obtain different constraints by choosing inequivalent polarizations for the $B$ field. This means that we have to consider a $B$ field which is polarized along the $r$ direction. However, since the physical constraints on the Maxwell frame are the same, it is simpler to just study a perturbation of the dual vector field $A_{\mu}$ of the form

\begin{equation}\label{eq:Aansatz}
	A = \phi(r) e^{-i \omega t + i k x^2} dx^1\, .
\end{equation}
One can see that the $H$ form obtained by dualizing this vector is not of the form \req{eq:Hformwave}, and in particular it has a term $\sim k dt\wedge dr\wedge dx^{3}\wedge\ldots\wedge dx^{d-1}$, indicating polarization of $B$ along the $r$ direction. The (linearized) modified Maxwell equations for this vector read

\begin{equation}
	\nabla_\mu \left[ \tensor{(\tilde{Q}^{-1})}{^{\mu\nu}_{\rho\sigma}}F^{\rho\sigma} \right] = 0\, ,
	\label{eq:EOMF}
\end{equation}
where $\tensor{(\tilde{Q}^{-1})}{^{\mu\nu}_{\rho\sigma}}$ is the inverse of the tensor in Eq.~\req{eq:Qspheric}. One can see the inverse is simply given by

\begin{equation}
\tensor{(\tilde{Q}^{-1})}{_{\mu\nu}^{\rho\sigma}}=\frac{1}{\gamma_{1}} \tensor{\rho}{_{[\mu}^{[\rho}}\tensor{\rho}{_{\nu]}^{\sigma]}}+\frac{2}{\gamma_{2}} \tensor{\rho}{_{[\mu}^{[\rho}}\tensor{\sigma}{_{\nu]}^{\sigma]}}+\frac{1}{\gamma_{3}} \tensor{\sigma}{_{[\mu}^{[\rho}}\tensor{\sigma}{_{\nu]}^{\sigma]}}\, .
\label{eq:Qsphericinvs}
\end{equation}
Now, the Maxwell equations for the ansatz in Eq.~\req{eq:Aansatz} are reduced to the single component $\nu=x^{1}$, which reads

\begin{equation}
\left(\frac{\omega^2L^2f_{\infty}}{r^2f \gamma_2}-\frac{k^2L^4}{r^4\gamma_{3}}\right)\phi+\frac{(d-1)f L^2}{\gamma_{2}r^3}\phi'+\frac{d}{dr}\left(\frac{f L^2\phi'}{r^2\gamma_2}\right)=0\, .
\end{equation}
Thus, in the short-wavelength limit we get

\begin{equation}\label{eq:omegak2}
\frac{\omega^2}{k^2}=\frac{L^2f(r)\gamma_{2}}{r^2f_{\infty}\gamma_{3}}\, ,
\end{equation}
and expanding this near infinity we have 

\begin{equation}
	\begin{aligned}
		\frac{\omega^2}{k^2}  = 1 - \frac{L^2 m \left[ d - 2 - (d - 1) (3d^2 - 14d + 4) f_{\infty}\alpha_1 - (d^2 - 6d + 4) f_{\infty}\alpha_2 \right]}{(d-1) (d-2) (2 - f_\infty) \alpha_\text{eff}^\text{EQG} r^d} + \mathcal{O} \left( \frac{1}{r^{2d}} \right)\, ,
	\end{aligned}
\end{equation}
where we plugged in the values of $\gamma_2$ and $\gamma_3$ given in \eqref{eq:gamma2} and \eqref{eq:gamma3}. In order for this perturbation to not violate causality, it is necessary that $\omega^2 / k^2 \leq 1$ as we move away from the boundary, and therefore, we obtain the constraint 

\begin{equation}\label{eq:causality2}
	d - 2 - (d - 1) (3d^2 - 14d + 4) f_{\infty}\alpha_1 - (d^2 - 6d + 4) f_{\infty}\alpha_2 \geq 0\, ,
\end{equation}
which is precisely the condition obtained by looking at the upper bound in the value of $a_2$, given in Eq.~\eqref{eq:a2c2}. By looking at the behavior of \req{eq:omegak2} in the bulk for several choices of the parameters, we seem to find that, whenever Eq.~\req{eq:causality2} is satisfied, then $\omega^2/k^2\le 1$ everywhere. However, it would again be interesting to perform a more thorough analysis in this regard. 

One can be convinced that there are no other inequivalent polarizations by counting the number of them captured by \req{eq:Bansatz} and \req{eq:Aansatz}. If we fix the direction of propagation, \req{eq:Bansatz} is the only possible  $B$ field orthogonal to the direction of propagation and with no $t$ and $r$ components, while there are $d-2$ polarizations of the type \req{eq:Aansatz} for $A$ obtained by exchanging $dx^{1}$ with $dx^{i}$, $i\neq 2$. In total we have $d-1=D-2$ different polarizations, which is the number of degrees of freedom of a massless vector field (and of a $(D-3)$-form) in $D$ dimensions.  Therefore, we conclude that there are no additional constraints from causality in the background of a neutral black brane.

It would be interesting to study as well the case of charged black branes, which would indeed be relevant if one wishes to perform holography in such backgrounds. In that case, gravitational and electromagnetic perturbations are linearly coupled, making the analysis of the speed of propagation a bit more involved. However, this could perhaps lead to even stronger constraints than the ones we have derived. 

Finally, let us note that there are other types of causality violations, like the ones found in Ref.~\cite{Camanho:2014apa} involving the graviton three-point vertex. One of the implications of that paper in the holographic context is that the Gauss-Bonnet coupling (in units of the AdS scale) must be very small: $|\lambda|<<1$. These bounds would be applicable in principle to any higher-order gravity that modifies the three-point function structure of Einstein gravity, but let us note that there are non-trivial higher-curvature terms that do not modify this three-point function, and one could not apply these results to them. In any case, we do not know of similar constraints for the $RH^2$ and $H^4$ terms in our theory \req{eq:EQTfour}. As a matter of fact, there are theories, as QCD, that have a large value of $a_2$, and in order to capture these holographically one needs bulk theories with non-minimal higher-derivative terms with $\sim{O}(1)$ couplings, as noted in \cite{Hofman:2008ar}.

\subsection{WGC and positivity of entropy corrections}\label{sec:WGC}
So far, we have been able to constrain three of the four parameters of our theory \req{eq:EQTfour} by imposing unitarity of the boundary theory, which is equivalent to causality in the bulk theory. However, the parameter $\beta$ is still unconstrained as it does not affect any 2- or 3-point function. Also, the existing constraints basically prevent the couplings from becoming too large, but they do not say anything about the sign of these parameters. Interestingly enough, additional constraints can be found by applying the mild form of the weak gravity conjecture (WGC) \cite{Cheung:2018cwt,Hamada:2018dde}, which has recently received a lot of attention in the context of higher-derivative theories \cite{Bellazzini:2019xts,Charles:2019qqt,Loges:2019jzs,Cano:2019oma,Cano:2019ycn,Andriolo:2020lul,Loges:2020trf,Cano:2021tfs,Arkani-Hamed:2021ajd,Aalsma:2021qga,Cano:2021nzo}.  In the case of AdS spacetime, the implications of the WGC were recently studied in Ref.~\cite{Cremonini:2019wdk} --- see also \cite{McInnes:2020jnm,McInnes:2021frb,McInnes:2022tut}. One of the heuristic ideas behind the WGC is that extremal black holes should be able to decay. This will happen if there exists a particle whose charge-to-mass ratio is larger than the one of an extremal black hole, which is the standard form of the WGC \cite{Arkani-Hamed:2006emk,Harlow:2022gzl}. However, the mild form involves only black holes and essentially it claims that the decay of an extremal black hole into a set of smaller black holes should be possible, at least from the point of view of energy and charge conservation. Since extremal black holes have a fixed mass for a given value of the charge, $M_{\rm ext}(Q)$, such decay process is only possible if
\begin{equation}\label{eq:MextWGC}
M_{\rm ext}(Q_1+Q_2)\ge M_{\rm ext}(Q_1)+M_{\rm ext}(Q_2)\, .
\end{equation}
For asymptotically flat black holes in Einstein-Maxwell theory we have $M_{\rm ext}(Q)\propto |Q|$, so the inequality above is saturated. On the other hand, higher-derivative corrections will modify the charge-mass relation, and by demanding that the deviations respect the property \req{eq:MextWGC} one obtains a constraint on the coefficients of the higher-derivative operators. In all cases, one can see that, in order to preserve \req{eq:MextWGC}, the corrections to the extremal mass must be negative, $\delta M_{\rm ext}<0$ \cite{Kats:2006xp}.

In Anti-de Sitter space, however, things work differently. As noted in \cite{Cremonini:2019wdk}, the bound \req{eq:MextWGC} is no longer saturated for extremal AdS black holes, and hence perturbative (arbitrarily small) higher-derivative corrections cannot violate it.\footnote{Note however, that in the limit of small size, AdS black holes behave as asymptotically flat ones, and in that limit \req{eq:MextWGC} could still be applied to constrain the higher-derivative corrections.} Instead, that reference makes use of the proposal of Ref.~\cite{Cheung:2018cwt} that the corrections to the entropy of black holes of arbitrary charge and mass should be positive as long as those black holes are thermodynamically stable. It is known \cite{Goon:2019faz} that, when applied to near-extremal black holes, the positivity of corrections to the entropy is connected to the negativity of the corrections to the extremal mass (see also \cite{McPeak:2021tvu}). Therefore, one can still use the condition $\delta M_{\rm ext}<0$ to bound the higher-order coefficients, just like in the asymptotically flat case.  However, the conditions studied in \cite{Cremonini:2019wdk} are more ambitious, as they demand $\delta S>0$ for arbitrary charge and mass (as long as the specific heats are positive), not only for near-extremal black holes.  Let us work out these conditions for our theory \req{eq:EQTfour}.

The Wald entropy of static black holes was computed in \eqref{eq:entropy1}, which we reproduce here for convenience,

\begin{equation}
	S = \frac{r_+^{d-1} V_{k, d-1}}{4 G} \left( 1 + \frac{2 L^2 Q^2 \alpha_1}{r_+^{2d-2}} + \frac{2 k L^2 (d-1) \lambda}{(d-3) r_+^2} \right) ~.
\end{equation}
This expression together with the relation \req{eq:massrplus} give us the exact value of the entropy $S(M,Q)$. However, here we only need the perturbative correction to the entropy at fixed charge and mass. It is useful to introduce the variable
\begin{equation}
x=\frac{r_{+}^{(0)}}{L}\, ,
\end{equation}
where $r_{+}^{(0)}$ is the zeroth-order value of the radius, which is obtained implicitly from \req{eq:massrplus} by setting to zero the higher-order terms. We also note that the extremal value of the charge in the two-derivative theory reads

\begin{equation}
Q_{\rm ext}^{(0)}=(Lx)^{d-2}\sqrt{\frac{d-1}{2}}\sqrt{d x^2+k(d-2)}\, ,
\end{equation}
and thus let us introduce the variable
\begin{equation}
\xi=\frac{Q}{Q_{\rm ext}^{(0)}}\, ,
\end{equation}
that ranges from $0$ to $1$. Since we are working at fixed $M$ and $Q$, the equation \req{eq:massrplus} allows us to obtain the correction to the horizon radius,
\begin{equation}
r_{+}=r_{+}^{(0)}+r_{+}^{(1)}+\ldots\, ,
\end{equation}
where the first-order correction reads

\begin{equation}
r_{+}^{(1)}=\frac{k^2 \lambda  L}{\left(\xi ^2-1\right) x \left((d-2) k+d x^2\right)}+\frac{3 \alpha_1 (d-1) k L \xi ^2}{\left(\xi ^2-1\right) x}+\frac{\alpha_2 k L \xi ^2}{\left(\xi ^2-1\right) x}-\frac{\beta (d-1) L \xi ^4 \left((d-2) k+d x^2\right)}{4 (3 d-4) \left(\xi ^2-1\right) x}\, .
\end{equation}
Inserting this into our expression for the entropy, we get the following shift at linear order,

\begin{equation}
\begin{aligned}
\delta S(M,Q)&= \frac{(d-1)L^{d-1}x^{d-3} V_{k, d-1}}{4 G} \Bigg[k \lambda  \left(\frac{k}{\left(\xi ^2-1\right) \left((d-2) k+d x^2\right)}+\frac{2}{d-3}\right)\\
&+\alpha_1 \xi ^2 \left(\frac{k \left(\xi^2(d-2)+2d-1\right)}{\xi ^2-1}+d x^2\right)+\frac{\alpha_2 k \xi ^2}{\xi ^2-1}-\frac{\beta (d-1) \xi ^4 \left((d-2) k+d x^2\right)}{4 (3 d-4) \left(\xi ^2-1\right)}\Bigg] \, .
\end{aligned}\label{eq:dSWGCgen}
\end{equation}
According to \cite{Cremonini:2019wdk}, we should then demand this correction to be positive for any black hole that is thermodynamically stable at zeroth order. Let us focus on spherically symmetric black holes $k=1$. The $k=0$ case is obtained as the limit of large size of spherical black holes, while the $k=-1$ case is somewhat different and we will comment on it below. We can consider first neutral black holes, $\xi=0$, in whose case only the Gauss-Bonnet term is relevant,  

\begin{equation}
\begin{aligned}
\delta S(M,Q)\big|_{\xi=0}&= \frac{(d-1)L^{d-1}x^{d-3} V_{1, d-1}}{4 G}  \lambda  \left(-\frac{1}{(d-2) +d x^2}+\frac{2}{d-3}\right)\, .
\end{aligned}
\end{equation}
The variable $x$ can range between $0$ and infinity, and for any of these values the quantity between parenthesis is positive for $d\ge 3$.\footnote{For $d=3$ one should redefine $\hat\lambda=\lambda/(d-3)$ and take the limit $d\rightarrow 3$ with fixed $\hat\lambda$. The correction to the entropy is topological and identical for any spherical black hole.} Now, neutral large black holes are known to be stable in AdS, and therefore, the WGC would imply that the GB coupling must be non-negative,
\begin{equation}
\lambda\ge 0\, .
\end{equation}
This actually makes sense, as the Gauss-Bonnet density arises explicitly from string theory effective actions and in many instances\footnote{However, Ref.~\cite{Bobev:2021qxx} showed that a negative $\lambda$ can also be achieved, indicating that $\lambda$ can actually have different signs depending on the setup.} this indeed has a positive coupling \cite{Metsaev:1987zx,Bergshoeff:1989de,Bachas:1999um,Schnitzer:2002rt} --- see also \cite{Cheung:2016wjt} and the discussion in the appendix B of \cite{Buchel:2008vz}. Next, we can look at the case of (near-) extremal black holes, which are also stable in the two-derivative theory. This corresponds to the limit $\xi\rightarrow 1$, and hence we get

\begin{equation}
\begin{aligned}
\delta S(M,Q)\big|_{\xi\rightarrow 1}&= \frac{(d-1)L^{d-1}x^{d-3} V_{1, d-1}}{4 G(1-\xi^2)} \Bigg[-\frac{\lambda}{(d-2) +d x^2}-3(d-1)\alpha_1 -\alpha_2\\
&+\frac{\beta (d-1) \left((d-2) +d x^2\right)}{4 (3 d-4)}\Bigg]\, .
\end{aligned}
\end{equation}
This correction has a non-trivial dependence on the radius of the black hole, and therefore imposing that it be positive implies several constraints on the coupling constants. For large black holes, the $\beta$ correction dominates and $\delta S\ge 0$ implies

\begin{equation}
\beta\ge 0\, .
\end{equation}
On the other hand, in the limit of small black holes $x\rightarrow 0$ we have

\begin{equation}\label{eq:wgc2}
-\frac{\lambda}{d-2}-3(d-1)\alpha_1-\alpha_2+\frac{\beta(d-1)(d-2)}{4(3d-4)}\ge 0\, .
\end{equation}
This is arguably the most reliable constraint we can produce from the WGC, as small black holes behave as asymptotically flat ones, and one recovers the argument of Eq.~\req{eq:MextWGC}. The condition above implies that the shift in the extremal mass is negative hence ensuring that \req{eq:MextWGC} is satisfied for black holes much smaller than the AdS scale. 

Finally, another interesting condition comes from large black holes $x\rightarrow\infty$ (or equivalently, black branes, $k=0$), of arbitrary charge. In that case we have
\begin{equation}
\begin{aligned}
\delta S(M,Q)\Big|_{x\rightarrow\infty}&= \frac{d(d-1)L^{d-1}x^{d-1} V_{1, d-1}}{4 G} \Bigg[\alpha_1  \xi ^2+\frac{\beta (d-1) \xi ^4 }{4 (3 d-4) \left(1-\xi ^2\right)}\Bigg]\, ,
\end{aligned}
\end{equation}
and in order for this quantity to remain positive for any value of $\xi\in [0,1)$, we must impose not only $\beta\ge0$, but also
\begin{equation}
\alpha_1\ge 0\, .
\end{equation}
This is a very powerful constraint, since, when combined with the unitarity bounds shown in Fig.~\ref{fig:Boundsalpha12}, it implies that $\alpha_1$ and $\alpha_2$ can only lie in a small compact set of the plane for $d=3,4,5$. The Gauss-Bonnet coupling is also bound to a small interval $0\le\lambda \leq \frac{(d - 2) (d - 3) (d^2 - d + 6)}{4 (d^2 - 3d + 6)^2}$, so only $\beta$ can take arbitrarily high values with the current constraints. It would be interesting to investigate whether different constraints could impose an upper bound on $\beta$. The results from our next section suggest indeed that $\beta$ should not be too large. 

Before closing this section, let us discuss what happens if one attempts to enforce the WGC bounds on hyperbolic black holes as well. For simplicity, we can consider neutral black holes, $\xi=0$. One can check that all of these solutions are thermally stable in the two-derivative theory, and therefore one should impose $\delta S\ge 0$. From \req{eq:dSWGCgen} we obtain

\begin{equation}
\begin{aligned}
\delta S(M,Q)\Big|_{k=-1,\xi=0}&= \frac{(d-1)L^{d-1}x^{d-3} V_{-1, d-1}}{4 G} (- \lambda)  \left(\frac{1}{d x^2-(d-2) }+\frac{2}{d-3}\right) \, ,
\end{aligned}
\end{equation}
and since hyperbolic black holes have $d x^2-(d-2)\ge 0$, the positivity of $\delta S$ implies in this case that $\lambda\le 0$, which is the opposite that what we found for spherical and planar black holes.\footnote{In $d=3$, Ref.~\cite{Bobev:2021oku} already noticed that the correction to the entropy associated to the GB term cannot have a definite sign, since one can have black holes of different topologies.} In principle, these constraints should hold at the same time for any choice of boundary geometry, since the dual CFT is always the same. However, this would lead to the conclusion that $\lambda=0$, which seems an unreasonably strong constraint.  Likewise we find similar stringent bounds on the other couplings if we combine the cases $k=1$ and $k=-1$. We do not know how to resolve this issue, but we feel more inclined to trust the constraints for spherical black holes, and ignore those for $k=-1$. On the one hand, spherical black holes make direct connection with the original motivation of the WGC regarding black hole evaporation, while the evaporation of a hyperbolic black hole is probably a meaningless problem (they are always stable). On the other, as we mentioned above, a positive GB coupling $\lambda>0$ is actually realized in many explicit string models (in particular, this is the case in the heterotic string effective action \cite{Metsaev:1987zx,Bergshoeff:1989de}). This suggests that the positivity-of-entropy bounds might not be applicable to hyperbolic black holes, but it would be interesting to understand why. For the rest of the paper, we will only make use of the constraints found for $k=1$.

\section{Thermodynamic phase space}
\label{sec:thermo}
The charged black hole solutions of the theory \req{eq:EQTfour}, that we studied in Sec.~\ref{sec:BHs}, describe, in the context of the AdS/CFT correspondence, CFT plasmas at finite temperature and chemical potential. These have different properties and interpretations depending on the geometry of the horizon. For the sake of clarity, let us repeat here the form of the black hole solutions we are considering. They are given by the following metric and $(d-1)$-form field strength

\begin{equation}\label{eq:ansatzBH}
\begin{aligned}
	ds^2 &= - N_{k}^2 f(r) dt^2 + \frac{dr^2}{f(r)} + r^2 d \Sigma^2_{k, (d-1)} \, ,\\
	H&=Q\, \omega_{k, (d-1)}\, ,
\end{aligned}
\end{equation}
where $N_{k}$ is a constant, $f(r)$ is the function given by \req{eq:fsolEQG} and $\omega_{k, (d-1)}$ is the volume form of the metric $d \Sigma^2_{k, (d-1)}$, corresponding to 

\begin{equation}\label{eq:transversespace}
	d \Sigma_{k, (d-1)}^2 = \left\lbrace 
	\begin{array}{ll}
		d\Omega_{(d-1)}^2 & \text{for } k = 1 \text{ (spherical),} \\[0.3em]
		\displaystyle \frac{1}{L^2} dx_{(d-1)}^2 & \text{for } k = 0 \text{ (flat),} \\[0.6em]
		d\Xi_{(d-1)}^2 & \text{for } k = -1 \text{ (hyperbolic).}
	\end{array} \right. 
\end{equation}
On the other hand, in the electromagnetic dual frame, we have a Maxwell field strength given by \req{eq:FstarH}, which leads to the following vector potential 

\begin{equation}
A=\Phi(r) dt\, ,
\end{equation}
where $\Phi(r)$ is given by Eq.~\req{eq:Electricpotentialgeneral}.  

We already discussed some thermodynamic properties of these black hole solutions in Section~\ref{sec:BHs}, but in order to make explicit contact with the dual CFT, it is convenient to obtain the free energy from the on-shell Euclidean action.

\subsection{Euclidean action and free energy}\label{sec:euclideanaction}
Let us work in the frame of the $(d-2)$-form $B$. We first perform a Wick rotation of our black hole solutions by writing $t=i\tau$. The Euclidean time $\tau$ has a periodicity $\tau\sim \beta+\tau$, with $\beta=1/T$, where the temperature is given by \req{eq:Temperature}. Our goal is to evaluate the Euclidean action, whose bulk part reads
\begin{equation}
\begin{aligned}
I_{\rm E}^{\rm bulk} = &- \frac{1}{16 \pi G}\int_{\mathcal{M}} d^{d+1}x\sqrt{g} \bigg[ R +\frac{d(d-1)}{L^2}-\frac{2}{(d-1)!}H^2+\frac{\lambda}{(d-2)(d-3)}L^2\mathcal{X}_{4}\\
&+ \frac{2 \alpha_1 L^2}{(d-1)!} \left( H^2R-(d-1)(2d-1)\tensor{R}{^{\mu\nu}_{\rho\sigma}}\tensor{\Hsq}{^{\rho\sigma}_{\mu\nu}}\right) + 
\\
&+ \frac{2 \alpha_2 L^2}{(d-1)!}  \left(\tensor{R}{^{\mu}_{\nu}} \tensor{\Hsq}{^{\nu}_{\mu}}-(d-1)\tensor{R}{^{\mu\nu}_{\rho\sigma}}\tensor{\Hsq}{^{\rho\sigma}_{\mu\nu}}\right)+\frac{\beta L^2}{(d-1)!^2}\Hsq^2\Bigg] \,.
\end{aligned}
\label{eq:EQTE}
\end{equation}
On top of this, we need to include generalized York-Gibbons-Hawking boundary terms to make the variational problem well posed \cite{York:1972sj,Gibbons:1976ue}, as well as counterterms, to make the action finite \cite{Emparan:1999pm}. The generalized YGH term for the Gauss-Bonnet density is known \cite{Myers:1987yn,Teitelboim:1987zz}, as well as the appropriate conterterms \cite{Yale:2011dq}. However, for the sake of simplicity, we can use instead the effective boundary terms proposed in Ref.~\cite{Bueno:2018xqc} (see also \cite{Araya:2021atx}),
\begin{equation}\label{eq:Ibdry}
I_{\rm E}^{\rm bdry}=-2C \int_{\partial \mathcal{M}}d^{d}x\sqrt{h}\left(K-\frac{d-1}{\tilde L}-\frac{\tilde L \Theta[d-3]}{2(d-2)}\mathcal{R}+\ldots\right)\, .
\end{equation}
Here, $K$ is the trace of the extrinsic curvature of the boundary, $\mathcal{R}$ is the Ricci scalar of the boundary metric, and $\Theta[d-3]=1$ for $d\ge 3$ and 0 otherwise. Additional $\mathcal{O}(\mathcal{R}^n)$ terms appear for $d\ge 5$. 
These are simply the same boundary terms as in Einstein gravity, but with a different proportionality constant, which reads

\begin{equation}
C=-\frac{\tilde L^2}{2d}\mathcal{L}\Big|_{\rm AdS}\, ,
\end{equation}
where $\mathcal{L}|_{\rm AdS}$ is the Lagrangian evaluated on the AdS vacuum to which the solutions asymptote. 
This prescription is valid for asymptotically AdS solutions (as in our case) and at least for theories that do not propagate additional degrees of freedom over AdS vacua (as in the case of Generalized Quasitopological theories), although we suspect this method actually works for general theories. For our Lagrangian, we have 

\begin{equation}
C=\frac{1}{16\pi G}\left(1-\frac{2(d-1)}{d-3}\lambda f_{\infty}\right)\, ,
\end{equation}
where we recall that $\tilde L^2=L^2/f_{\infty}$.  On the other hand, the variation of the terms $RH^2$ with respect to the metric decays very fast at infinity, so one does not need to include boundary terms. Also, they behave at infinity as the $H^2$ term, so no counterterms are needed either.

In order to compute the Euclidean action, we note that the Lagrangian becomes an explicit total derivative when evaluated on \req{eq:ansatzBH} (this is actually the defining property of the Electromagnetic Quasitopological theories). We find 

\begin{equation}
16\pi G \mathcal{L}\big|_{\rm on-shell}=\frac{1}{r^{d-1}}\frac{d \mathcal{I}(r)}{dr}\, ,
\end{equation}
where 

\begin{equation}
\begin{aligned}
\mathcal{I}(r)=&-r^{d-1} f'(r)-(d-1) r^{d-2}(f(r)-k)+\frac{(d-1) r^d}{L^2}+\frac{2 Q^2 r^{2-d}}{d-2}\\
&-2 \alpha _1 L^2 Q^2 r^{-d} \left(3 (d-1) (f(r)-k)+r f'(r)\right)-2 \alpha _2 L^2 Q^2 r^{-d} (f(r)-k)\\
&+\frac{(d-1) \lambda}{d-3}  r^{d-4} (f(r)-k) \left((d-3) (f(r)-k)+2 r f'(r)\right)+\frac{\beta  L^2 Q^4 r^{4-3 d}}{4-3 d}\, .
\end{aligned}
\end{equation}
Therefore, the bulk part of the Euclidean action is given by

\begin{equation}
I_{\rm E}^{\rm bulk} =-\frac{\beta N_{k} V_{k,(d-1)}}{16\pi G}\int_{r_{+}}^{\infty} r^{d-1}\mathcal{L}=\frac{\beta N_{k} V_{k,(d-1)}}{16\pi G}\Big[\mathcal{I}(r_{+})-\mathcal{I}(r\rightarrow\infty)\Big]\, .
\end{equation}
The evaluation at infinity $\mathcal{I}(r\rightarrow\infty)$ is divergent, but one can check all these divergencies are exactly cancelled by the boundary contributions \req{eq:Ibdry}. Furthermore, the boundary terms do not introduce any meaningful finite terms to the on-shell action.\footnote{In odd $d$ some counterterms can introduce contributions of the form $I_{\rm E}\rightarrow I_{\rm E}+c\beta$, for a constant $c$, but this simply represents a global shift in the free energy. We will simply assume that these finite counterterms have been chosen so that pure AdS has zero free energy.} Hence, we get 

\begin{equation}
I_{\rm E} =I_{\rm E}^{\rm bulk}+I_{\rm E}^{\rm bdry}=\frac{\beta N_k V_{k,(d-1)}}{16\pi G}\mathcal{I}(r_{+})\, .
\end{equation}

Now, let us note that the fact that we are computing the Euclidean action in the frame of the $B$-form has a non-trivial effect. As we observed in Section~\ref{subsec:dual}, when we dualize the $B$-form into a vector field, we generate a boundary term in the Maxwell frame, that in the thermodynamic context corresponds to working in the canonical ensemble. This implies that the Euclidean action we have computed corresponds to the Helmholtz free energy $F=T I_{\rm E}$, which is a function of the temperature and of the charge. From the result above, we have

\begin{equation}
\begin{aligned}
F =\frac{ N_k V_{k,(d-1)}}{16\pi G}\mathcal{I}(r_{+})=& \frac{ N_k V_{k,(d-1)}}{16\pi G} \Bigg[ \frac{(d-1) r_{+}^d}{L^2}-r_{+}^{d-1} \frac{4\pi T}{N_k}+k(d-1) r_{+}^{d-2}+\frac{2 Q^2 r_{+}^{2-d}}{d-2}\\
&-2 \alpha _1 L^2 Q^2 r_{+}^{-d} \left(-3k (d-1) +\frac{4\pi T r_{+}}{N_{k}}\right)+2 k \alpha _2 L^2 Q^2 r_{+}^{-d}\\
&+(d-1) \lambda  r_{+}^{d-4} k\left( k-\frac{8 \pi T r_{+}}{(d-3)N_{k}}\right)+\frac{\beta  L^2 Q^4 r_{+}^{4-3 d}}{4-3 d} \Bigg] \, ,
\end{aligned}
\end{equation}
and we recall that the temperature is given by 

\begin{equation}
	\begin{aligned}
		T = & \frac{N_k}{4 \pi r_+ \left(1 - 2 L^2 Q^2 \alpha_1 r_+^{-2(d-1)} + 2 k L^2 \lambda r_+^{-2} \right) } \Bigg[ \left( (d-2) k + d \frac{r_+^2}{L^2} + (d-4) k^2 \lambda \frac{L^2}{r_+^2} \right) \\
		& - \frac{2 Q^2}{(d-1) r_+^{2(d-1)}} \left(r_+^2 + d k L^2 (3(d-1)\alpha_1 + \alpha_2) \right) + \frac{\beta L^2 Q^4}{(d-1) r_+^{2(2d-3)}} \Bigg] \, .
	\end{aligned}
 	\label{eq:Temperature2}
\end{equation}
We also introduce the chemical potential $\mu=\lim_{r\rightarrow\infty} \ell_{*}^{-1}A_{t}$,  and from \req{eq:phiinf} we read

\begin{equation}
	\mu= \frac{N_k Q}{\ell_{*}} \Bigg[ \frac{1}{(d-2) r_+^{d-2}} + \frac{L^2 \alpha_1}{r_+^d} \left( 3 (d-1)k - r_+ \frac{4 \pi T}{N_k} \right) + \frac{L^2 \alpha_2 k}{r_+^d} - \frac{L^2 Q^2 \beta}{(3d - 4) r_+^{3d-4}} \Bigg] \, .
	\label{eq:potential2}
\end{equation}
We then check that this free energy satisfies the usual first law,

\begin{equation}
dF=-S dT+\mu\, d\mathcal{N}\, ,
\end{equation}
where $S$ is Wald's entropy given by \req{eq:entropy1}, and where $\mathcal{N}=q\ell_{*}$, where $q$ is the physical charge introduced in \req{eq:chargedef}, \textit{i.e.}, 

\begin{equation}
\mathcal{N}=\frac{V_{k,(d-1)}\ell_{*}Q}{4\pi G}\, ,
\end{equation}
and it represents the number of charged particles under the current $J$ in the boundary theory. 

We wish to work in the grand canonical ensemble (\textit{i.e.}, at fixed chemical potential), so instead of $F$ we are interested in the grand potential (or grand free energy), defined as

\begin{equation}
\Omega=F-\mu\, \mathcal{N}\, .
\end{equation}
This can also be obtained directly from the Euclidean action by adding or removing appropriate boundary terms (depending of whether we are in the Maxwell or $B$-form frames). By construction, this satisfies

\begin{equation}
d\Omega=-S dT-\mathcal{N}d\mu\, ,
\label{eq:1stlawomega}
\end{equation}
and it is to be understood as a function of $T$ and $\mu$. Explicitly, it reads

\begin{equation}
\begin{aligned}
\Omega=&\frac{N_{k}V_{k,(d-1)}}{16\pi G}\Bigg[\frac{(d-1) r_{+}^d}{L^2}-\frac{2 Q^2 r_{+}^{2-d}}{d-2}+r_{+}^{d-2} \left((d-1) k-\frac{4
   \pi T r_{+}}{N_k}\right)+\frac{3 \beta L^2 Q^4 r_{+}^{4-3 d}}{3 d-4}\\
   &+(d-1) \lambda k
   L^2 r_{+}^{-4+d} \left(k-\frac{8 \pi  T r_{+}}{N_{k}(d-3)}\right) +2  \alpha _1 L^2 Q^2 r_{+}^{-d}
   \left(-3 (d-1) k+\frac{4 \pi T r_{+}}{N_{k}}\right)\\
   &-2 k \alpha _2 L^2 Q^2 r_{+}^{-d} \Bigg]\, .
\end{aligned}
\label{eq:omegagen}
\end{equation}
We study next the properties of these thermal states in the case of a flat boundary. The hyperbolic case will be considered in Section~\ref{sec:renyi} in the context of R\'enyi entropies. 

\subsection{Flat boundary: black branes}
\label{sec:Thermoflat}
Let us consider the case of planar black holes, $k=0$. The boundary geometry is Minkowski space, and therefore these black hole solutions probe the properties of thermal CFTs in flat space. More precisely, the boundary metric is conformal to 
\begin{equation}\label{eq:bdrymetric}
ds^2_{\rm bdry}=-N_{0}^2f_{\infty}dt^2+dx_{(d-1)}^2\, ,
\end{equation}
and therefore, we set 

\begin{equation}
N_{0}=\frac{1}{\sqrt{f_{\infty}}}\, ,
\end{equation}
which is equivalent to working in units in which the speed of light is one. Also note that, on account of our definition of transverse space in Eq.~\req{eq:transversespace}, the volume reads, 
\begin{equation}
V_{0,(d-1)}=\frac{V_{\mathbb{R}^{d-1}}}{L^{d-1}}\, ,
\end{equation}
where now $V_{\mathbb{R}^{d-1}}$ is the volume of the constant-$t$ spatial slices of the boundary metric \req{eq:bdrymetric}.
We can then define the grand potential, entropy, mass, and number densities as 

\begin{equation}
\omega=\frac{\Omega}{V_{\mathbb{R}^{d-1}}}\, ,\quad s=\frac{S}{V_{\mathbb{R}^{d-1}}}\, ,\quad \rho=\frac{M}{V_{\mathbb{R}^{d-1}}}\, ,\quad N=\frac{\mathcal{N}}{V_{\mathbb{R}^{d-1}}}\, .
\end{equation}
It also proves useful to replace the charge parameter $Q$ in terms of a new dimensionless parameter $p$, defined by 

\begin{equation}\label{eq:ppardef}
Q=p\frac{r_{+}^{d-1}}{L}\, .
\end{equation}
In this way, we get the following expressions for the grand potential density, entropy density, temperature and chemical potential,

\begin{align}
\omega=&-\frac{r_{+}^d}{16 \pi G L^{d+1} \sqrt{f_{\infty}} (d-1) } \left( \frac{\left(d^2-3 d+2+2 p^2\right)}{d-2}-\frac{\beta p^4}{3d-4}\right)\, ,\\
s=&\frac{r_{+}^{d-1}}{4L^{d-1}G}\left(1+2p^2\alpha_1\right)\, ,\\\label{eq:tempplanar}
T =& \frac{r_{+}}{4 \pi \sqrt{f_{\infty}}L^2} \frac{d (d-1) - 2p^2 + \beta p^4}{(d-1) (1 - 2 p^2 \alpha_1)}\, ,\\
\begin{split}
	\mu = & \frac{r_+}{\sqrt{f_\infty}\ell_{*} L} \frac{p}{(d-1) ( 1 - 2 \alpha_1 p^2 )} \Bigg[ (d-1) \left( \frac{1}{d-2} - d \alpha_1 \right) - \left( \frac{2 \alpha_1}{d-2} + \frac{(d-1) \beta}{3d-4} \right) p^2 \\
	& - \frac{(d-2) \alpha_1 \beta p^4}{3d-4} \Bigg] \, ,
\end{split}
\end{align}
and interestingly, the thermodynamic quantities are independent of $\lambda$ and $\alpha_2$.\footnote{In the case of $\lambda$ there is an implicit dependence through $f_{\infty}$, but the effect is a trivial rescaling of some thermodynamic potentials.} 
Note that all of these quantities have a certain scaling with $r_{+}$, and observe in particular that $r_{+}\propto T$. In the neutral limit $(p=0)$ this gives rise to the well-known relation

\begin{equation}
s\big|_{\mu=0}=C_{S} T^{d-1}\, ,
\end{equation}
where the $C_S$ is the thermal entropy charge, that in our case reads

\begin{equation}
C_{S}=\frac{(4\pi L\sqrt{f_{\infty}})^{d-1}}{4 G d^{d-1}}\, .
\end{equation}
It is known that for holographic Gauss-Bonnet gravity (and Lovelock gravity in general), this charge does not receive corrections besides the appearance of the $f_{\infty}$ factor, but there are other theories, such as Einsteinian cubic gravity \cite{Bueno:2018xqc} and Generalized Quasitopological gravities \cite{PabloPablo3,Mir:2019ecg,Mir:2019rik} that do introduce non-trivial corrections to $C_{S}$. Using this constant, and replacing $r_{+}$ in terms of $T$ and $p$, we can write

\begin{align}
\omega&=-\frac{C_{S}}{d}T^{d}\left(\frac{1-2\alpha_1 p^2}{1+\frac{\beta p^4-2p^2}{d(d-1)}}\right)^{d} \left(1+\frac{2p^2}{(d-1)(d-2)}-\frac{\beta p^4}{(d-1)(3d-4)}\right)\, ,\\
s&=C_{S}T^{d-1}\left(\frac{1-2\alpha_1 p^2}{1+\frac{\beta p^4-2p^2}{d(d-1)}}\right)^{d-1} (1+2\alpha_1 p^2)\, .
\end{align}
Now, the variable $p$ is a function of the dimensionless ratio between the chemical potential and the temperature. The precise relation reads

\begin{align}\notag
\hat{\mu} \equiv \frac{\ell_{*}\mu}{4 \pi L T}  &= \frac{p}{\left( d (d-1) - 2 p^2 + \beta p^4 \right)} \Bigg[ (d-1)  \left( \frac{1}{d-2} - d \alpha_1 \right) - \left( \frac{2 \alpha_1}{d-2} + \frac{(d-1)\beta}{3d-4} \right) p^2 \\
&-  \frac{(d-2) \alpha_1 \beta p^4}{3d-4} \Bigg]\, .
\label{eq:muhat}
\end{align}
Note the appearance of the ratio $\ell_{*}/L$ whose value is an input of the holographic duality. This expression defines implicitly $p$ as a function of $\hat \mu$, and therefore it means that 
\begin{equation}
\hat\omega\equiv\frac{\omega}{T^{d}}
\end{equation} 
is only a function of this variable. Studying the dependence of $\hat\omega$ on $\hat\mu$ is the same as studying how the grand free energy varies as we change the chemical potential and keep fixed the temperature --- or, equivalently, studying the free energy as a function of the temperature at fixed chemical potential. Thanks to the scaling properties of the problem, only the ratio between chemical potential and temperature $\hat\mu$ is relevant. We can wonder in particular about the monotonicity of the grand canonical potential or the existence of phase transitions. 

Before performing a more in-depth analysis, let us remark a few points. First, since in most cases we are not able to invert \req{eq:muhat} in a simple way, it is useful to use $p$ in order to obtain the curves $(\hat\mu, \hat\omega)$ parametrically. Second, we note that, in order to have black hole solutions, the temperature $T$ in \req{eq:tempplanar} must be non-negative.  It follows that the extremal limit $T=0$ is reached for

\begin{equation}
 d (d-1) - 2 p^2 + \beta p^4=0 \, ,
\end{equation}
which has real solutions only for $\beta\le 1/(d(d-1))$.  
It is also important to notice that black hole solutions only exist if $1-2\alpha_1 p^2\ge 0$. Even though, looking at \req{eq:tempplanar}, one could have $T>0$ when this quantity is negative, it turns out that $1-2\alpha_1 p^2< 0$ implies the existence of a naked singularity for $r>r_{+}$, as can be seen in \req{eq:fsolEQG}, so this case must be ruled out.

In the case of Einstein-Maxwell theory it is possible to invert \req{eq:muhat} explicitly to obtain
\begin{equation}
p=\frac{1}{4\hat\mu}\left[-\frac{d-1}{d-2}+\sqrt{\left(\frac{d-1}{d-2}\right)^2+8d(d-1)\hat\mu^2}\right]\, .
\end{equation}
One can check that this is a one-to-one relation, with $p(\mu\rightarrow \pm\infty)= \pm \sqrt{d (d-1)/2}$, and that the relation $\hat\omega(\hat\mu)$ is monotonic as well. Therefore, there are no phase transitions for charged plasmas in holographic Einstein-Maxwell theory. 

To simplify the study of the relation $\hat\omega(\hat\mu)$, we consider first the cases $\beta=0$ and $\alpha_1=0$ separately. 

\subsubsection{Phase space with $\alpha_1 \neq 0$ and $\beta = 0$}
\label{sec:beta0}

\begin{figure}
	\centering
	\includegraphics[width=0.6\linewidth]{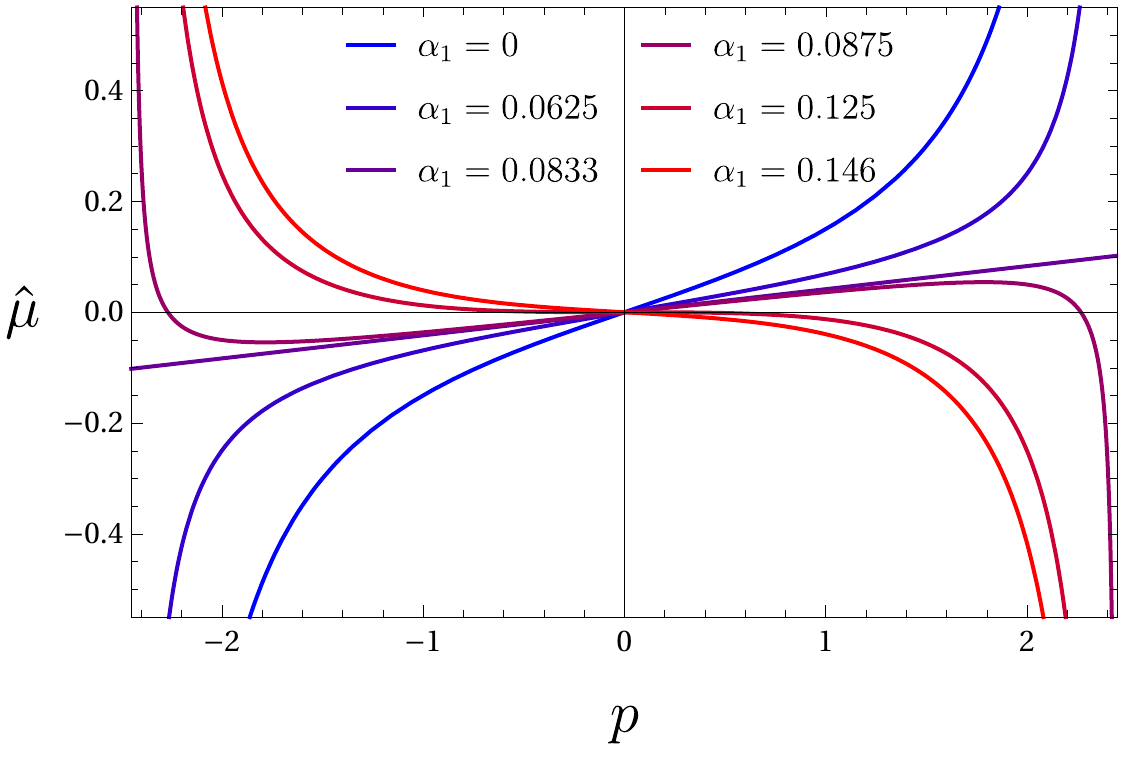}
	\caption{Dimensionless chemical potential $\hat{\mu}$ as a function of the parameter $p$ for different values of the coupling $\alpha_1$ and $\beta = 0$, in $d = 4$ dimensions. Notice that in this case the range \eqref{eq:alpha1trans} where $\hat{\mu} (p)$ is non-invertible becomes $\alpha_1 \in (1/12, \, 1/8)$.}
	\label{fig:hatmup}
\end{figure}

\begin{figure}
	\centering
	\begin{subfigure}{0.49\textwidth}
		\includegraphics[width=\linewidth]{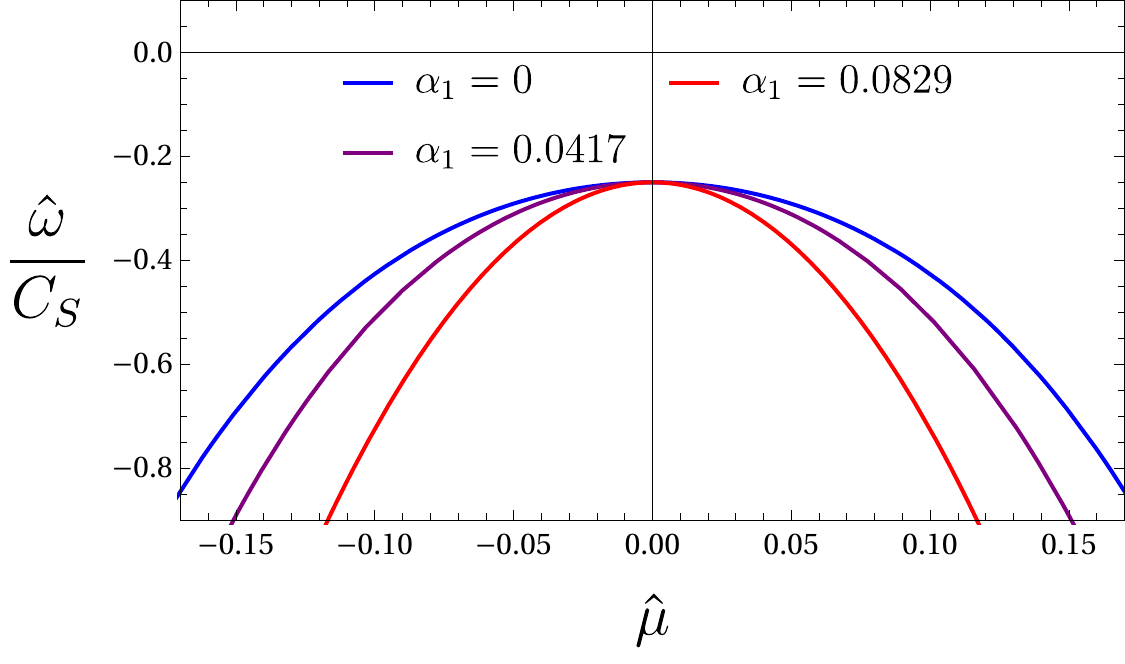}
		\caption{Values of alpha outside the range \eqref{eq:alpha1trans}.}
	\end{subfigure}
	\centering
	\begin{subfigure}{0.49\textwidth}
		\includegraphics[width=\linewidth]{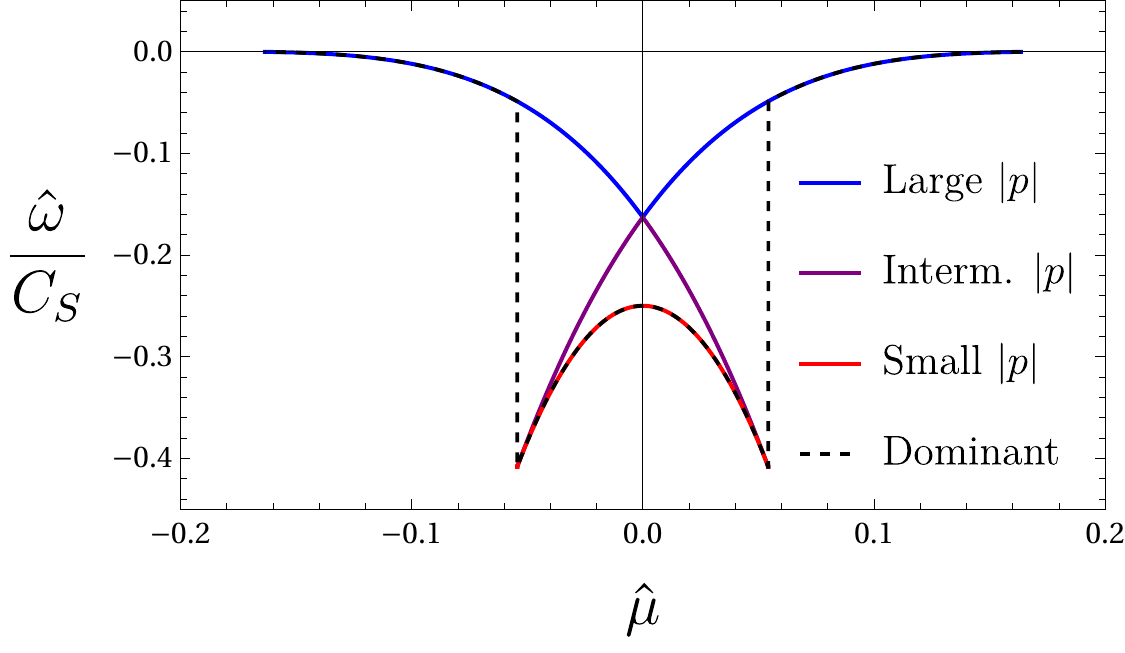}
		\caption{$\alpha_1 = 0.0875$, inside the range \eqref{eq:alpha1trans}.}
	\end{subfigure}
	\caption{We show the rescaled grand potential $\hat{\omega}$ with respect to $\hat{\mu}$ with $\beta = 0$ in $d = 4$ dimensions. Left plot: we take $\alpha_1\le 1/12$, so that there is a single phase for every value of $\hat\mu$. Right plot: we take $\alpha_1$ within the interval \req{eq:alpha1trans}, and we see that there appear three phases with different values of $p$. The dominant configuration is denoted with a black dashed line. We observe a zeroth-order phase transition. Also note that there are no solutions beyond a maximum value of $\hat\mu$ (corresponding to $\hat\mu_{\rm max}=\sqrt{\alpha_1/2}$.)}
	\label{fig:Omegamu}
\end{figure}

Let us start by setting $\beta = 0$, to isolate the contribution of $\alpha_1$ to the phase space of the system. We will consider $\alpha_{1}\ge 0$, as a negative $\alpha_1$ is discarded by the weak gravity conjecture as we discussed in Section~\ref{sec:WGC}.
As was shown above, in order to have meaningful solutions we must have $T\ge 0$, but also $1-2\alpha p^2\ge 0$. Hence, the range of allowed values of $p$ is actually

\begin{equation}
|p|\le \min\left\{p_{\rm ext},\frac{1}{\sqrt{2\alpha_1}}\right\}\, ,\quad p_{\rm ext}=\sqrt{\frac{d (d-1)}{2}}\, .
\end{equation}
The transition between the two possible maximum values of $p$ happens for $\alpha_1=1/(d(d-1))$. It turns out that we have to distinguish four different cases. 
\begin{enumerate}
\item For $\alpha_1<1/(d(d-1))$, the relation $\hat\mu(p)$ (shown in Fig.~\ref{fig:hatmup}) is one-to-one  and $\hat\mu$ takes values in all of the real line, diverging to $\pm\infty$ for $p=\pm p_{\rm ext}$, which corresponds to the extremal limit. The curves for $\hat\omega(\hat\mu)$ --- shown in the plot (a) of Fig.~\ref{fig:Omegamu} --- behave qualitatively as in Einstein-Maxwell theory. 

\item The case $\alpha_1=1/(d(d-1))$ is somewhat special: here the relation $\hat\mu(p)$ becomes linear, and the temperature is independent of $p$:
\begin{equation}
\hat\mu=\frac{p}{d(d-1)(d-2)}\, ,\quad T=\frac{r_{+}d}{4\pi \sqrt{f_{\infty}} L^2}\, .
\end{equation}
Thus, there are solutions for arbitrary values of $\hat\mu$, but there is no extremal limit. We also have in this case a simple expression for the grand potential density,

\begin{equation}
\omega=-\frac{C_{S}}{d}T^{d}\left(1+2d^2(d-1)(d-2)\left(\frac{\ell_{*}\mu}{4\pi T L}\right)^2\right) \, .
\end{equation}

\item A new behaviour takes place for $\alpha_{1}>1/(d(d-1))$. Since the extremal value $p_{\rm ext}$ is not allowed in this case,  we see from Eq.~\req{eq:muhat} that only a finite range of values of $\hat\mu$ is allowed. In particular, it is not difficult to see that $|\hat\mu|<\sqrt{\alpha_1/2}$. There are no black hole solutions for larger values of $\hat\mu$. In addition, we see that, if $\alpha_1$ is within the interval
\begin{equation}
	\alpha_1 \in \left( \frac{1}{d (d-1)} , \, \frac{1}{d (d-2)} \right)\, ,
	\label{eq:alpha1trans}
\end{equation}
then the relation $\hat\mu(p)$ becomes non-invertible. In this range, the function $\hat\mu(p)$ has a maximum and a minimum (see Fig.~\ref{fig:hatmup}), and some values of $\hat\mu$ are reached by three distinct values of $p$, meaning that we have three different phases, with either small, intermediate or large values of $p$. 
To explore the existence of phase transitions we must look at the phase diagram $\hat\omega$ vs $\hat\mu$ for all the branches of solutions. We  show this in Fig.~\ref{fig:Omegamu} (b) in the case of $d=4$ for a value of $\alpha_1$ inside this interval. As we can see, for small $\hat\mu$ there are three phases, and the one with the smallest $p$ has the lowest free energy (and hence dominates) and it has the usual quadratic behavior, with $\partial^2\omega/\partial\mu^2<0$. However, this phase and the one with intermediate values of $p$ merge at certain $\hat\mu_{\rm cr}$ and cease to exist for larger $\hat\mu$. At that point, a zeroth-order phase transition takes place towards the solution with large $p$, as illustrated in Fig.~\ref{fig:Omegamu}. The new phase is quite exotic, as it has $N=-\partial\omega/\partial\mu<0$ for $\hat\mu>0$, so a positive chemical potential generates a negative number density, and viceversa. It nevertheless still satisfies $\partial N/\partial\mu>0$. This phase ceases to exist at $|\hat\mu_{\rm max}|=\sqrt{\alpha_1/2}$, and no solutions exist for larger $\hat\mu$. 

\item If we set $\alpha_1 = 1 /(d (d-2))$ the maximum and the minimum of $\hat\mu(p)$ collapse at $p=0$. Therefore, for $\alpha_1\ge \frac{1}{d (d-2)}$ there exists a single phase, and it always has $N<0$ for $\hat\mu>0$, but $\partial N/\partial\mu>0$. Also, this phase only exist for a limited range for values of $\hat\mu$. 

\end{enumerate}

An interesting question is whether these phase transitions, and all the special features that happen for $\alpha_{1}>1/d(d-1)$, are allowed by the physical constraints we derived in Sec.~\ref{sec:constraints}. If only $\alpha_1$ is active, \textit{i.e.},  $\alpha_2=\lambda=0$,  then according to Eq.~\req{eq:a1alone} we have
\begin{equation}
\alpha_1\le \frac{1}{7d^2-9d+2}\, ,
\end{equation}
which is smaller than $1/(d(d-1))$ for $d\ge 3$, and therefore these phase transitions are not allowed. It is most interesting to ask what happens if we consider non-vanishing $\alpha_2$ and $\lambda$, which although do not affect directly the thermodynamic relations, do change the allowed range of values of $\alpha_1$. We have seen that in $d=3,4,5$ the unitarity constraints impose an upper bound on $\alpha_1$ given by Eq.~\req{eq:a1abs}. This constraint still allows for phase transitions to take place, however, we find a stronger bound if we take into account the constraint \req{eq:wgc2} from the WGC. In $d=3$ (in which $\lambda=0$), combining \req{eq:wgc2} with \req{eq:a2c1} and \req{eq:a2c2}, one can see that $\alpha_1$ is bound from above by 
\begin{equation}
\alpha_1\big|_{d=3,\beta=0}\le \frac{1}{8}\, ,
\end{equation}
while the limit to have phase transitions is $\alpha_{1}^{\rm tr}=1/(d(d-1))=1/6$. The $d=4$ case is more involved, as now the Gauss-Bonnet term also has an effect. The upper bound on $\alpha_1$ is obtained from the intersection of \req{eq:wgc2} and \req{eq:a2c2} at saturation and it yields
\begin{equation}
\alpha_1\big|_{d=4,\beta=0}\le \frac{1-f_{\infty}\lambda}{12 f_{\infty}}\, .
\end{equation}
Now, since $\lambda\ge 0$ by virtue of the WGC, this value is always smaller (or equal, when $\lambda=0$) than $1/12$, which is precisely the threshold value to produce phase transitions. 
Thus, the different physical constraints conspire quite impressively to avoid the existence of phase transitions, at least in the cases $d=3,4$, which are the most relevant ones. This also avoids the nonphysical situation of absence of solutions for large $\hat\mu$. 

In $d=5$ the same constraints lead to an absolute maximum $\alpha_1\big|_{d=5,\beta=0}\le 1/16$, which is not enough to rule out the phase transition, whose threshold value is $\alpha_1^{\rm tr}=1/20$.  For $d\ge 6$ the value of $\alpha_1$ can be arbitrarily large (as long as $\alpha_2<0$ is also large), so these exotic phase transitions cannot be avoided for large dimensions on the basis of our current  constraints. 

\subsubsection{Phase space with $\alpha_1 = 0$ and $\beta \neq 0$}
\label{sec:alpha10}
Let us now study the effect of the $H^4$ term by setting $\alpha_1=0$. We also take $\beta\ge 0$, as this is implied by the WGC constraints studied in Section~\ref{sec:WGC}. We distinguish two main scenarios depending on the roots of
\begin{equation}
 d (d-1) - 2 p^2 + \beta p^4=0 \, ,
\end{equation}
which determine extremality. The solutions to these equation are $p=\pm p^{\pm}_{\rm ext}$ (the two signs are independent), where 

\begin{equation}\label{eq:pmax2}
p^{\pm}_{\rm ext}=\sqrt{\frac{1}{\beta} \left( 1 \pm \sqrt{1 - d (d-1) \beta} \right) }\, .
\end{equation}

\begin{figure}
	\centering
	\includegraphics[width=0.6\textwidth]{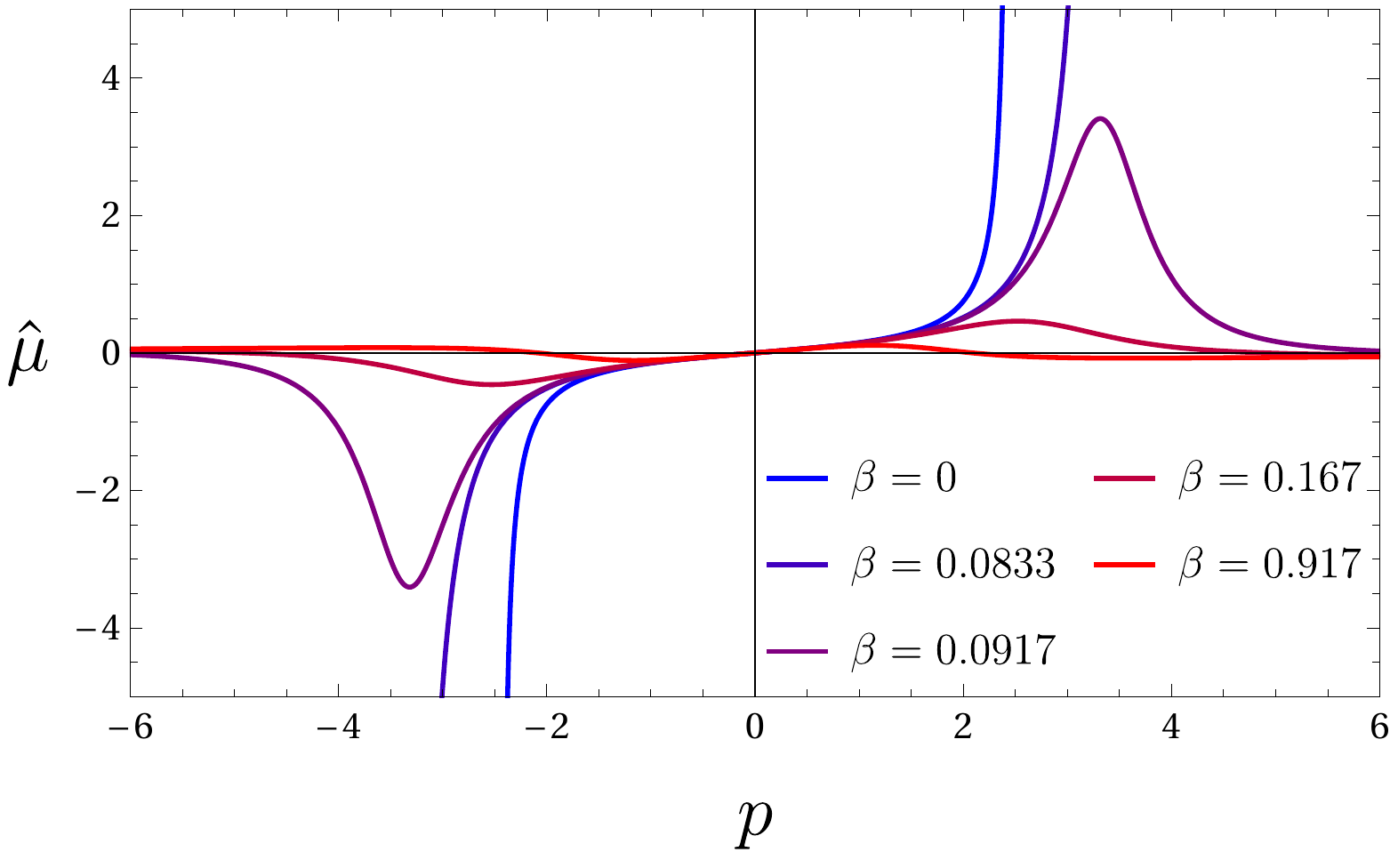}
	\caption{Dimensionless potential $\hat{\mu}$ with respect to the parameter $p$ for different values of the coupling $\beta$ and $\alpha_1 = 0$, in $d = 4$ dimensions. In this case, the condition for $\hat{\mu}(p)$ to become non-invertible is $\beta > 1/12$.}
	\label{fig:mupbeta}
\end{figure}

\begin{figure}
	\centering
	\includegraphics[width=0.6\textwidth]{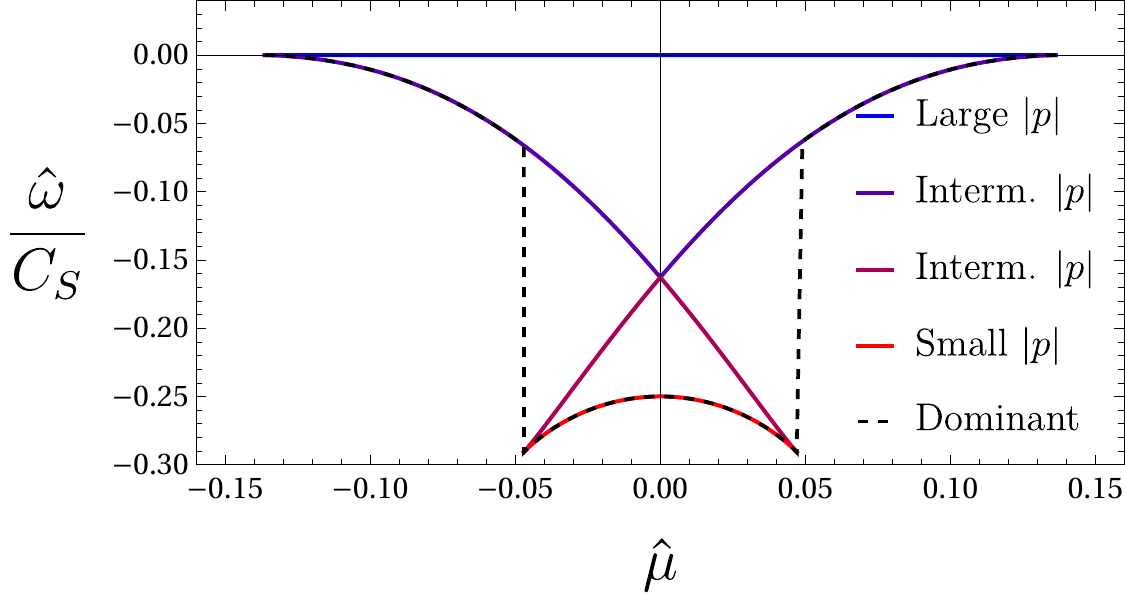}
	\caption{We show $\hat{\omega}$ with respect to $\hat{\mu}$ for $\beta= 4.333>\beta_{\rm thr}$ and $\alpha_1 = 0$, in $d = 4$ dimensions.  There is a region where four phases with different $p$ coexist, and the dominant one is drawn with a black dashed line. At some point there is a zeroth-order phase transition, but there is a maximum value of $|\hat\mu|$ above which no solutions exist.}
	\label{fig:omegahatbeta}
\end{figure}

\begin{itemize}
\item If $\beta\le 1/(d(d-1))$, then these roots are real and we have three families of solutions.  The first family has $|p|\le p_{\rm ext}^{-}$ and it is connected to the solutions of Einstein-Maxwell theory. It exists for arbitrary values of $\hat\mu$ and reaches the extremal limit for $p=p_{\rm ext}^{-}$. A second branch happens for 
\begin{equation}
p_{\rm ext}^{+}\le |p|\le \sqrt{\frac{3d-4}{(d-2)\beta}}\, 
\end{equation}
(the upper limit corresponding to $\hat\mu=0$). This one also exists for arbitrary $\hat\mu$ and has a extremal limit for $p=p_{\rm ext}^{+}$. However, it can be seen that it is not possible to find a value of $p$ such that $\hat{\mu}'(p) = 0$ in this range, and hence there is only one thermodynamic phase. Finally, there is a third type of solutions, with $|p|>\sqrt{\frac{3d-4}{(d-2)\beta}}$, which only exist for very small values of $\hat\mu$. However, there are no phase transitions either, as the solution in the Einstein-Maxwell branch always has the smallest grand-canonical free energy.

\item If $\beta > 1/(d(d-1))$, the roots $p^{\pm}_{\rm ext}$ become complex, implying that the extremal limit does not exist. In this case, the relation $\hat\mu(p)$ has a shape as shown in Fig.~\ref{fig:mupbeta}, with a maximum and a minimum for $p>0$ (and viceversa for $p<0$). This means that for small $\hat\mu$ there are up to four different phases.  It is also clear that there exists a maximum value of $\hat\mu$ for which any phase exists. 
If we keep increasing $\beta$, we find that there is one point where the maximum value of $\hat\mu$ for $p>0$ becomes smaller than its maximum for $p<0$ (and respectively with the minima at $p<0$ and $p>0$). This happens for
\begin{equation}
	\beta_{\rm thr} = \frac{(3d - 4)^2}{d (d-1) (d-2)^2}\, .
	\label{eq:betatr}
\end{equation}
When $\beta<\beta_{\rm thr}$ there are no phase transitions, and the dominant phase exists up to a certain $\hat\mu_{\rm max}$, beyond which simply there are no solutions.  When $\beta>\beta_{\rm thr}$, on the other hand, there is another solution that extends beyond the dominant phase, and hence, when this one finds its endpoint there is a zeroth-order phase transition. However, the new phase once again only extends up to a maximum value of $\hat\mu$.  This is illustrated in Fig.~\ref{fig:omegahatbeta} in the case of $d=4$. Notice that the phase with large $|p|$ is always subdominant.
\end{itemize}

In general, the scenario with $\beta > 1/(d(d-1))$ looks rather unphysical, as the theory should allow for solutions of arbitrary $\hat\mu$. However, unlike the case of $\alpha_1$, we have no additional constrains on $\beta$ that could rule out this case from basic principles. The coupling $\beta$ is the less constrained one in our theory, as it does not affect, \textit{e.g.}, the linearized equations on neutral backgrounds or the correlators $\langle JJ\rangle$, $\langle TJJ\rangle$. At the level of correlators, it would make its first appearance at $\langle JJJJ\rangle$, and it will also be non-trivial when studying the propagation of electromagnetic waves on charged backgrounds. It would be interesting to study those cases and investigate possible unitarity or causality constraints on $\beta$. Perhaps  such constraints could remove the unwanted values $\beta > 1/(d(d-1))$.


\subsubsection{Phase space with $\alpha_1 \neq 0$ and $\beta \neq 0$}
\label{sec:thermoab}
We have seen that, separately, both couplings $\alpha_1$ and $\beta$ have qualitatively similar effects: they produce simultaneous phases that can lead to zeroth-order phase transitions when the couplings are large enough. More worrying is the fact that, whenever this happens,  there exist no phases beyond a maximum value of $|\hat\mu|$, which seems rather unphysical. When $\beta=0$ this situation is forbidden in $d=3,4$ by the physical constraints, and if one further assumes $\alpha_2=0$ (which does not affect the thermodynamics) then this behavior is ruled out in every dimension. When $\beta\neq 0$ we cannot avoid this strange behavior because $\beta$ can be arbitrarily large, but we have argued that it is likely that there are other constraints that will bound the possible values of $\beta$ to a finite interval. 

Of course, if both $\beta$ and $\alpha_1$ are non-vanishing and large enough, we find again these exotic situations, and even more involved ones. However, in view of our previous discussion, we are interested in the cases in which $\hat\mu$ can take arbitrary values, since the opposite seems pathological. Let us determine which values of $\alpha_1$ and $\beta$ satisfy this property. For that, note that, whenever $\alpha_1>0$, $p$ has a maximum value,  since we must fulfill $1 - 2 p^2 \alpha_1 \ge 0$, as we saw earlier. Then, from \req{eq:muhat} it follows that, in order for $\hat\mu$ to take arbitrarily high values, the extremal limit must exist. This means that the roots \req{eq:pmax2} must be real and therefore 
\begin{equation}\label{eq:betamu}
\beta\le \frac{1}{d(d-1)}\, .
\end{equation}
In addition we must demand

\begin{equation}\label{eq:alphamu}
\alpha_1\le \frac{1}{2(p_{\rm ext}^{-})^2}\, ,
\end{equation}
so that $1 - 2 p^2 \alpha_1 \ge 0$ holds. 
Whenever these two conditions are satisfied, it is guaranteed that there are solutions for any value of $\hat\mu$. 
The question we want to answer then is whether there are any special features in the phase space within this reasonable set of couplings. 

We note that the solutions with $|p|\le p_{\rm ext}^{-}$ always exist and that for these $\hat\mu$ take values in the whole real line. Could there be multiple phases within these values of $p$? For that, it should happen that $\hat{\mu}'(p^*) = 0$ for some $p = p^*$ in this interval. However, we have observed that this never happens as long as $\beta$ and $\alpha_1$ are constrained by \req{eq:betamu} and \req{eq:alphamu}, respectively. Therefore, $|p|\le p_{\rm ext}^{-}$ generates a unique phase for which $\hat\mu$ ranges from $-\infty$ to $+\infty$.  Furthermore,  this phase has the same qualitative behavior as the one found for Einstein-Maxwell theory, that we showed in Fig.~\ref{fig:Omegamu} (a).

Interestingly, there can be other phases if $\alpha_1$ is small enough. This happens if $\alpha_1\le 1/(2(p_{\rm ext}^{+})^2)$, in which case there are additional solutions for
\begin{equation}
p_{\rm ext}^{+}\le|p|\le \frac{1}{2\sqrt{\alpha_1}}\, .
\end{equation}
In general these have a quite similar profile to those in the interval $|p|\le p_{\rm ext}^{-}$, but however they never dominate.
The conclusion seems to be that, whenever we have the reasonable situation in which $\hat\mu$ is unbounded, then no phase transitions take place, and everything works more or less as in Einstein-Maxwell theory.

\section{Holographic hydrodynamics: shear viscosity}
\label{sec:hydro}
Charged black branes, whose thermodynamic properties we have explored in Section~\ref{sec:Thermoflat} above, describe holographic CFTs at finite temperature and chemical potential. In this state, the theory behaves as a plasma, and in the hydrodynamic limit we can study it as a fluid  \cite{Policastro:2002se,Herzog:2002fn,Kovtun:2003wp,Baier:2007ix,Hartnoll:2009sz}. Thus, we can ask about the propagation of sound waves or about different transport coefficients, which determine the response of the plasma under perturbations. In this respect, one of the quantities that attracted more attention from the early times of holographic hydrodynamics was the shear viscosity $\eta$. As it turns out, the ratio between this quantity and the entropy density takes the following value for Einstein gravity \cite{Kovtun:2003wp,Buchel:2003tz,Kovtun:2004de},
\begin{equation}
\frac{\eta}{s}\Big|_{\text{EG}}=\frac{1}{4\pi}\, ,
\end{equation}
which holds also in the presence of a chemical potential \cite{Mas:2006dy,Son:2006em,Maeda:2006by,Benincasa:2006fu}. This led to the conjecture that such result was universal in holographic CFTs and to the Kovtun-Son-Starinets (KSS) bound \cite{Kovtun:2004de}, which claims that $\eta/s\ge 1/(4\pi)$ for any fluid in nature. 
However, if one considers higher-curvature corrections in the bulk theory, the holographic prediction for $\eta/s$ is indeed modified, showing it is not a truly universal result \cite{Buchel:2004di,Brigante:2007nu,Kats:2007mq,Myers:2008yi,Buchel:2008vz,Banerjee:2009fm}. This ratio has even been computed in a non-perturbative fashion in theories such as Gauss-Bonnet \cite{Brigante:2007nu,Brigante:2008gz,deBoer:2009pn}, Lovelock gravity \cite{Ge:2009ac,Shu:2009ax,Camanho:2010ru}, cubic \cite{Myers:2010jv} and quartic \cite{Dehghani:2013ldu} Quasitopological gravity  and Generalized Quasitopological gravities \cite{Bueno:2018xqc,Mir:2019ecg,Mir:2019rik}, among other models. Those examples showed that, even when the theories were constrained by physical requirements, such as unitarity, positivity of energy and causality, the KSS bound could be lowered, although without reaching zero.\footnote{Nevertheless, it can be made arbitrarily small for Lovelock gravity by taking $d\rightarrow\infty$ \cite{Camanho:2010ru}.} The status of the question about how much the KSS bound can be consitently lowered is not clear, as there are arguments, as those of Ref.~\cite{Camanho:2014apa}, that constrain the higher-order couplings to be perturbatively small. However, once again the arguments of Ref.~\cite{Camanho:2014apa} cannot be directly applied to all types of higher-derivative interactions. So, the common belief is that a bound for $\eta/s$, lower than the KSS one, does exist \cite{Cremonini:2011iq}.

On the other hand, the effect of the chemical potential on the shear viscosity to entropy density ratio has been less explored. As we mentioned, $\eta/s$ remains independent of $\mu$ in holographic Einstein-Maxwell theory, but this is no longer the case when one introduces higher-derivative corrections. Refs.~\cite{Cremonini:2009sy,Myers:2009ij}  computed the effect of perturbative corrections on the viscosity in the case of $N=2$ supergravity and in a general $d=4$ EFT --- see also \cite{Cai:2011uh}. Regarding non-perturbative calculations, we may highlight Refs.~\cite{Cai:2008ph,Ge:2008ni,Ge:2009ac}, that considered the case of Lovelock gravity minimally coupled to a vector field. 

However, a more general analysis is still missing, and in particular, the effect of non-minimally coupled operators has not yet been studied at a non-perturbative level. 
Thus, here we will perform the first exact and analytic computation of $\eta/s$ with a chemical potential in a higher-derivative theory with all kinds of interactions. This will give us a more complete perspective on the effect of the higher-derivative terms on this ratio, and will allow us to investigate how much the KSS bound can be lowered while satisfying the physical constraints of Section~\ref{sec:constraints}.

In order to set out the computation of the shear viscosity, let us write down our charged planar black hole solutions, which are given by

\begin{align}\label{eq:metricvisco1}
ds^2 &= - \frac{f(r)}{f_\infty} dt^2 + \frac{dr^2}{f(r)} + \frac{r^2}{L^2} dx^2_{(d-1)}\, ,\\
H&=\frac{Q}{L^{d-1}}dx^1\wedge\ldots\wedge dx^{d-1} \, ,
\end{align}
where we fixed the value of the constant function $N_0 = 1 / \sqrt{f_\infty}$ so that the speed of light at the conformal boundary is equal to one, and where $f(r)$ is given by \req{eq:fsolEQG} with $k=0$. Let us replace the radial coordinate by

\begin{equation}
z = 1 - \frac{r_+^2}{r^2}\, ,
\label{eq:Definition coordinate z for viscosity}
\end{equation}
defined in such a way that the horizon is located at $z = 0$ and the boundary at $z = 1$. Also, let us introduce the function $\tilde f$ as 
\begin{equation}
f = \frac{r^2}{L^2} \tilde{f}\, , \qquad \lim_{r \rightarrow \infty} \tilde{f}(r) = f_\infty ~.
\end{equation}
With these definitions, the metric \eqref{eq:metricvisco1} reads
\begin{equation}
ds^2 = \frac{r_+^2}{L^2 (1-z)} \left( - \frac{\tilde{f}(z)}{f_\infty} dt^2 + dx^2_{(d-1)} \right) + \frac{L^2}{4 \tilde{f}(z)} \frac{dz^2}{(1-z)^2} ~.
\label{eq:metricvisco2}
\end{equation}
Since the horizon is located at $z = 0$ in these coordinates, we can expand the function $\tilde{f}(z)$ around it as
\begin{equation}
\tilde{f}(z) = \tilde{f}_+' z + \frac{1}{2} \tilde{f}_+'' z^2 + \frac{1}{6} \tilde{f}_+''' z^3 + \cdots ~.
\label{eq:Expansion ftilde around the horizon}
\end{equation}

The shear viscosity can be computed from the retarded Green function of the stress-energy tensor by means of Kubo's formula,

\begin{equation}
\eta=-\lim_{\omega\rightarrow 0}\frac{1}{\omega}\text{Im}\, G^{R}_{12,12}(\omega,\vec{k}=0)\, ,
\end{equation}
where $G^{R}_{12,12}(\omega,\vec{k}=0)$ is a component of said Green function at zero momentum,

\begin{equation}
G^{R}_{ab,cd}(\omega,\vec{k}=0)=-i\int dt d^{d-1}\vec{x}\theta(t)\langle\left[T_{ab}(\vec{x}),T_{cd}(0)\right]\rangle\, .
\end{equation}
This correlator can then be computed holographically by considering a metric perturbation $g_{\mu\nu}\rightarrow g_{\mu\nu}+h_{\mu\nu}$, with $h_{12}\neq 0$ \cite{Policastro:2002se,Son:2002sd}. A rigorous computation of this quantity for higher-derivative theories (at perturbative order) can be found, for instance, in \cite{Myers:2009ij}. However, the method can be summarized in the much simpler recipe found in Ref.~\cite{Paulos:2009yk}, later applied to several cases \cite{Myers:2010jv,Dehghani:2013ldu,Bueno:2018xqc,Mir:2019ecg,Mir:2019rik}. Nonetheless, as a check we will compare our results in the perturbative limit with those of \cite{Myers:2009ij}.

According to the recipe provided by \cite{Paulos:2009yk}, it is enough to consider a metric perturbation obtained by performing
\begin{equation}
dx^2 \rightarrow dx^2 + \varepsilon e^{-i \omega t} dx^1
\end{equation}
in the background \req{eq:metricvisco2}, where $\varepsilon$ is a small parameter. By evaluating the Lagrangian $\sqrt{|g|} \Lag$ in this perturbed metric and expanding it to second order in $\varepsilon$, one finds that this produces a pole at $z = 0$. Then, the shear viscosity can be read off from the residue of this pole,

\begin{equation}\label{eq:polemethod}
\eta = - 8 \pi T \lim_{\omega, \varepsilon \rightarrow 0} \frac{\text{Res}[\sqrt{|g|} \Lag, z = 0]}{\omega^2 \varepsilon^2} ~.
\end{equation}

This computation is straightforward with the help of a computer algebra system, and it yields the following value for the viscosity in our theory \eqref{eq:EQTfour}, 

\begin{equation}
\eta = \frac{1}{16 \pi G (d-2) L^{d-1} r_+^{d+1}} \left[ r_+^{2d} (d - 2 - 4 \tilde{f}_+' \lambda) + 2 L^2 Q^2 r_+^2 \left( (5d-4) \alpha_1 + 2 \alpha_2 \right) \right]\, .
\end{equation}
On the other hand, the entropy density in the boundary theory was obtained in Section~\ref{sec:Thermoflat} and it reads

\begin{equation}
s = \frac{1}{4 G} \left( \frac{r_+}{L} \right)^{d-1} \left( 1 + 2 L^2 Q^2 r_+^{-2(d-1)} \alpha_1 \right)\, .
\end{equation}
Therefore the ratio of the shear viscosity to the entropy density is given by
\begin{equation}
\frac{\eta}{s} = \frac{1}{4 \pi} \frac{r_+^{2d} ( d - 2 - 4 \tilde{f}_+' \lambda ) + 2 L^2 r_+^2 Q^2 \left( (5d-4) \alpha_1 + 2 \alpha_2 \right) }{(d-2) \left[ r_+^{2d} + 2 L^2 Q^2 r_+^2 \alpha_1 \right]} \, .
\label{eq:etas0}
\end{equation}
The coefficient of the expansion $\tilde{f}_+'$ is related to the temperature in Eq.~\eqref{eq:Temperature} and it reads
\begin{equation}
\tilde{f}_+' = \frac{2 \pi \sqrt{f_\infty} L^2}{r_+} T = \frac{d(d-1) r_+^{4d} - 2 L^2 Q^2 r_+^{2(d+1)} + L^4 Q^4 r_+^4 \beta}{2 (d-1) r_+^{2d} \left[ r_+^{2d} - 2 L^2 Q^2 r_+^2 \alpha_1 \right]}\, .
\end{equation}
Finally, we can write everything in terms of the parameter $p$ introduced in Eq.~\req{eq:ppardef}. After some massaging, we can write $\eta/s$ in an appealing way as follows,

\begin{equation}
\frac{\eta}{s} = \frac{1}{4 \pi} \left[ 1 + \frac{4 p^2 \left( (2d-1) \alpha_1 + \alpha_2 \right) }{(d-2) \left( 1 + 2 p^2 \alpha_1 \right) } - \lambda\frac{2 \left( d (d-1) - 2 p^2 + p^4 \beta \right)}{(d-1) (d-2) \left( 1 - 4 p^4 \alpha_1^2 \right) }  \right]\, ,
\label{eq:etas1}
\end{equation}
which is a function of the ratio between the chemical potential and the temperature $\hat\mu=\mu \ell_{*}/(4\pi L T)$, whose relation to $p$ we repeat here for convenience,

\begin{align}\notag
\hat{\mu}  &= \frac{p}{\left( d (d-1) - 2 p^2 + \beta p^4 \right)} \Bigg[ (d-1)  \left( \frac{1}{d-2} - d \alpha_1 \right) - \left( \frac{2 \alpha_1}{d-2} + \frac{(d-1)\beta}{3d-4} \right) p^2 \\
&-  \frac{(d-2) \alpha_1 \beta p^4}{3d-4} \Bigg]\, .
\label{eq:muhat2}
\end{align}

Let us first of all compare our result with that of Ref.~\cite{Myers:2009ij}. Working at linear order in the couplings, it is enough to invert \req{eq:muhat2} at zeroth order, and thus we get, in $d=4$,
\begin{equation}
\frac{\eta}{s}\Big|_{d=4} = \frac{1}{4 \pi} \left[ 1  - 4\lambda+\frac{8(8\hat\mu)^2}{\left(1+\sqrt{1+\frac{2}{3}\left(8\hat\mu\right)^2}\right)}\left(  7 \alpha_1 + \alpha_2  + \frac{\lambda}{3}\right)  +\ldots\right]\, .
\label{eq:etaspert}
\end{equation}
Now, we have to take into account that our $\hat\mu$ is related to their $\bar\mu$ by $\bar\mu=8\hat\mu$ (one factor of $4$ comes from our definition as $\hat\mu=\mu \ell_{*}/(4\pi L T)$, and an additional factor of 2 is due to the different normalization of the vector fields). In addition, it is clear that $\lambda$ is related to the coupling $c_1$ of   \cite{Myers:2009ij} by $\lambda/2=c_1$, as these are the coefficients of the Riemann square term. We also see, by using the Maxwell-frame action of our theory given by \req{eq:dualEQGfour},  that the coefficient of the $R_{\mu\nu\rho\sigma}F^{\mu\nu}F^{\rho\sigma}$ term, called $c_2$ in \cite{Myers:2009ij}, is precisely $4c_2=7\alpha_1+\alpha_2$ (the factor of $4$ accounts for the different normalization of the vector field). Thus, we reproduce Eq.~(3.25) of \cite{Myers:2009ij}, which serves as a consistency check of the pole method \req{eq:polemethod} we employed.
Let us now study the properties of $\eta/s$ in a fully non-perturbative level.

It is clear from Eq.~\req{eq:etas1} that there are two contributions to $\eta/s$: one proportional to $\lambda$ and another one proportional to the combination $ (2d-1) \alpha_1 + \alpha_2$. Interestingly, this is the quantity that appears in the numerator of $a_2$ in Eq.~\req{eq:a2EQG}. The GB coupling also controls the energy flux parameter $t_2$, and therefore, in these holographic CFTs it follows that 
\begin{equation}\label{eq:a2t2connection}
a_2=t_2=0\,\, \Rightarrow \frac{\eta}{s}=\frac{1}{4\pi}\, .
\end{equation}
It is tantalizing to speculate about a possible relation between having ``trivial" three-point functions $\langle TTT\rangle$, $\langle TJJ\rangle$ and the absence of corrections to $\eta/s$ for more general CFTs. However, this is probably just an accident because our theory has only a few parameters. Note that in fact, the combination $(2d-1) \alpha_1 + \alpha_2$ corresponds to the coupling of the term $\tensor{\Hsq}{^{\mu\nu}_{\rho\sigma}} \tensor{R}{^{\rho\sigma}_{\mu\nu}}$ in the action, so it is no surprise that it controls both $a_2$ and the corrections to $\eta/s$. Additional higher-derivative terms will probably spoil this connection, but it is nevertheless quite remarkable that \req{eq:a2t2connection} holds non-perturbatively for our theories. 
We also observe that in the extremal limit, which happens when $ d (d-1) - 2 p^2 + p^4 \beta=0$, the GB contribution to $\eta/s$ vanishes, and one is left with the contribution coming from the non-minimal couplings. This feature was already observed by Refs.~\cite{Cai:2008ph,Myers:2009ij}. 

Let us study the dependence of the shear viscosity to entropy density ratio on the chemical potential. We focus on the values of the $\alpha_1$ and $\beta$ parameters that give rise to a single phase with an extremal limit, as studied in Section~\ref{sec:thermoab}. All the parameters are additionally constrained by the bounds found in Section~\ref{sec:constraints}. 
Let us rewrite \req{eq:etas1} as follows
\begin{equation}
\frac{\eta}{s} = \frac{1}{4 \pi} \left[ 1 + \left( (2d-1) \alpha_1 + \alpha_2 \right)f_1(p) - \lambda f_2(p)  \right]\, ,
\label{eq:etas2}
\end{equation}
with
\begin{equation}
f_1(p)=\frac{4 p^2}{(d-2) \left( 1 + 2 p^2 \alpha_1 \right) }\, ,\qquad f_2(p)=\frac{2 \left( d (d-1) - 2 p^2 + p^4 \beta \right)}{(d-1) (d-2) \left( 1 - 4 p^4 \alpha_1^2 \right) }\, .
\end{equation}
Since $\alpha_1\ge 0$, it is straightforward to see that $f_1(p)$ is a monotonically growing function, ranging from $f_1(0)=0$ to its maximum value at extremality $p=p_{\rm ext}$. On the other hand, $f_2(p)$ has just the opposite behavior: it takes its maximum value at $p=0$ and then it decreases to zero at extremality.\footnote{We recall that we take $\alpha_1$ to satisfy $1-2\alpha_1 p_{\rm ext}^2>0$.}

Since $\lambda\ge 0$ by virtue of the WGC, the GB contribution is always negative, but monotonically increasing as we increase the chemical potential. Then, the global behavior of $\eta/s$ depends strongly on the sign of $a_2$, so we have to distinguish both cases.

\subsection*{Case 1: $a_2\le 0$}

It follows that, whenever $a_2\le 0$ (so that $(2d-1) \alpha_1 + \alpha_2\ge 0$), the shear viscosity to entropy density ratio is a growing function of $\hat\mu$. It therefore reaches its minimum value for $\hat\mu=0$, 

\begin{equation}
\frac{\eta}{s}\Big|_{\hat \mu=0} = \frac{1}{4 \pi} \left[ 1  - \lambda \frac{2  d (d-1)}{(d-1) (d-2)}  \right]\, .
\end{equation}
In addition, the largest value of $\lambda$ is given by the upper bound in \eqref{eq:lambdaconstr}, and therefore we find that the absolute minimum value of $\eta/s$ whenever $a_2\le 0$ is

\begin{equation}
\min\left[\frac{\eta}{s}\right] = \frac{1}{4 \pi} \left[ 1 - \frac{d (d-3) ( d^2 - d + 6 )}{2 (d^2 - 3d + 6)^2} \right]\, . 
\end{equation}
This is the lower bound found for GB gravity  \cite{Brigante:2007nu,Brigante:2008gz,deBoer:2009pn,Buchel:2009sk}. One can see that $\eta/s$ never reaches zero, and its minimum is reached for $d=8$, which gives $4\pi \eta/s\ge \frac{219}{529}\approx 0.41399$. 
Likewise, for $a_2\le 0$ the maximum value of $\eta/s$ as a function of the chemical potential is reached at extremality, \textit{i.e}, for
\begin{equation}
p_{\rm ext}=\sqrt{\frac{1}{\beta} \left( 1 - \sqrt{1 - d (d-1) \beta} \right) } \, .
\end{equation}
This gives

\begin{equation}
\frac{\eta}{s}\bigg|_{\rm ext}= \frac{1}{4 \pi} \left[ 1 + \left( (2d-1) \alpha_1 + \alpha_2 \right)f_1(p_{\rm ext})\right]\, .
\end{equation}
Furthermore, since $f_1(p)$ is a growing function, this value will be larger if $p$ is larger. Now, the maximum value of $p_{\rm ext}$ is reached for $\beta=1/(d(d-1))$, which is the largest value of $\beta$ compatible with the existence of an extremal limit, and for this we obtain $p_{\rm ext}^2=d(d-1)$. Therefore, we can already put the following upper bound on $\eta/s$,

\begin{figure}
	\centering
	\includegraphics[width=0.6\textwidth]{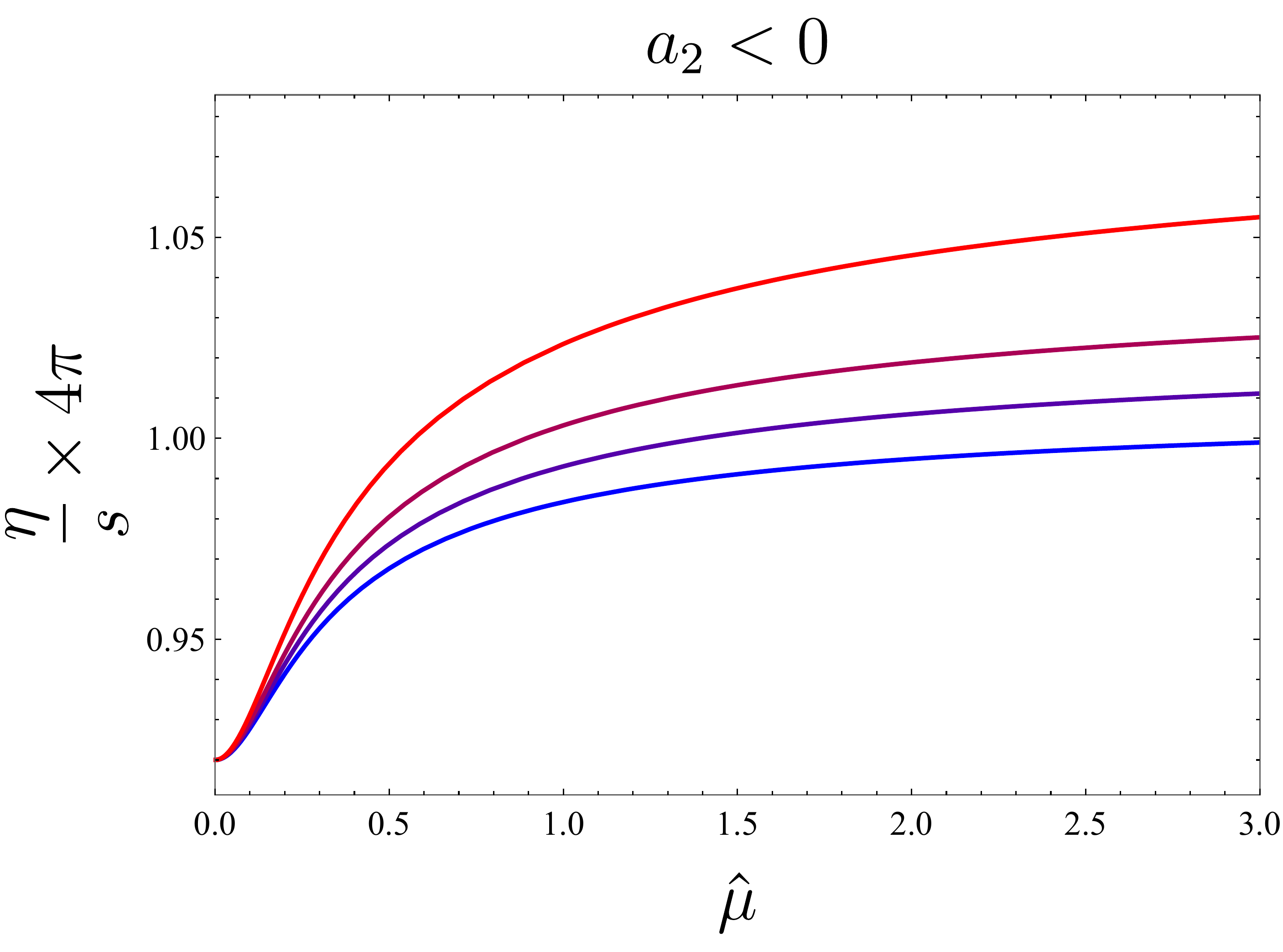}
	\caption{Shear viscosity over entropy density as a function of the chemical potential in the case in which $a_2<0$. We take $\lambda=0.02$, and from blue to red, the different curves correspond to $\{\alpha_1,\alpha_2,\beta\}=\{0.0024,-0.016,0.082\}$, $\{0.0019,-0.012,0.080\}$,$\{0.0006,-0.0023, 0.077\}$,$\{0,0.0038,0.083\}$, which satisfy all the physical constraints. As a general rule,  $\eta/s$ is a growing function of $\hat\mu$ and its value does not depart much from $1/(4\pi)$. }
	\label{fig:etas1}
\end{figure}

\begin{equation}\label{eq:etasmax1}
\frac{\eta}{s}\le \frac{1}{4 \pi} \left[ 1 + \frac{4 d(d-1)\left( (2d-1) \alpha_1 + \alpha_2 \right)}{(d-2) \left( 1 + 2 d(d-1) \alpha_1 \right) }\right]\, .
\end{equation}
Now we can further explore what are the values of the parameters $\alpha_1$ and $\alpha_2$, compatible with the constraints in Section~\ref{sec:constraints}, that maximize this value. First of all, we have $\alpha_1\ge 0$ on account of the WGC. Then, we have the unitarity constraints \req{eq:a2c1} and \req{eq:a2c2} that only involve the rescaled couplings $f_{\infty}\alpha_{1,2}$. Since \req{eq:etasmax1} has no local maximum inside the region delimited by \req{eq:a2c1} and \req{eq:a2c2}, the maximum must occur at the boundary (where the constraints are saturated). By studying what happens for a few values of $d$, it is not hard to see that the maximum is reached at the intersection of $\alpha_1=0$ and the boundary given by \req{eq:a2c1}. We obtain
\begin{equation}\label{eq:maxetas1}
\frac{\left( (2d-1) \alpha_1 + \alpha_2 \right)}{ 1 + 2 d(d-1) \alpha_1  }\Bigg|_{\req{eq:a2c1}}\le \frac{1}{(3d-2)f_{\infty}}\, ,
\end{equation}
and the maximum happens for $\alpha_2=\frac{1}{(3d-2)f_{\infty}}$. Since $f_{\infty}\ge 1$, this value is larger when $\lambda=0$ $(f_{\infty}=1)$.
On the other hand, we can consider the WGC constraint given by Eq.~\req{eq:wgc2}. The looser bound on the $\alpha_{1,2}$ couplings is precisely achieved for $\lambda=0$ and the maximum allowable value of $\beta$, this is, $\beta=1/(d(d-1))$. This yields
\begin{equation}\label{eq:wgc6}
3(d-1)\alpha_1+\alpha_2\le \frac{(d-2)}{4d(3d-4)}\, .
\end{equation}
Now, when we combine this together with $\alpha_1\ge 0$, we find

\begin{equation}\label{eq:maxetas2}
\frac{\left( (2d-1) \alpha_1 + \alpha_2 \right)}{ 1 + 2 d(d-1) \alpha_1  }\Bigg|_{\req{eq:wgc6}}\le \frac{d-2}{4 d (3 d-4)}\, ,
\end{equation}
and the maximum is reached for $\alpha_1=0$, $\alpha_2=\frac{d-2}{4 d (3 d-4)}$. We can see that, in every dimension $d\ge 3$, the bound \req{eq:maxetas2} is stronger than \req{eq:maxetas1}, and therefore \req{eq:maxetas2} is the relevant one. 
Therefore, we conclude that the maximum possible value for $\eta/s$ is
\begin{equation}
\max\left[\frac{\eta}{s}\right] = \frac{1}{4 \pi} \left[ 1 +\frac{d-1}{3d-4} \right]\, . 
\end{equation}
Among these, the larger value happens for $d=3$, which gives $(7/5)/(4\pi)$.

In conclusion, in the case $a_2\le 0$, the shear viscosity to entropy density ratio is a growing function of the dimensionless chemical potential, and it has absolute lower and upper bounds, given by
\begin{equation}\label{eq:etasmaxmin}
\frac{219}{529}\le  4\pi \frac{\eta}{s}\le \frac{7}{5}\, .
\end{equation}
Remarkably, these hold in arbitrary dimension, for arbitrary chemical potential and for any value of the higher-order couplings that satisfy the physical conditions that we have discussed. Thus, whenever $a_2\le 0$, the ratio $\eta/s$ cannot depart a lot from the Einstein-Maxwell prediction $1/(4\pi)$. We illustrate this in Fig.~\ref{fig:etas1}, where we show the profile of $\eta/s$ as a function of $\hat\mu$ for several values of the parameters compatible with all the physical constraints. 

\subsection*{Case 2: $a_2>0$}

The situation is very different when we consider $a_2>0$, this is, $(2d-1) \alpha_1 + \alpha_2 < 0$. In this case, both corrections to $\eta/s$ are negative, so this quantity is always smaller than $1/(4\pi)$. On the other hand, the $a_2$ contribution to $\eta/s$ is now a decreasing function of $\hat\mu$, but the GB contribution is still growing, so the overall character of $\eta/s$ depends on the case. By expanding Eq.~\req{eq:etas2} near $p=0$, one can see that this becomes a decreasing function of the chemical potential (at least for small $\hat\mu$) if
\begin{equation}
2 (2 (d-1)\alpha_1 +\alpha_2)+\frac{\lambda  \left(2-4 \alpha_1^2 (d-1) d\right)}{(d-1)}<0\, .
\end{equation}
Since the GB correction on its own is already able to lower the KSS bound, whenever the inequality above is satisfied, we may obtain an even lower bound by turning on the chemical potential. 
In order to answer how much the value of $\eta/s$ can be diminished we need to take into account all physical constraints on the higher-order couplings. However, a general analysis is more complicated than before, since we do not know anymore for which value of $\hat\mu$ the ratio $\eta/s$ reaches its minimum. 

In spite of this, it will be enough to consider a simple example to show that we can lower the value of $\eta/s$ (and therefore that of $\eta$) all the way down to zero. Let us consider $\alpha_1=\lambda=\beta=0$, so that the only active coupling is $\alpha_2$, which furthermore has to be negative, $\alpha_2\le 0$, on account of the WGC bound \req{eq:wgc2}. In that case, the shear viscosity to entropy density ratio reads

\begin{figure}
	\centering
	\includegraphics[width=0.6\textwidth]{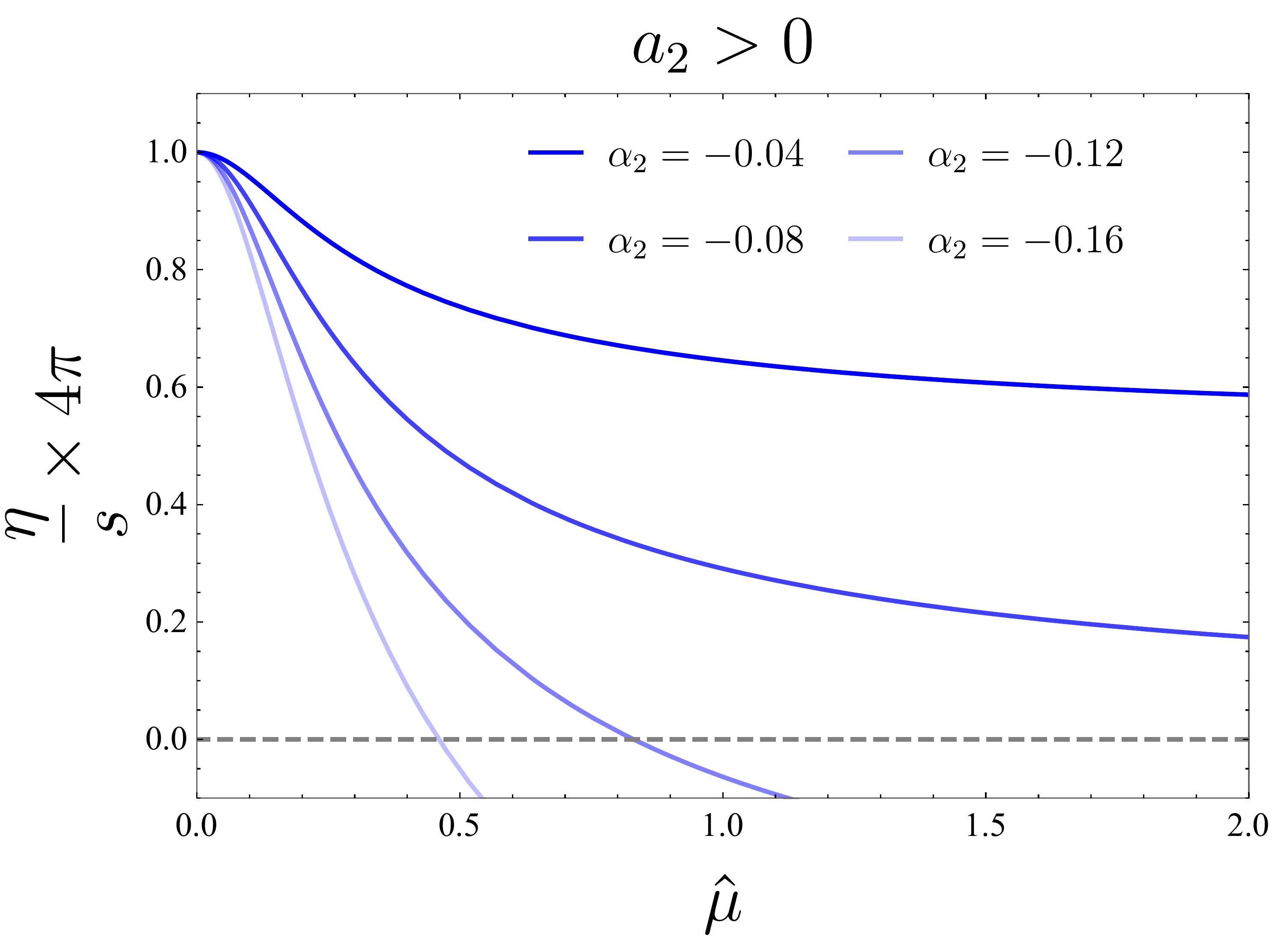}
	\caption{Shear viscosity over entropy density ratio as a function of the chemical potential in the case in which $a_2>0$. We show the representative case $\lambda=\alpha_1=\beta=0$ and $\alpha_{2}<0$ in $d=4$. Even when $\alpha_2$ satisfies all the physical constraints we have studied, it is possible to achieve $\eta/s=0$ providing that the chemical potential is sufficiently large. }
	\label{fig:etas2}
\end{figure}

\begin{equation}
\frac{\eta}{s} = \frac{1}{4 \pi} \left[ 1 + \frac{4p^2}{d-2} \alpha_2 \right]\, ,
\end{equation}
and we show it as a function of $\hat\mu$ in Fig.~\ref{fig:etas2}. 
The parameter $p$ ranges from $0$ to $\sqrt{d(d-1)/2}$ when $\hat\mu$ goes from $0$ to infinity, and therefore, the minimum value of $\frac{\eta}{s}$ (for negative $\alpha_2$) is reached at extremality

\begin{equation}
\frac{\eta}{s}\bigg|_{\rm ext} = \frac{1}{4 \pi} \left[ 1 + \frac{2d(d-1)}{d-2} \alpha_2 \right]\, .
\end{equation}
Then we have to look at the bounds on $\alpha_2$, which are given by Eqs.~\req{eq:a2c1}, \req{eq:a2c2} besides \req{eq:wgc2}, which we are already taking into account. One can see that Eq.~\req{eq:a2c1} does not impose any additional constraint since it is always satisfied for $\alpha_2\le 0$, while \req{eq:a2c2} only sets a lower bound on $\alpha_2$ for $d=3,4,5$, namely

\begin{equation}\label{eq:constralpha2}
\alpha_2\ge -\frac{d-2}{6d-d^2-4} \, . \quad (d=3,4,5)
\end{equation}
In higher dimensions there is no limit to how negative $\alpha_2$ can be, and $\eta/s$ can certainly be taken to zero. 
Focusing on the lower dimensions, which are also the most relevant ones, we see that the minimum value of $\eta/s$ in this example is

\begin{equation}
\min\left[\frac{\eta}{s}\right] = \frac{1}{4 \pi} \left[ 1 - \frac{2d(d-1)}{6d-d^2-4} \right]\, .
\end{equation}
This is in fact negative for the three values $d=3,4,5$. A negative $\eta$ is of course meaningless and it indicates that the plasma is unstable. There is a critical value of $\hat\mu$ at which the viscosity vanishes and above this value it makes no sense to talk about hydrodynamics. From \req{eq:muhat2} we see that this value is
\begin{equation}
\hat\mu_{\rm cr}=\frac{\sqrt{-\frac{(d-2)}{\alpha_2}}}{2d(d-1)+\frac{(d-2)}{\alpha_2}}\, ,
\end{equation}
and its minimum value is precisely reached for the smallest $\alpha_2$ in \req{eq:constralpha2},

\begin{equation}
\hat\mu_{\rm cr}^{\rm min}=\frac{\sqrt{6d-d^2-4}}{3d^2-8d+4}\, . \quad (d=3,4,5)
\end{equation}
This can probably be lowered by considering other couplings besides $\alpha_2$. This result means that the viscosity cannot vanish as long as $\hat\mu<\hat\mu_{\rm cr}^{\rm min}$, for any choice of parameters satisfying the physical constraints. 
For $d\ge 6$ we have $\hat\mu_{\rm cr}^{\rm min}=0$ as $\alpha_2$ is in principle allowed to be arbitrarily negative, so a vanishing $\eta$ can be achieved with an arbitrarily small chemical potential. 

It would be interesting to investigate if other types of constraints, such as those coming from plasma instabilities \cite{Ge:2008ni,Ge:2009ac,Ge:2009eh} or causality deep into the bulk \cite{Camanho:2010ru}, can be used to rule out values of $\eta/s$ very close to zero.  

\section{Charged R\'enyi entropies and generalized twist operators}
\label{sec:renyi}
Entanglement entropy (EE) \cite{Calabrese:2004eu} and R\'enyi entropies (RE) \cite{renyi1961,renyi1} --- as well as  their holographic counterparts \cite{Ryu:2006bv,Ryu:2006ef,Hung:2011nu} --- constitute a very useful way to probe the amount of entanglement in quantum field theories \cite{Klebanov:2011uf,Laflorencie:2015eck}. Given a biparition of the Hilbert space into two subspaces $A$ and $B$, R\'enyi entropies are defined as
\begin{equation}
S_{n}(A)=\frac{1}{1-n} \log \Tr \rho_{A}^{n}\, ,
\end{equation}
where $\rho_{A}=\Tr_{B}\rho$ is the reduced density matrix of the subsystem $A$, obtained by taking the partial trace over the subsystem $B$ of the total density matrix. Here, we are interested in the case in which $A$ and $B$ correspond to the subsystems associated to two spatial regions (at a fixed time) separated by an entangling surface $\Sigma$. The R\'enyi index $n$ is usually considered an integer, which allows one to compute these entropies by using the replica trick \cite{Calabrese:2004eu}. However, if one is able to continue $n$ to an arbitrary real number, then one can recover the entanglement entropy  as the limit $n\rightarrow 1$,

\begin{equation}
S_{\rm EE}(A)=-\Tr\left[\rho_{A}\log(\rho_A)\right]=\lim_{n\rightarrow 1} S_n(A)\, .
\end{equation}

Now, these entropies can be generalized to the case in which the QFT is charged under a global symmetry. The appropriate generalization, proposed in Ref.~\cite{Belin:2013uta}, reads

\begin{equation}
S_{n}(\mu)=\frac{1}{1-n}\log\Tr \left[\bar\rho_{A}(\mu)\right]^n\, ,
\end{equation}
where 
\begin{equation}
\bar\rho_{A}(\mu)=\frac{\rho_{A}e^{\mu Q_A}}{\Tr\left[\rho_{A}e^{\mu Q_A}\right]}\, 
\end{equation}
is a new density matrix that depends on the chemical potential $\mu$, conjugate to the charge $Q_A$ enclosed in the region $A$. 

Let us focus on the case in which the quantum theory is defined in flat space and the entanglement surface $\Sigma$ is a sphere of radius $R$, namely $\Sigma=\mathbb{S}^{d-2}(R)$. For a CFT, one can then prove, by using the Casini-Huerta-Myers map \cite{Casini:2011kv}, that these charged R\'enyi entropies are related to the thermal entropy of the same theory placed on the hyperbolic cylinder $\mathbb{S}^1\times\mathbb{H}^{d-1}(R)$. The precise relation reads \cite{Hung:2011nu,Belin:2013uta}
\begin{equation}\label{eq:renyithermo}
S_n(\mu)=\frac{n}{n-1}\frac{1}{T_0} \int_{T_0/n}^{T_0}S_{\rm thermal}(T,\mu)dT\, ,
\end{equation}
where 

\begin{equation}
T_0=\frac{1}{2\pi R}\, .
\end{equation}
We remark that this is a formula that applies to a CFT, but from here it is evident how to compute these quantities holographically. In fact, the thermal entropy of a holographic CFT on $\mathbb{S}^1\times\mathbb{H}^{d-1}(R)$ is nothing but the Wald's entropy of a black hole with a hyperbolic horizon. In this section we explore the properties of the holographic RE for the theory \req{eq:EQTfour}, and afterwards we also analyze a couple of related quantities: the scaling dimension and the magnetic response of generalized twist operators \cite{Belin:2013uta}.

\subsection{R\'enyi entropies}
In order to compute charged R\'enyi entropies for the holographic CFTs dual to \req{eq:EQTfour}, we have to consider charged black hole solutions with hyperbolic horizons, which for our theories take the form
\begin{equation}
ds^2=-N_{-1}^2 f(r)dt^2+\frac{dr^2}{f(r)}+r^2d\Xi^2\, ,
\end{equation}
where $N_{-1}$ is a constant, $d\Xi^2$ is the hyperbolic space of unit radius and $f(r)$ is given by Eq.~\req{eq:fsolEQG} with $k=-1$. Since $f(r)$ behaves asymptotically as $f(r)\sim r^2f_{\infty}/L^2$, we set the constant $N_{-1}$ to
\begin{equation}
N_{-1}=\frac{L}{\sqrt{f_{\infty}} R}\, .
\end{equation}
In this way, the boundary metric is conformal to
\begin{equation}
ds^2_{\rm bdry}=-dt^2+R^2d\Xi^2\, ,
\end{equation}
so that the spatial slices are hyperbolic spaces of radius $R$. The R\'enyi entropies across a spherical region are then computed through the integral \req{eq:renyithermo}, where $S_{\rm thermal}$ is the black hole entropy, given by \req{eq:entropy1}. Notice that it is important that $S_{\rm thermal}$ is considered as a function of $T$ and $\mu$, so that the integration is carried out at constant $\mu$. Although at first sight the integration may look tricky, it is nonetheless straightforward, since the first law \req{eq:1stlawomega} implies

\begin{equation}
S=-\frac{\partial\Omega(T,\mu)}{\partial T}\, .
\end{equation}
We then obtain 

\begin{equation}
S_n=\frac{n}{n-1} \frac{1}{T_0} \left ( \Omega(T_0/n,\mu)-\Omega(T_0,\mu) \right)\,.
\label{eq:renyiomega}
\end{equation}
Back in Eq. \eqref{eq:omegagen} we already obtained the expression for the grand canonical potential $\Omega$ in terms of the horizon radius and the charge for spherical, planar or hyperbolic horizon topologies. Setting $k=-1$, defining $x=r_+/L$ and $Q=p x^{d-1} L^{d-2}$ and writing $T=T_0/n=(2 \pi R n)^{-1}$, the expression for $\Omega$ reduces to
\begin{equation}
\begin{aligned}
\Omega=&\frac{L^{d-1} V_{-1,(d-1)}}{16\pi G \sqrt{f_\infty} R}\Bigg[(d-1) x^d-\frac{2 p^2 x^{d}}{d-2}-x^{d-2} \left((d-1) +\frac{2 x\sqrt{f_\infty}}{n}  \right)+\frac{3 \beta p^4 x^{d}}{3 d-4}\\
   &-(d-1) \lambda 
  x^{-4+d} \left(1+\frac{4  x \sqrt{f_\infty}  }{n(d-3)}\right) +2  \alpha _1  p^2 x^{d-2}
   \left(3 (d-1) +\frac{2 x\sqrt{f_\infty} R}{n} \right)+2  \alpha _2 p^2 x^{d-2} \Bigg]\, .
\end{aligned}
\label{eq:omegahyp}
\end{equation}
However, on account of \eqref{eq:renyiomega}, we need to write $\Omega$ in terms of $n$ and $\mu$, so that we have to find the relations $x=x(n,\mu)$ and $p=p(n,\mu)$. For that, it is convenient to present the expressions of $n$ and $\mu$ in terms of $x$ and $p$, which follow after setting $k=-1$ in Eqs. \eqref{eq:Temperature2} and \eqref{eq:potential2}: 
\begin{align}
\nonumber
		\frac{1}{n} = & \frac{1}{2 x \sqrt{f_\infty}    \left(1 - 2 p^2 \alpha_1 - 2   \lambda x^{-2} \right) } \Bigg[ \left( -(d-2)  + d x^2 + (d-4)  \lambda x^{-2}\right) \\\label{eq:nrel}
		& - \frac{2 p^2}{(d-1)} \left(x^2 - d   (3(d-1)\alpha_1 + \alpha_2) \right) + \frac{\beta  x^2 p^4}{(d-1)} \Bigg] \, ,\\
	\mu=&  \frac{L p}{\ell_{*} \sqrt{f_\infty} R } \Bigg[ \frac{x}{(d-2)} - \frac{\alpha_1}{x} \left( 3 (d-1) + \frac{2 x  \sqrt{f_\infty}}{n}  \right) - \frac{ \alpha_2 }{x} - \frac{ x p^2 \beta}{(3d - 4)} \Bigg] \, .\label{eq:prel}
\end{align}
The equations \req{eq:omegahyp}, \req{eq:nrel} and \req{eq:prel} allow us to study the R\'enyi entropies \req{eq:renyiomega} exactly. A useful intermediate expression for  $\Omega_n(\mu)\equiv \Omega(T_0/n,\mu)$, is the following one,
\begin{equation}
\begin{aligned}
\Omega_n(\mu)=\frac{L^{d-1} V_{-1,(d-1)}}{16\pi G \sqrt{f_\infty} R}\Bigg[&(d-1) x^{d-4}(x^4-x^2+\lambda)+ \frac{\beta p^4 x^d}{(3d-4)}\\&-2 \frac{\sqrt{f_\infty}}{n} x^{d-1} \left ( 1- \frac{2 \lambda(d-1)}{(d-3) x^2} \right) -2 \frac{\ell_{*} R \sqrt{f_\infty}}{L} \mu  p x^{d-1} \Bigg ] \, ,
\end{aligned}
\label{eq:omegasim}
\end{equation}  
which is a bit simpler, but it still depends on $x=x(n,\mu)$ and $p=p(n,\mu)$. In practice, it seems extremely challenging (if not impossible) to analytically invert the equations \req{eq:nrel} and \req{eq:prel} to encounter $x=x(n,\mu)$ and $p=p(n,\mu)$ explicitly. To circumvent this impediment, we focus next in two limiting regimes, namely, small $\mu$ and $\mu\rightarrow \infty$.

\subsubsection{Small $\mu$}
In order to reduce the clutter, let us introduce the notation

\begin{equation}
\bar\mu=\frac{\ell_{*} R \sqrt{f_\infty}}{L} \mu\, ,
\end{equation}
as this combination appears everywhere. 
We consider here the case in which $\bar\mu<<1$, so that carrying out the inversion procedure of Eqs.~\req{eq:nrel} and \req{eq:prel} in a perturbative expansion in $\bar\mu$ suffices. Furthermore, as an attempt to make explicit computations and capture the effects produced by the non-minimal couplings, we are going to set $\lambda=0$ all along this section (so $f_\infty=1$). After all, the effect of the GB coupling on (uncharged) R\'enyi entropies is known \cite{Hung:2011nu} --- see also  \cite{Galante:2013wta,Dey:2016pei,Puletti:2017gym,Bueno:2018xqc} for other studies of holographic RE in higher-order gravities.

Consequently, we can expand $x(n,\mu)$ and $p(n,\mu)$ as
\begin{equation}
\label{eq:xqpert}
x(n,\mu)=\hat{x}_n+ \delta \hat{x}_n\, \bar\mu^2+\mathcal{O}{\left (\bar\mu^4 \right)} \,, \qquad p(n,\mu)=\delta p_n \bar\mu+ \mathcal{O}\left (\bar\mu^3 \right)\,.
\end{equation}
By solving Eqs.~\req{eq:nrel} and \req{eq:prel}, the coefficients $\hat{x}_n$, $\delta \hat{x}_n$ and $\delta \tilde{p}_n$ can be found to be
\begin{align}
\hat{x}_n=&\frac{n^{-1}+ \sqrt{n^{-2}+d(d-2)} }{d}\, , \\
\delta \hat{x}_n=& -\frac{2(d-2)^2 \hat{x}_n^3 (2(d^2-1)\alpha_1+\hat{x}_n^2(d(d-1)\alpha_1-1)+d \alpha_2)}{(d-1)(d(\hat{x}_n^2+1)-2)(\hat{x}_n^2(d(d-2)\alpha_1-1)+(d-2)((2d-1)\alpha_1+\alpha_2))^2}\,,\\
\delta p_n=&\frac{(d-2) \hat{x}_n}{\alpha_{\mathrm{eff}}^{\mathrm{EQG}}-(\hat{x}_n^2-1)(d(d-2)\alpha_1-1)}\, .
\end{align}
We recall that $\alpha_{\mathrm{eff}}^{\mathrm{EQG}}$, given in Eq.~\req{eq:alphaeffEQG}, is the combination that appears in the denominator of the central charge $C_J$ in Eq.~\req{eq:CJgen}. Taking into account these perturbative expansions, $\Omega_n$ can be written in the following explicit form:

\begin{equation}
\begin{aligned}
\Omega_n(\mu)=&-\frac{L^{d-1} V_{-1,(d-1)}}{16 \pi G R}\left[\hat{x}_n^{d-2}(\hat{x}_n^2+1)+\frac{2(d-2) \hat{x}_n^d}{\alpha_{\mathrm{eff}}^{\mathrm{EQG}}-(\hat{x}_n^2-1)(d(d-2)\alpha_1-1)} \bar\mu^2 \right]\\&+\mathcal{O}{\left (\bar\mu^4 \right)}\,.
\end{aligned}
\end{equation}
Now, noting that $\hat{x}_1=1$, we have
 
\begin{align}
\Omega_1(\mu)=&-\frac{L^{d-1} V_{-1,(d-1)}}{8 \pi G R}\left[1+\frac{d-2}{\alpha_{\mathrm{eff}}^{\mathrm{EQG}}}\bar\mu^2 \right]+\mathcal{O}{\left (\bar\mu^4 \right)}\,.
\end{align}
Form here, we can infer the following form for the  $n$-th R\'enyi entropy:

\begin{equation}
\begin{aligned}
S_n=&\frac{n L^{d-1} V_{-1,(d-1)}}{4 (n-1) G } \left[ \frac{2- \hat{x}_n^{d-2}(\hat{x}_n^2+1)}{2}+\frac{(d-2)}{\alpha_{\mathrm{eff}}^{\mathrm{EQG}}}\left (1-\frac{\hat{x}_n^d}{1-\frac{(\hat{x}_n^2-1)}{\alpha_{\mathrm{eff}}^{\mathrm{EQG}}}(d(d-2)\alpha_1-1)} \right) \bar\mu^2 \right]\\&+\mathcal{O}{\left (\bar\mu^4 \right)}\, . 
\end{aligned}
\end{equation}
Let us remark at this point that the volume $V_{-1,(d-1)}$ is a (diverging) function of the ratio between the radius of the entangling surface $R$ and a cut-off $\delta$. In fact, the leading term gives an area law,

\begin{equation}
V_{-1,(d-1)}=\frac{V_{\mathbb{S}^{d-2}}}{d-2}\frac{R^{d-2}}{\delta^{d-2}}+\ldots\, , \quad \text{where} \quad V_{\mathbb{S}^{d-2}}=\frac{2 \pi^{(d-1)/2}}{\Gamma[(d-1)/2]}\, .
\end{equation}

It is interesting to keep only the universal part in this expansion, which will provide us with the regularized RE. In even $d$, the series expansion of the volume contains a term $\log(R/\delta)$, and it is clear that the coefficient of this term is universal as it is invariant under rescalings of the cut-off. On the other hand, for odd $d$ the series contains a constant term. The universality of this term is less clear, as it could be shifted by performing a rescaling of $R$ of  the form $R\rightarrow R(1+c \delta)$, but we will not worry about this issue here.\footnote{A natural regulator in that case is provided by the mutual information \cite{Casini:2015woa}, which is a UV finite quantity.}
Taking this into account, one can see that the universal part of the volume reads \cite{Casini:2011kv}

\begin{equation}
\label{eq:regvol}
V_{-1,d-1}^{\rm univ.}=\frac{\nu_{d-1}}{4 \pi} V_{\mathbb{S}^{d-1}}\, , \quad \text{where} \quad \nu_{d-1}=\left \lbrace \begin{matrix}
(-)^{\frac{d-2}{2}} 4  \log  (R/\delta) & & & d \text{ even}\, ,\\ & & & \\
(-)^{\frac{d-1}{2}} 2  \pi & & & d \text{ odd}\,.
\end{matrix} \right. 
\end{equation}
We will use this regularized volume from now on. It is also useful to introduce the following quantity,
\begin{equation}
a^*=\frac{L^{d-1}}{8G} \frac{\pi^{(d-2)/2}}{\Gamma(d/2)}\,,
\end{equation}
which represents the universal contribution to the regularized EE in holographic Einstein gravity. This parameter can also be easily computed for higher-curvature gravities \cite{Myers:2010tj,Myers:2010xs} and in general it coincides with the $a$-type trace-anomaly charge in the case of even $d$, while in odd dimensions it is proportional to the free energy of the corresponding theory evaluated on $\mathbb{S}^{d}$ \cite{Casini:2011kv}.
Using this parameter, we can finally write our holographic REs as

\begin{equation}
\begin{aligned}
S_n=\frac{n a^* \nu_{d-1}}{(n-1) } &\Bigg [ \frac{2- \hat{x}_n^{d-2}(\hat{x}_n^2+1)}{2}+\frac{(d-2)}{\alpha_{\mathrm{eff}}^{\mathrm{EQG}}}\left (1-\frac{\hat{x}_n^d}{1-\frac{(\hat{x}_n^2-1)}{\alpha_{\mathrm{eff}}^{\mathrm{EQG}}}(d(d-2)\alpha_1-1)} \right)\bar\mu^2 \Bigg]\\&+\mathcal{O}{\left (\bar\mu^4 \right)}\, .
\end{aligned}
\label{eq:renyin}
\end{equation}

Let us then explore the properties of these entropies, starting with the relevant case of the entanglement entropy $n\rightarrow 1$.  This limit yields

\begin{equation}
S_{\small \mathrm{EE}}=\underset{n \rightarrow 1}{\lim} S_n= a^* \nu_{d-1} \left[1+\frac{(d-2)^2 (1-3d(d-1)\alpha_1-d \alpha_2)}{(d-1) \left ( \alpha_{\mathrm{eff}}^{\mathrm{EQG}}\right )^2} \bar\mu^2 \right]+\mathcal{O}{\left (\bar\mu^4 \right)}\, .
\label{eq:see}
\end{equation}


\noindent 
It is interesting to wonder about the sign of the coefficient of $\mu^2$ in \eqref{eq:see}, or more precisely, of the quantity

\begin{equation}
\frac{\partial_{\bar\mu}^2S_{\rm \small EE}}{S_{\rm \small EE}}\bigg|_{\mu=0}=\frac{2(d-2)^2 (1-3d(d-1)\alpha_1-d \alpha_2)}{(d-1) \left ( \alpha_{\mathrm{eff}}^{\mathrm{EQG}}\right )^2}\, .
\end{equation}
In Einstein-Maxwell theory we can see it is positive, so that the holographic entanglement entropy grows when we turn on a chemical potential. Could this be different in other theories?
If the parameters $\alpha_1$ and $\alpha_2$ were arbitrary, this coefficient could have either sign, but we must take into account the constraints in Sec.~\ref{sec:constraints}. In fact, it suffices to consider the unitarity constraints in \ref{sec:TJJconstr}.  Let us first note that the unitarity constraint \req{eq:a2c1} can be expressed as
\begin{equation}
-\frac{2}{d-2}+2d\alpha_1+\frac{3d-2}{d(d-2)} \alpha_{\mathrm{eff}}^{\mathrm{EQG}} \geq 0\, .
\label{eq:ineuni}
\end{equation}
Then, we have
\begin{equation}
1-3d(d-1)\alpha_1-d \alpha_2=-\frac{2}{d-2}+2 d \alpha_1+\frac{d}{d-2}\alpha_{\mathrm{eff}}^{\mathrm{EQG}} > -\frac{2}{d-2}+2d\alpha_1+\frac{3d-2}{d(d-2)} \alpha_{\mathrm{eff}}^{\mathrm{EQG}} \geq 0\,,
\end{equation}
where we simply used that $\alpha_{\mathrm{eff}}^{\mathrm{EQG}} >0$ and that $\frac{3d-2}{d(d-2)}<\frac{d}{d-2}$ for $d\ge 3$. Note that the result we obtain is a strict inequality, since $\alpha_{\mathrm{eff}}^{\mathrm{EQG}}=0$ is not allowed. Thus, this result implies that 
\begin{equation}
\frac{\partial_{\bar\mu}^2S_{\rm \small EE}}{S_{\rm \small EE}}\bigg|_{\mu=0}>0
\label{eq:seepos}
\end{equation}
for all the (unitary) holographic CFTs dual to our bulk theories. Given the robustness of this result, it is very tempting to conjecture that the entanglement entropy should always grow with the chemical potential for any unitary CFT at zero temperature.\footnote{This is in line with the results of Ref.~\cite{Kundu:2016dyk} for the holographic EE of an infinite rectangular strip in the case of $\mu\neq 0$ and $T=0$.}

Remarkably, it is possible to extend this result to prove that the coefficient of $\mu^2$ for all R\'enyi entropies (in Eq.~\req{eq:renyin}) with $n \geq 1$ is strictly positive. For that, let us note that for $n >1$ we have $\sqrt{(d-2)/d}<x_n< 1$. On noting the inequality
\begin{equation}
1- \frac{(3d-2)}{2d} (1-\hat{x}_n^2) < \hat{x}_n^d\, , \quad d \geq 3\, , \, n>1
\end{equation}
we observe that, defining $\xi=d(d-2)\alpha_1-1$, for $n>1$ we have

\begin{equation}\label{eq:proofpos}
1-\frac{\hat{x}_n^d}{1-\frac{(\hat{x}_n^2-1)}{\alpha_{\mathrm{eff}}^{\mathrm{EQG}}}\xi } > 1- \frac{1- \frac{(3d-2)}{2d} (1-\hat{x}_n^2)}{1-\frac{(\hat{x}_n^2-1)}{\alpha_{\mathrm{eff}}^{\mathrm{EQG}}}\xi }=\frac{\frac{3d-2}{2d}(1-\hat{x}_n^2)\alpha_{\mathrm{eff}}^{\mathrm{EQG}}+(1-\hat{x}_n^2)\xi }{\alpha_{\mathrm{eff}}^{\mathrm{EQG}}+(1-\hat{x}_n^2)\xi}\geq 0\,.
\end{equation}
The inequalities here follow from the fact that both the numerator and the denominator in the last term are positive:

\begin{equation}
\begin{aligned}
\frac{3d-2}{2d}(1-\hat{x}_n^2)\alpha_{\mathrm{eff}}^{\mathrm{EQG}}+(1-\hat{x}_n^2)\xi&=\frac{(1-\hat{x}_n^2)(d-2)}{2} \left( \frac{(3d-2)}{d(d-2)}\alpha_{\mathrm{eff}}^{\mathrm{EQG}}+2d \alpha_1-\frac{2}{d-2} \right) \geq 0\,,\\
\alpha_{\mathrm{eff}}^{\mathrm{EQG}}+(1-\hat{x}_n^2)\xi &> \frac{3d-2}{2d}(1-\hat{x}_n^2)\alpha_{\mathrm{eff}}^{\mathrm{EQG}}+(1-\hat{x}_n^2)\xi \geq 0\,,
\end{aligned}
\end{equation}
where have used \eqref{eq:ineuni} and taken into account that $ 1 > 1-\hat{x}_n^2 \geq  0$  and that $\frac{(3d-2)}{2d} (1-\hat{x}_n^2)<1$ for every $d \geq 3$.
By applying \req{eq:proofpos} in \req{eq:renyin} and taking into account \eqref{eq:seepos}, it follows that

\begin{equation}
\frac{\partial_{\bar\mu}^2S_{n}}{S_{n}}\bigg|_{\mu=0}>0\, , \quad n \geq 1\,.
\end{equation}
Therefore, we have proven that, as long as unitarity is respected, the R\'enyi entropies (with $n\ge 1$) always grow when a chemical potential is turned on. Again, it is interesting to speculate about the possible validity of this result beyond our current holographic setup.

Regarding the case $n<1$, we can consider the limit $n\rightarrow 0$, which yields

\begin{align}
\underset{n \rightarrow 0}{\lim} S_n=&\frac{2^{d-1} a^* \nu_{d-1} }{n^{d-1} d^d}\left[ 1+\frac{(d-1) n^2 d^2}{4}+\frac{(d-2) n^2 d^2}{2(1-d(d-2)\alpha_1) }\bar\mu^2 \right]+\mathcal{O}{\left (\bar\mu^4 \right)}\, .
\end{align}
For $n\rightarrow 0$ the effect of the chemical potential becomes irrelevant, as it scales with $n^2$ relative to the leading term, but we observe that the coefficient of $\bar\mu^2$ is necessarily positive in $d=3,4,5$, on account of the bound \req{eq:a1abs}, coming again only from unitarity constraints. However, this fails to be true in higher-dimensions, since $\alpha_1$ can take arbitrarily large values in $d\ge 6$.

\begin{figure}[t!]
	\centering
	\begin{subfigure}{0.49\linewidth}
		\includegraphics[width=\linewidth]{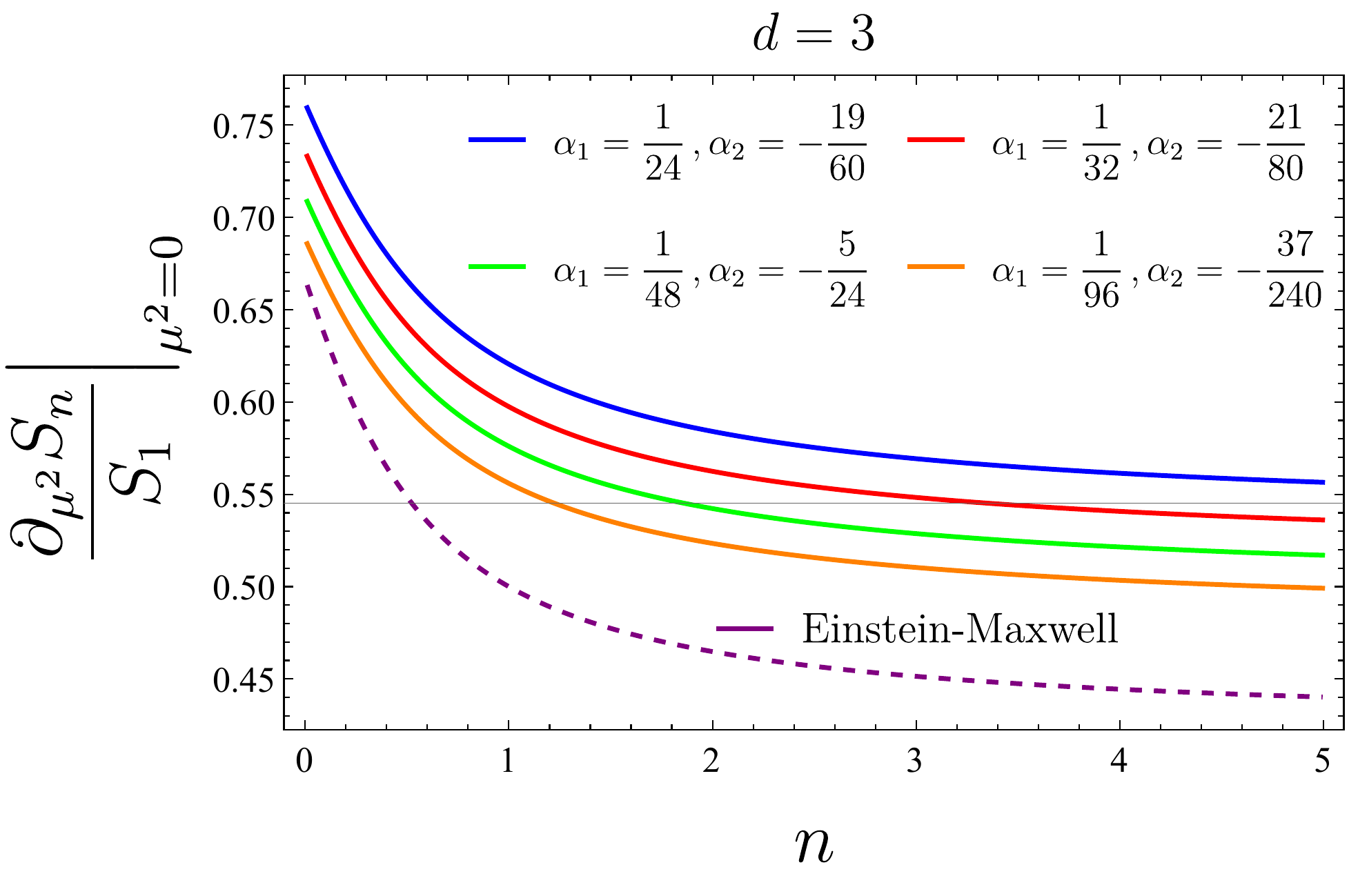}
	\end{subfigure}
	\centering
	\begin{subfigure}{0.49\linewidth}
		\includegraphics[width=\linewidth]{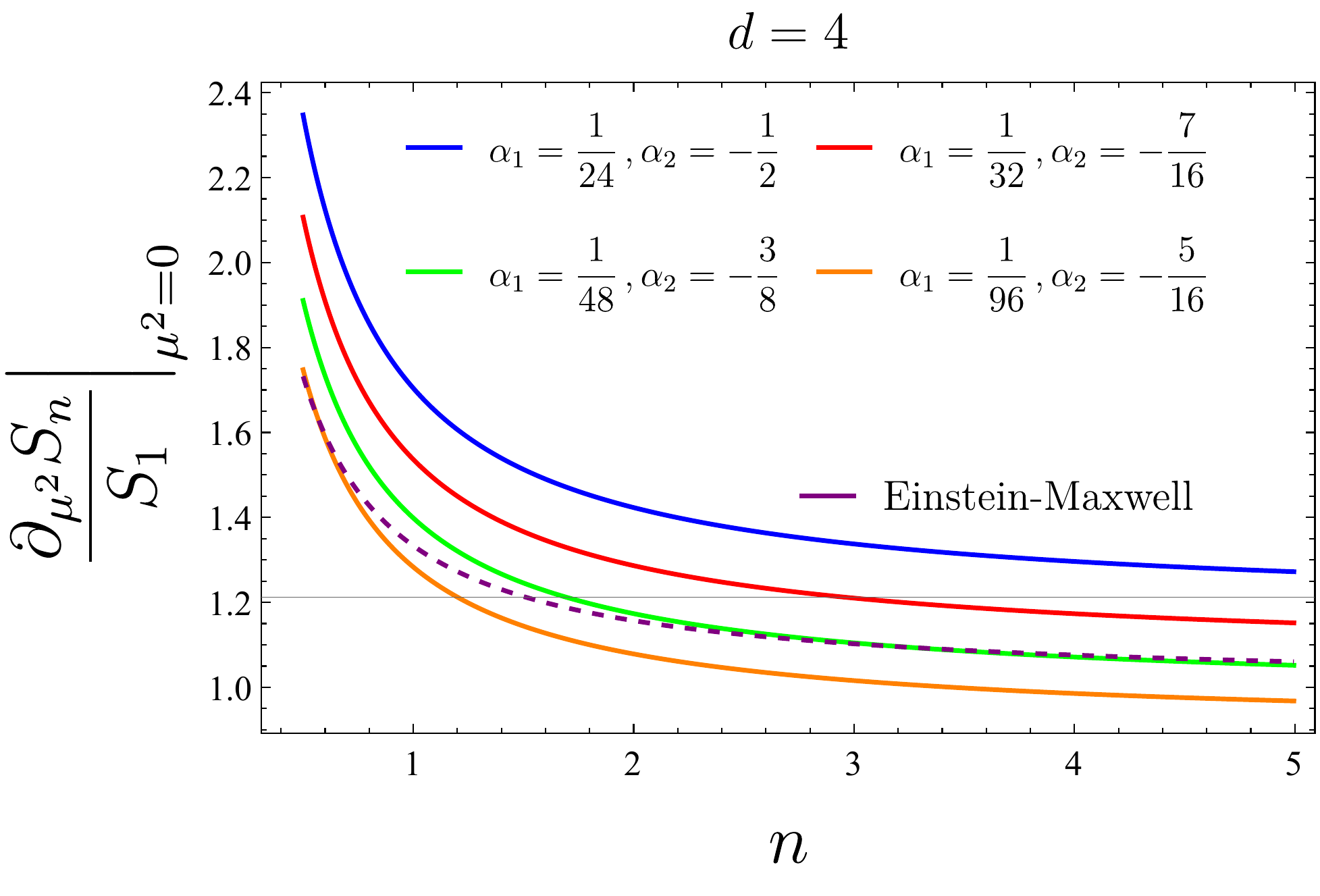}
	\end{subfigure}
	\centering
	\begin{subfigure}{0.49\linewidth}
		\includegraphics[width=\linewidth]{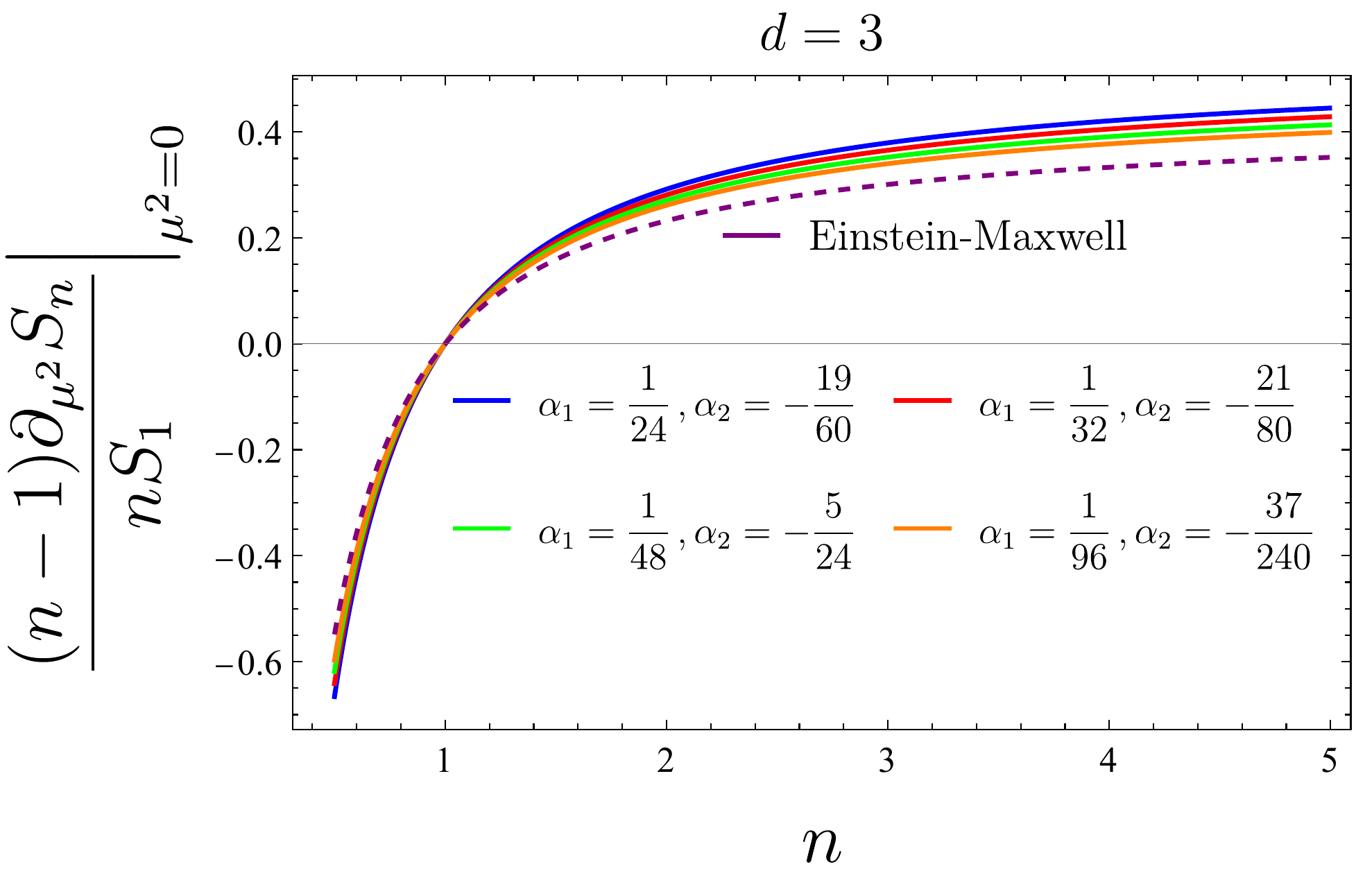}
	\end{subfigure}
	\centering
	\begin{subfigure}{0.49\linewidth}
		\includegraphics[width=\linewidth]{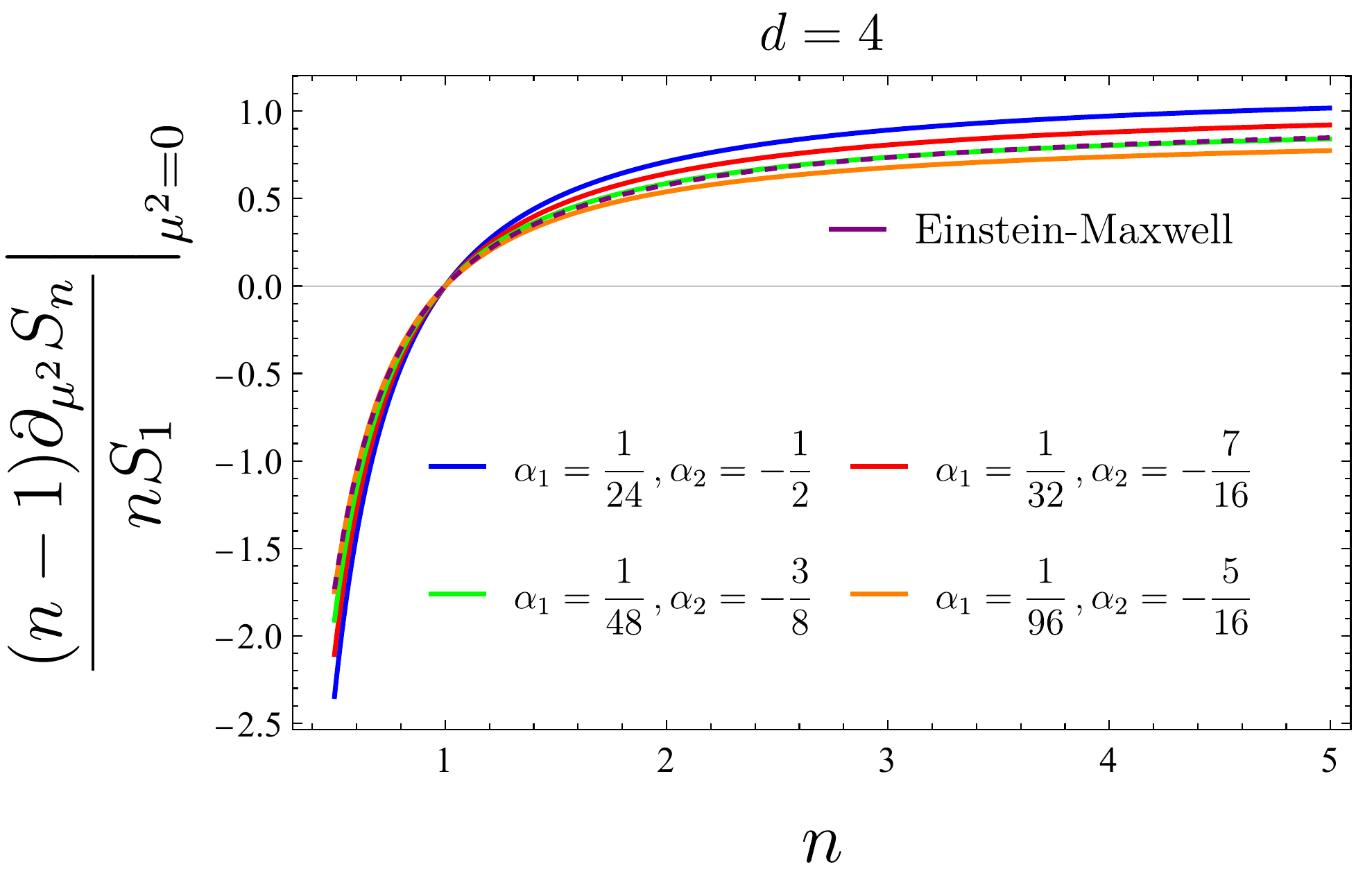}
	\end{subfigure}
	\centering
	\begin{subfigure}{0.49\linewidth}
		\includegraphics[width=\linewidth]{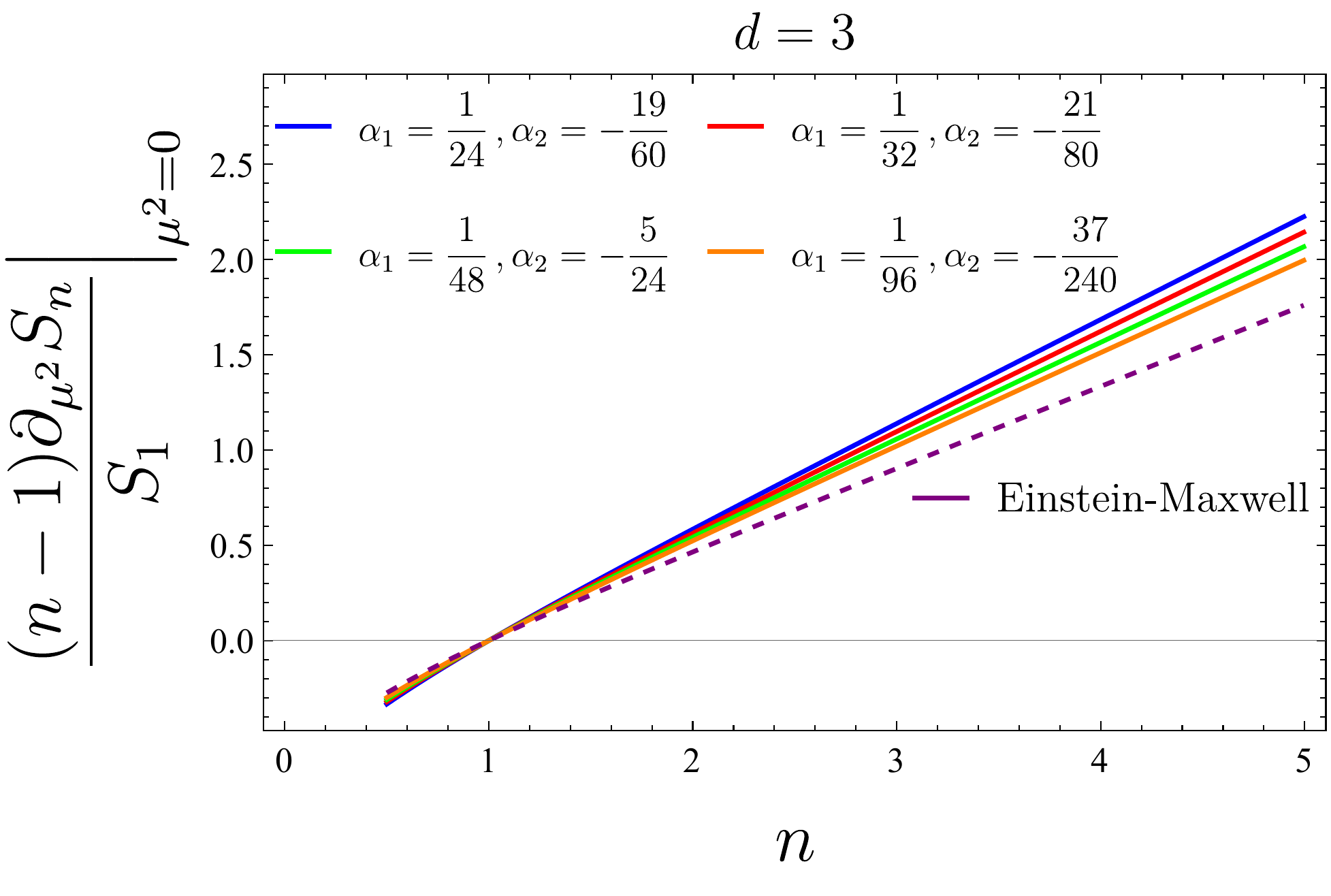}
	\end{subfigure}
	\centering
	\begin{subfigure}{0.49\linewidth}
		\includegraphics[width=\linewidth]{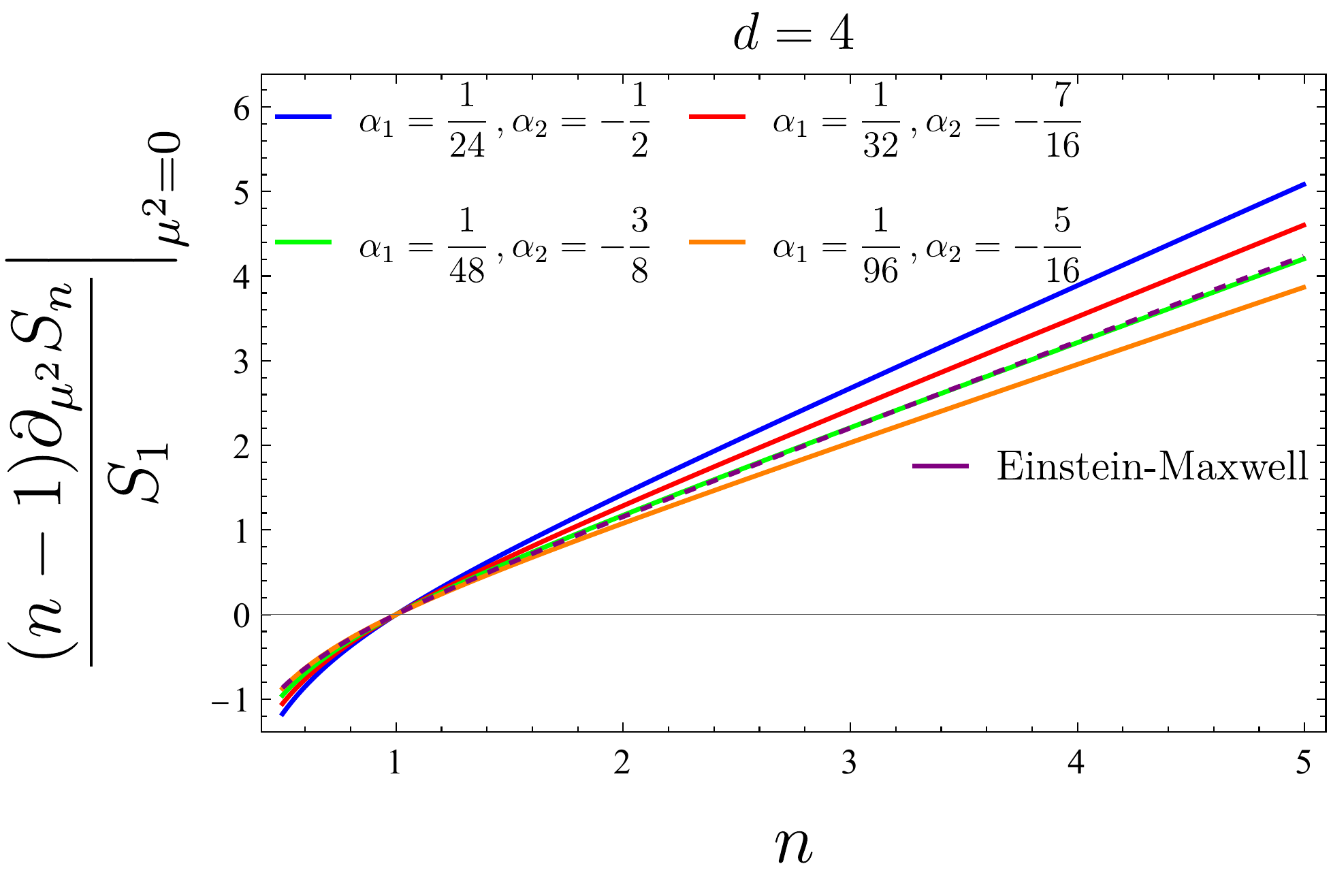}
	\end{subfigure}
	\caption{The coefficient of $\bar \mu^2$ in the R\'enyi entropies for $d=3,4$. We have used different values for the couplings $\alpha_1$ and $\alpha_2$ which are compatible with unitarity and the WGC. Remarkably enough, we find that all inequalities \eqref{eq:inerenyi} are satisfied.}
	\label{fig:ineqrenyi}
\end{figure}

We can finally study the dependence of the REs on the index $n$. It is known that standard (\textit{i.e.}, at zero chemical potential) REs must satisfy the following inequalities \cite{Hung:2011nu}:
\begin{equation}
\label{eq:inerenyi}
\begin{aligned}
\frac{\partial}{\partial n} S_n &\leq 0\, , \quad \frac{\partial}{\partial n} \left ( \frac{n-1}{n} S_n\right)  \geq 0 \, ,\\
\frac{\partial}{\partial n} ((n-1) S_n) &\geq 0\, , \quad \frac{\partial^2}{\partial n^2} ((n-1) S_n) \leq 0\,.
\end{aligned}
\end{equation}
It was shown in Ref.~\cite{Belin:2013uta} that these inequalities are also satisfied by the holographic charged R\'enyi entropies in Einstein-Maxwell theory.  It is therefore interesting to check whether these inequalities still hold for our holographic higher-derivative theories, assuming that the values of the couplings satisfy the unitarity and WGC constraints in Sec.~\ref{sec:constraints}. Since the uncharged R\'enyi entropies for holographic Einstein gravity (obtained by setting  $\bar \mu=0$ in Eq. \eqref{eq:renyin}) already satisfy such inequalities \cite{Hung:2011nu}, it suffices to check that the coefficient of $\bar \mu^2$ in Eq. \eqref{eq:renyin} fulfills them. This will guarantee that the charged RE also satisfy those inequalities, at least in the regime where the $\mathcal{O}(\mu^4)$ terms are subleading. To this aim, we show in Fig \ref{fig:ineqrenyi} the profile of $\partial_{\bar\mu}^2S_{n}/S_{1}\big|_{\mu=0}$  for a few values of $\alpha_1$ and $\alpha_2$ which are allowed by the physical constraints in $d=3,4$. We check that all the previous inequalities indeed hold for our EQG theories. 

It is quite impressive that all of the properties one expects to find in R\'enyi entropies are satisfied whenever the parameters of the bulk theory are taken to satisfy a minimal set of physical constraints. We remark that for arbitrary values of $\alpha_1$ and $\alpha_2$ one could obtain very different results, and even divergencies in the RE. In fact, we have been able to observe that choosing values of these couplings that do not satisfy all the constraints obtained from causality/unitarity and the WGC does lead to these problems. Instead, for the physically sensible values of these parameters, the chemical potential always increases the amount of entanglement and the REs have the same qualitative features found for Einstein-Maxwell theory.

\subsubsection{Large $\mu$}
Let us now study the opposite limit, $\mu\rightarrow \infty$. First it is convenient to revise this limit in the case of Einstein-Maxwell theory \cite{Belin:2013uta}, since this will inspire us to properly generalize the study for our EQG \req{eq:EQTfour}. For that, let us write the temperature $T=(2\pi R n)^{-1}$ and the chemical potential $\bar \mu$ in this particular case:
\begin{equation}
\left. \frac{1}{n} \right \vert_{\mathrm{EM}}=\frac{1}{2x} \left[ d x^2-(d-2) -\frac{2 p^2 x^2}{d-1} \right] \, , \qquad \left. \bar{\mu}\right \vert_{\mathrm{EM}}=\frac{px}{d-2}\,.
\end{equation}
From here, one can solve for $x$ to find:
\begin{equation}
\left.x \right\vert_{\mathrm{EM}}=\frac{1}{nd} \left [1+ \sqrt{1+ d(d-2) n^2+ \frac{2d (d-2)^2 n^2}{(d-1)} \bar \mu^2} \right]\,.
\end{equation}
In the limit $\bar \mu \rightarrow \infty$, we infer that
\begin{equation}
\begin{aligned}
\left.x\right\vert_{\mathrm{EM}}&=(d-2) \sqrt{\frac{2}{d(d-1)}} \bar \mu+\frac{1}{nd}+\mathcal{O}\left ( \frac{1}{\bar \mu} \right)\,, \\
\left. p\right\vert_{\mathrm{EM}} &= \sqrt{ \frac{d(d-1)}{2}} - \frac{d-1}{2n (d-2) \bar \mu} +\mathcal{O}\left ( \frac{1}{\bar \mu^2} \right)\,.
\end{aligned}
\end{equation}
Given this structure for the perturbative expansions of $x$ and $p$ as $\bar \mu \rightarrow \infty$  in the Einstein-Maxwell limit, we expect the corresponding perturbative expansions for our EQG theories to keep the same form:
\begin{equation}
x=x_1 \bar \mu+ x_0 + \mathcal{O}\left ( \frac{1}{\bar \mu} \right)\,, \qquad p=p_0+ \frac{p_{-1}}{\bar \mu}+\mathcal{O}\left ( \frac{1}{\bar \mu^2} \right)\,.
\end{equation}
In fact, taking these ansätze into Eqs. \eqref{eq:nrel} and \eqref{eq:prel}, we find:
\begin{equation}
\begin{aligned}
p_0&=\pm \sqrt{\frac{1-\sqrt{1-\beta d(d-1)}}{\beta}}\, ,\qquad x_1= \frac{(d-2)(3d-4)p_0}{d(p_0^2+(d-1)(d-2))}\, ,\\
p_{-1}&=-\frac{(d-1)\sqrt{f_\infty}(1-2 p_0^2 \alpha_1)p_0}{2n x_1(d(d-1)-p_0^2)}\,, \\ x_0&=\frac{((3d-8)p_0^2-3d(d-1)(d-2))p_{-1} x_1 + 2/n (3d-4)(d-2)\sqrt{f_\infty} p_0^3 \alpha_1}{d p_0(p_0^2+d(d-3)+2)}\,,
\end{aligned}
\end{equation}
where the sign of $p_0$ coincides with that of $\bar\mu$.
Taking into account the previous relations and Eqs. \eqref{eq:renyiomega} and \eqref{eq:omegasim}, the R\'enyi entropy in the limit $\mu \rightarrow \infty$ turns out to be:
\begin{equation}
\begin{aligned}
\lim_{\bar \mu \rightarrow \infty} S_n= \nu_{d-1}\frac{(\ell_{*}R\mu)^{d-1}\pi^{(d-2)/2}}{8G\Gamma(d/2)} (1+2 \alpha_1 p_0^2 ) \left (\frac{(d-2)(3d-4) p_0 \sqrt{f_{\infty}}}{d(p_0^2+(d-1)(d-2))} \right)^{d-1}\, . 
\end{aligned}
\end{equation}
In analogous fashion to the Einstein-Maxwell case, we observe that R\'enyi entropies are independent of $n$ as $ \mu \rightarrow \infty$ and they scale with $\mu^{d-1}$. Let us note that, since the dependence with $n$ becomes trivial for large $\mu$, it is very likely that the inequalities \req{eq:inerenyi}, that we showed to hold for small $\mu$, are actually satisfied for every $\mu$.  
Regarding the sign of the corrections, we note that this is not definite. Since $\alpha_1>0$ on account of the WGC, this coupling always has the effect of increasing the value of the RE. On the other hand, by looking at the dependence of $x_1$ on $\beta$ (which again must be non-negative) we see that it is a decreasing function for $d=3,4$ and a non-monotonic function for $d\ge 5$. Hence, the corrections can either increase or decrease the value of the RE depending on the relative values of the couplings and on the dimension. In spite of this, we notice that this quantity is always positive providing that the WGC is satisfied. If, contrarily, one were not to impose the WGC bounds, then $\alpha_1$ could get arbitrarily negative, since this behavior is allowed by the unitarity constraints as shown in Fig.~\ref{fig:Boundsalpha12}. Therefore, in order to avoid the RE to become negative at large chemical potential, unitarity is not enough, but we also need to impose the WGC.

\subsubsection{Exact result: an example in connection to the WGC}

Let us finally take a look at the exact value of the R\'enyi entropy as a function of $\mu$ and $n$.  Performing a thorough analysis would require a separate work due to the large number of parameters and variables involved, so let us study a quite illustrative example. First, we set $d=4$, since this is the most interesting case. Then, for a given choice of couplings $\{\lambda,\alpha_1,\alpha_2,\beta\}$ we need to solve \req{eq:nrel} and \req{eq:prel} in order to obtain $S_n(\mu)$ according to \req{eq:renyiomega}. However, it can happen (and we have observed that this is the case for some values of the couplings) that the equations \req{eq:nrel} and \req{eq:prel} have several admissible solutions for the same $n$ and $\mu$. If this happens, it denotes the existence of multiple phases, and in that case one must choose the one with smallest value of $\Omega$, which is the dominant one. 
Then, we wish to study the profile of $S_n(\mu)$ when we take into account the physical constraints in Sec.~\ref{sec:constraints} as well as those in Sec.~\ref{sec:thermoab}, that as we saw ensure the existence of a large $\mu$ limit. 

\begin{figure}[t!]
	\centering
	\begin{subfigure}{0.49\linewidth}
		\includegraphics[width=\linewidth]{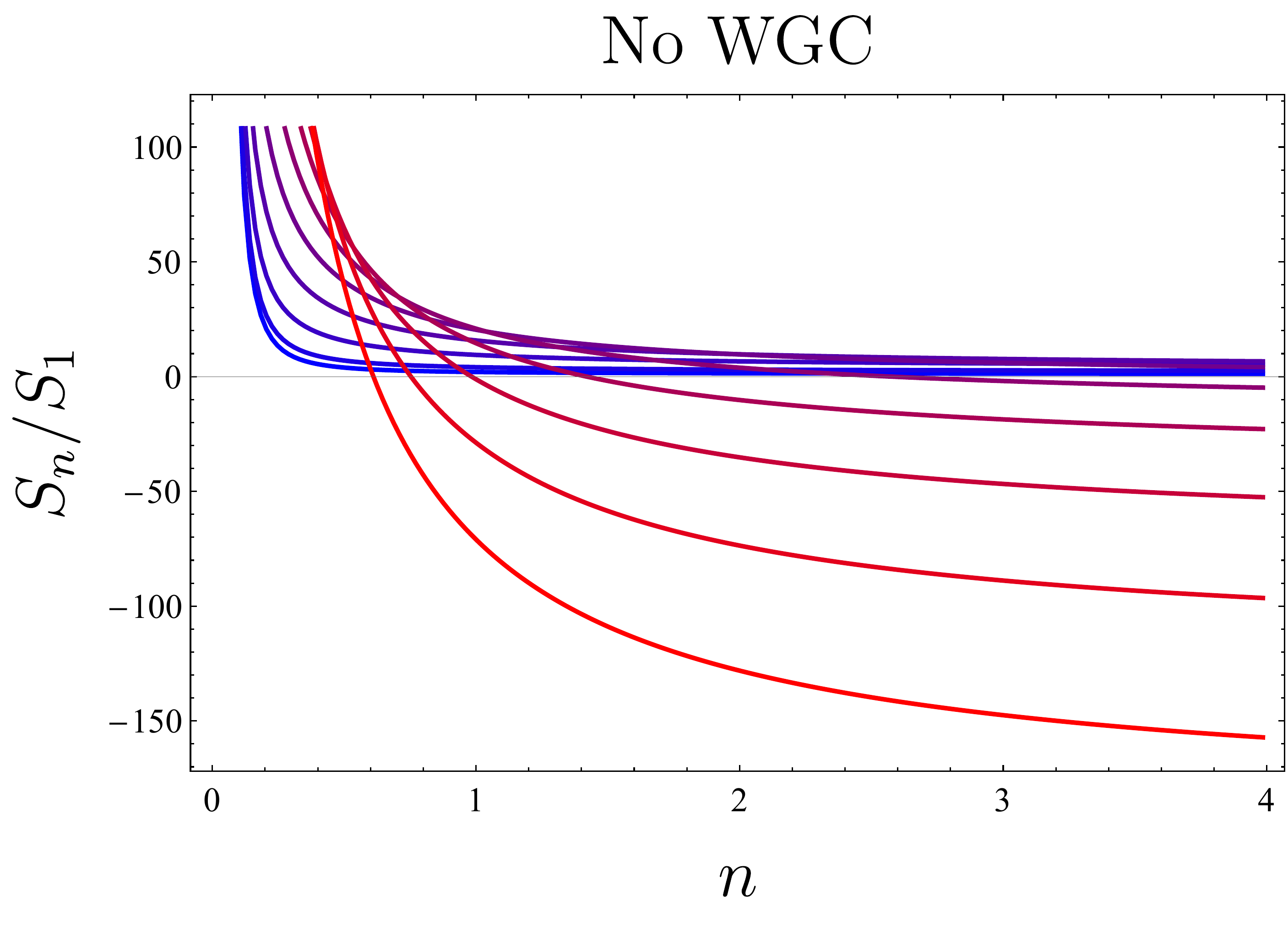}
	\end{subfigure}
	\centering
	\begin{subfigure}{0.49\linewidth}
		\includegraphics[width=\linewidth]{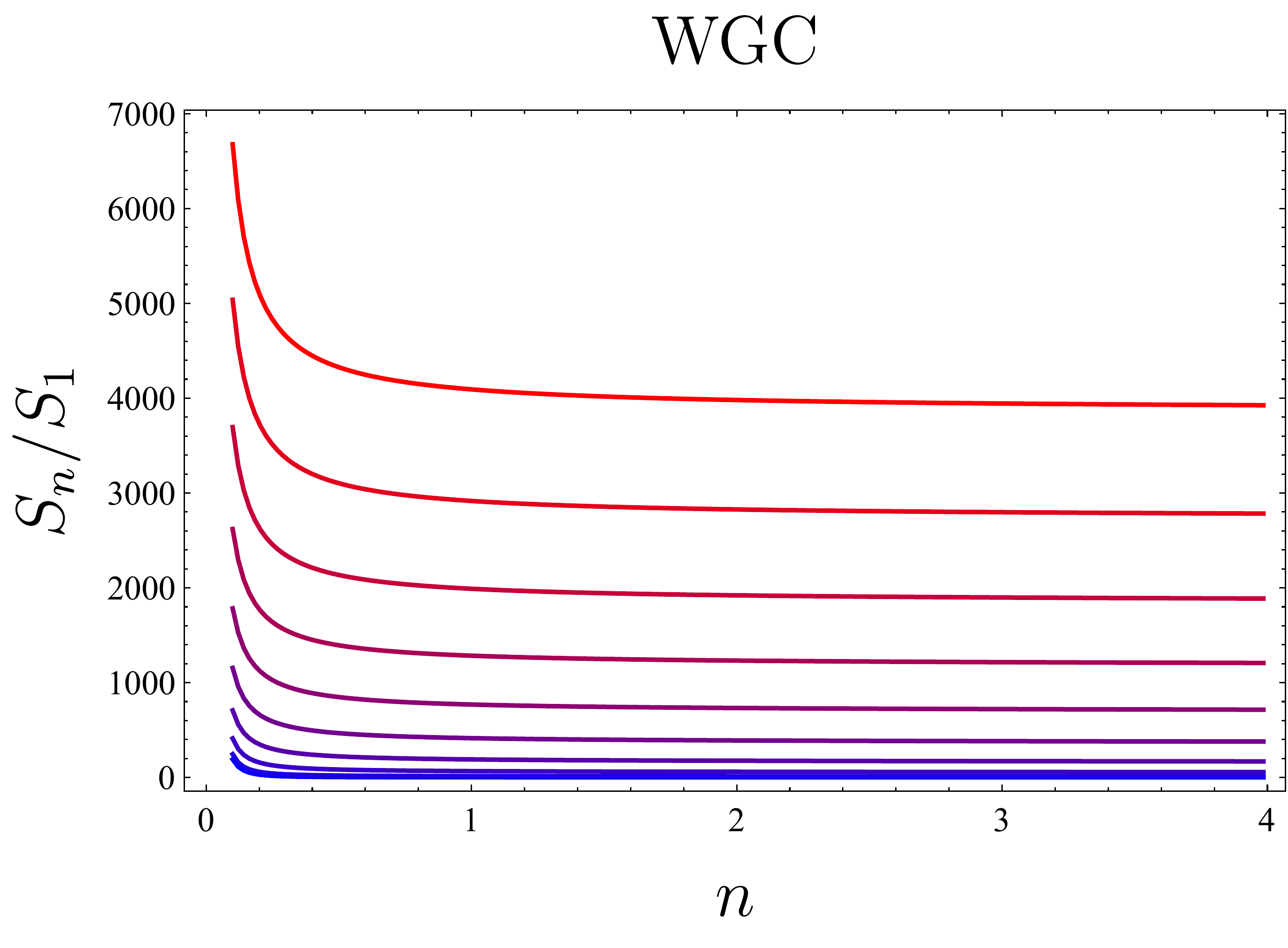}
	\end{subfigure}
	\centering
	\begin{subfigure}{0.49\linewidth}
		\includegraphics[width=\linewidth]{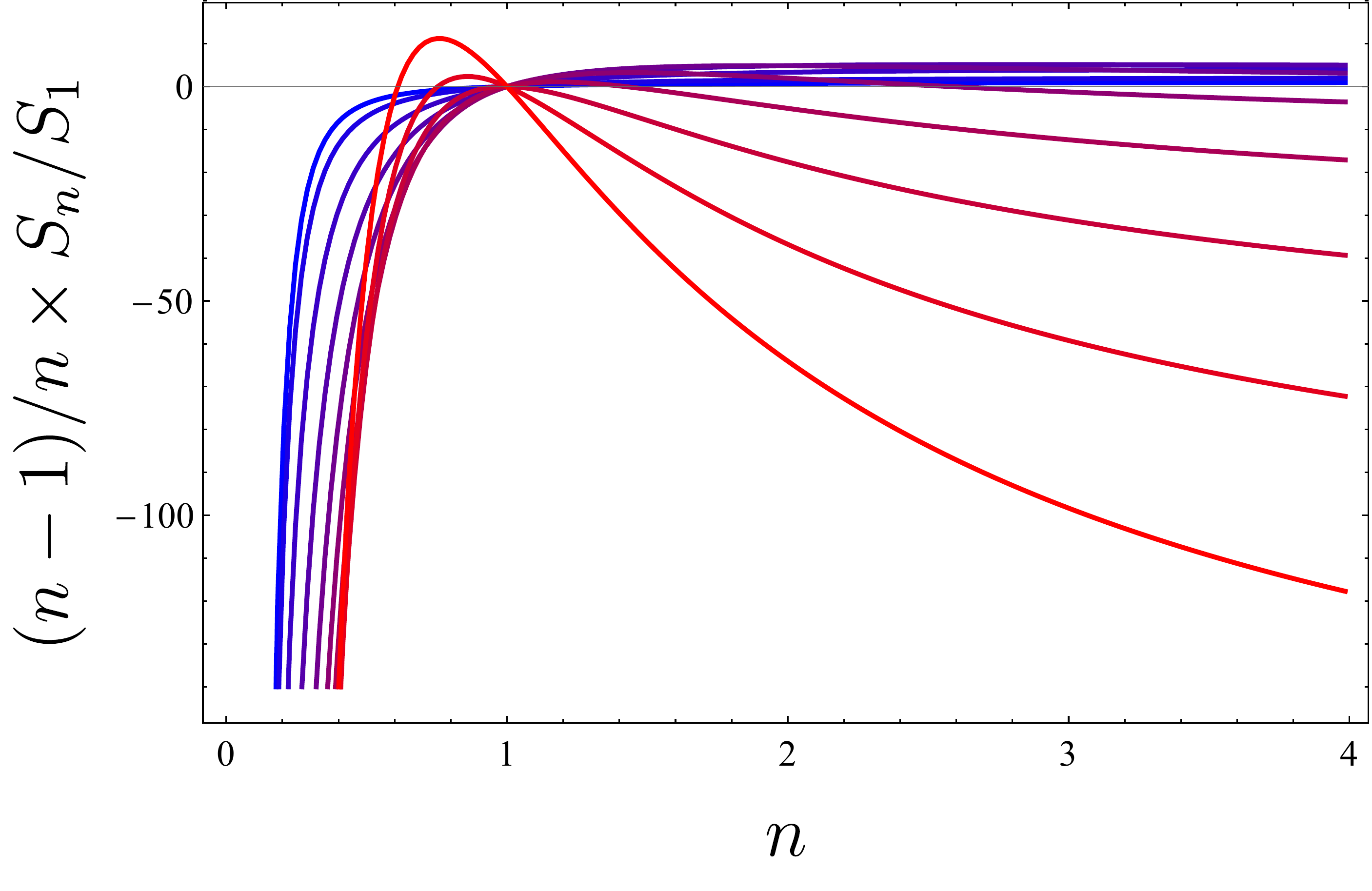}
	\end{subfigure}
	\centering
	\begin{subfigure}{0.49\linewidth}
		\includegraphics[width=\linewidth]{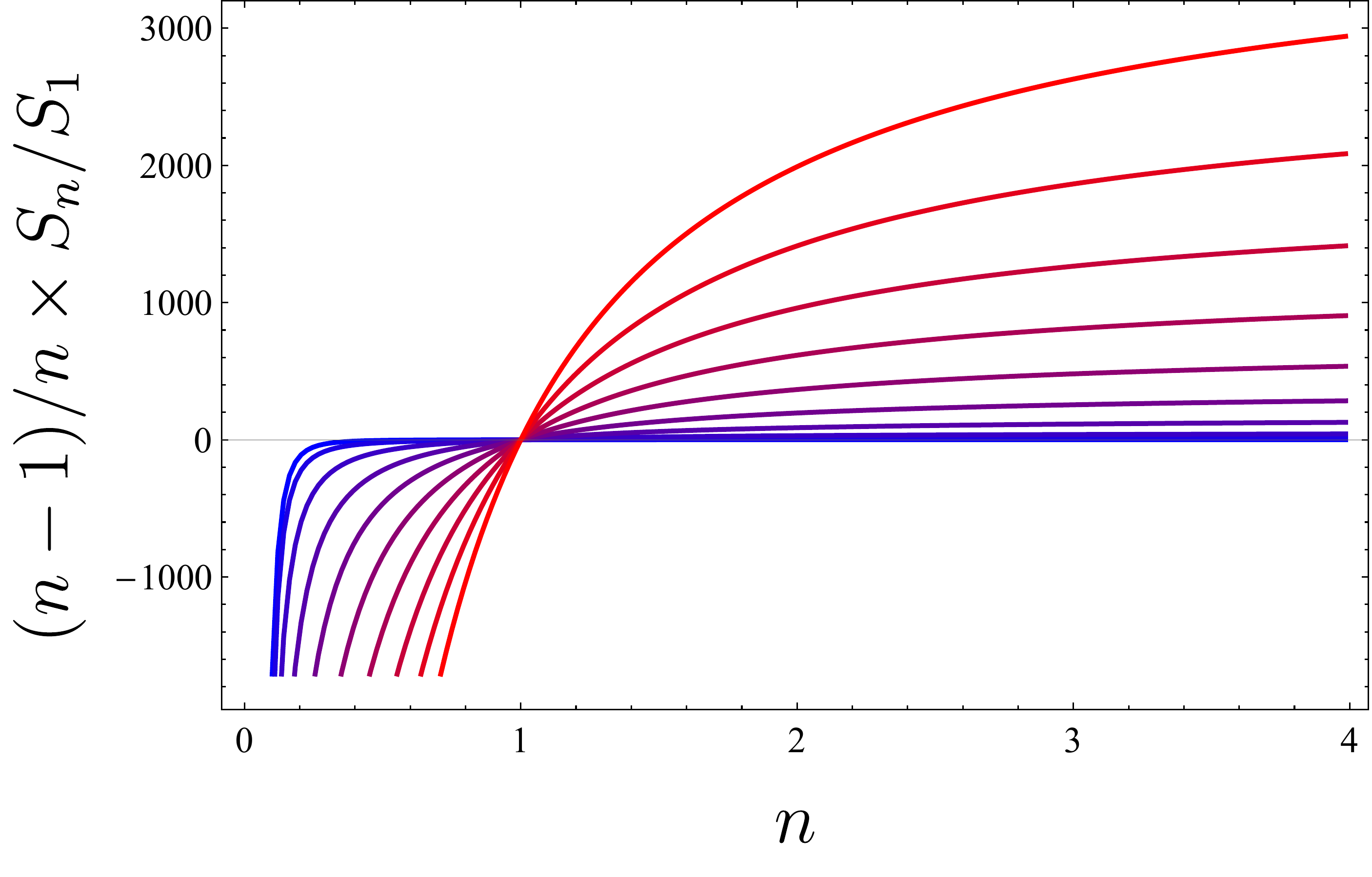}
	\end{subfigure}
	\centering
	\begin{subfigure}{0.49\linewidth}
		\includegraphics[width=\linewidth]{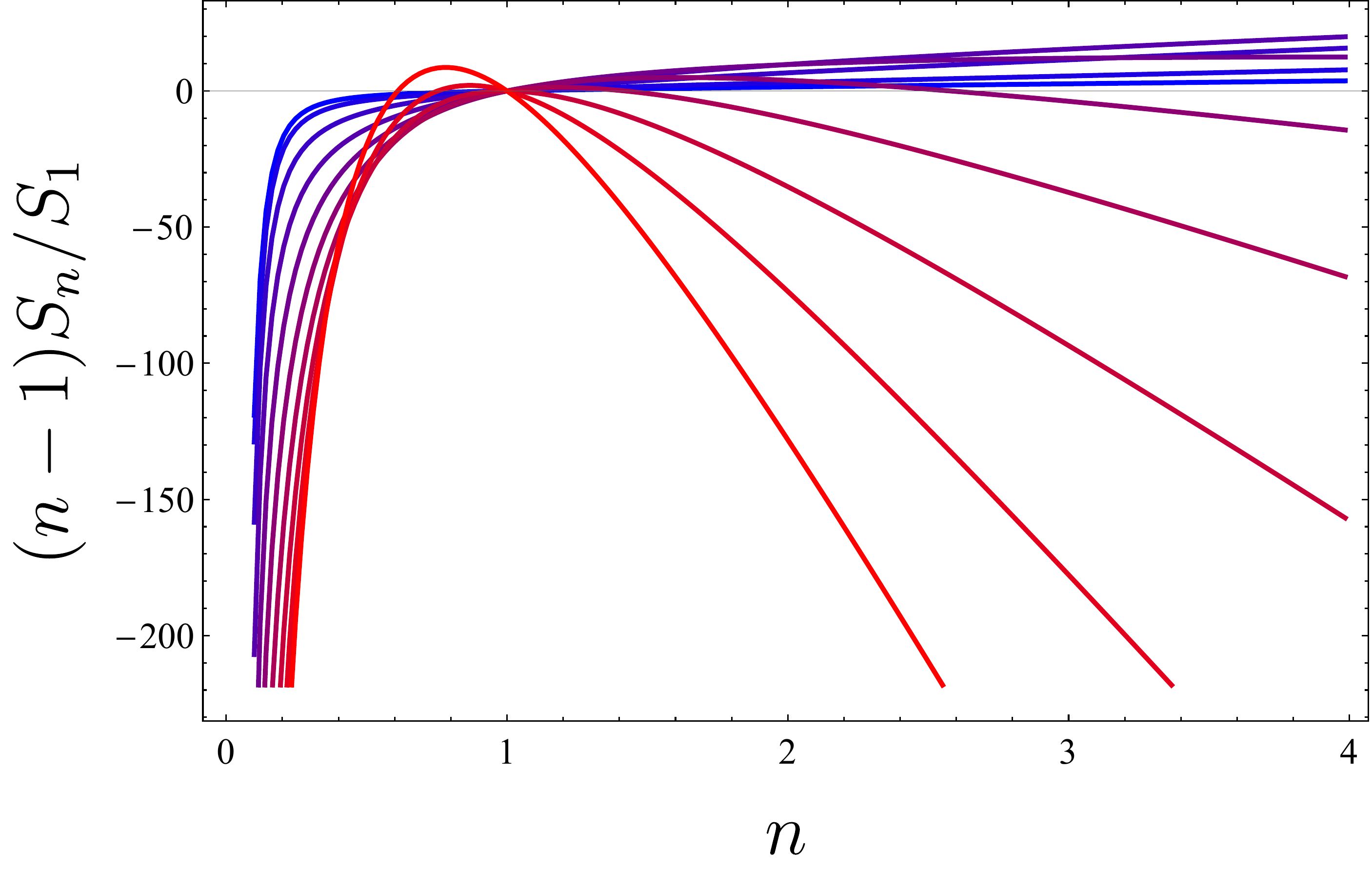}
	\end{subfigure}
	\centering
	\begin{subfigure}{0.49\linewidth}
		\includegraphics[width=\linewidth]{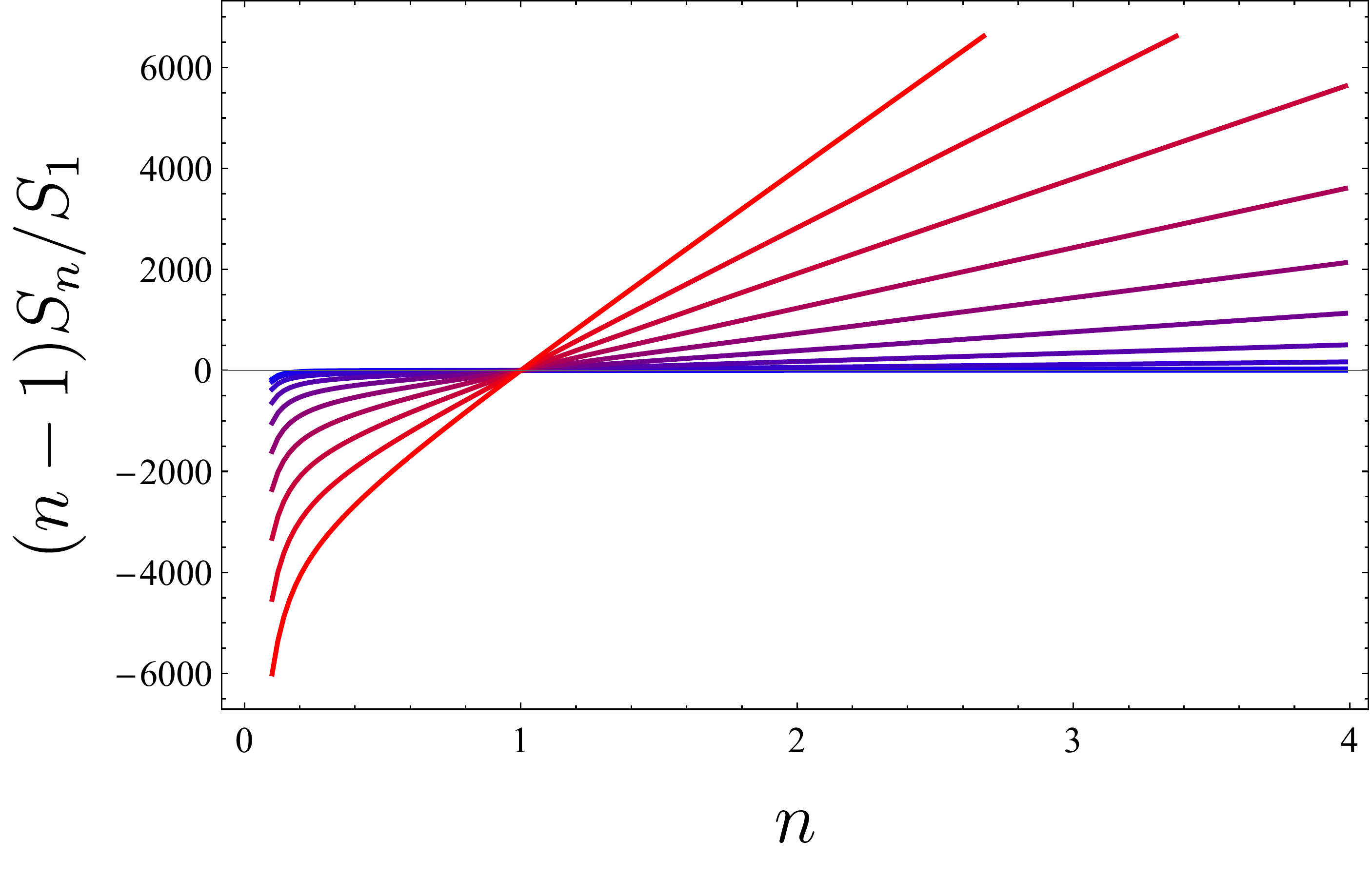}
	\end{subfigure}
	\caption{R\'enyi entropies as function of $n$ for two particular choices of couplings in $d=4$. From blue to red, the different curves correspond to $\bar\mu/\sqrt{f_{\infty}}=0,1,\ldots 10$. Left: a theory that satisfies unitarity constraints but not the WGC: $\{\lambda,\alpha_1,\alpha_2,\beta\}=\{0.052,-0.100,0.875,0.0049\}$. Right: a theory that satisfies both unitarity and WGC bounds: $\{\lambda,\alpha_1,\alpha_2,\beta\}=\{0.077,0.057,-0.596,0.023\}$. The standard properties of R\'enyi entropies may be violated if the WGC is not satisfied. }
	\label{fig:renyiwgc}
\end{figure}

We perform the following experiment: we take a set of random values of the couplings satisfying both WGC and unitarity constraints, and a second set of couplings that satisfy unitarity but not the WGC. We then study the properties of the RE for each set. The details of course depend on the particular values of the couplings, but we show a representative example in Fig.~\ref{fig:renyiwgc}. In the left column we have represented $S_n/S_1$ (and also $(n-1)/n (S_n/S_1)$ and $(n-1) S_n/S_1$) as a function of $n$, for several values of $\bar\mu$, for a set of couplings that do not satisfy the WGC (but that do respect unitarity). In the right column, we show the same quantities for a different choice of couplings that now respect both unitarity and the WGC. The differences are stark. 
While in the right column, the RE is always positive and respects the inequalities \req{eq:inerenyi}, the RE for the theory that breaks the WGC violates the second and third of them when $\bar\mu$ becomes large enough. Furthermore, the RE even becomes negative in that case. 

Certainly, this is only an example, but looking at randomly generated couplings we have not found any instance of a theory that satisfies the WGC and unitarity and behaves as in the left column. In fact, in all those cases we obtain plots similar to those in the right column of  Fig.~\ref{fig:renyiwgc}. Thus, it seems that the WGC bounds are key to produce a sensible dual CFT.

\subsection{Generalized twist operators}

A very interesting notion in the context of R\'enyi entropies is that of twist operators, which possess a great deal of information about the CFT. 
Let us remember that in the computation of R\'enyi entropies for some region $A$ via the replica trick one uses the following result
\begin{equation}
\Tr \rho_{A}^{n}=\frac{Z_n}{Z_1^n}\, ,
\end{equation}
where $Z_n$ is the partition function of an $n$-fold cover of Euclidean space in which cuts have been introduced in $A$. Along these cuts the $k$-th geometry must be glued to the $(k+1)$-th one by implementing appropriate boundary conditions \cite{Calabrese:2004eu}.

However, an alternative route to compute this quantity involves the insertion of dimension-$(d-2)$ operators $\sigma_n$ (the twist operators) extending over the entangling surface $\Sigma=\partial A$ \cite{Calabrese:2004eu,Hung:2011nu,Hung:2014npa,Swingle:2010jz}.  Then, the path integral over the replicated geometry can be replaced by a path integral for the symmetric product of $n$ copies of the CFT, with the $\sigma_n$ inserted, on a single copy of the geometry. One can then obtain the desired trace of $\rho_{A}^{n}$ as the expectation value of these twist operators, $\Tr \rho_{A}^{n}=\braket{\sigma_n}$, computed in the $n$-fold symmetric product CFT.

It is possible to define a generalized notion of conformal dimension for the twist operators by performing an insertion of the stress-energy tensor $T_{ab}$ at a small distance $y$ from $\Sigma$.  In particular, the leading singularity of  the  correlator $\braket{T_{ab}\sigma_n}$ takes the form \cite{Hung:2011nu,Hung:2014npa},
\begin{equation}
\braket{T_{ab}\sigma_n}=-\frac{h_n}{2\pi} \frac{b_{ab}}{y^d}\, ,
\end{equation}
where $b_{ab}$ is a fixed tensorial structure and $h_n$ is the conformal dimension of $\sigma_n$.  
In the case of a spherical entangling surface, and with a finite chemical potential, the conformal mapping from flat space to the hyperbolic cylinder allows one to show that \cite{Hung:2011nu,Belin:2013uta}

\begin{equation}\label{eq:hndef}
h_{n}(\mu)=\frac{2\pi n}{d-1}R^{d}\left(\E(T_0,\mu=0)-\E(T_0/n,\mu)\right)\, ,
\end{equation}
where $\E(T,\mu)$ is the thermal energy density of the theory placed on $\mathbb{S}^1\times \mathbb{H}^{d-1}(R)$.
 
Likewise, when a chemical potential is present, we also have at hand its associated current $J^a$, and one can also study the correlator $\braket{J_a\sigma_n(\mu)}$. In this case, the leading singularity takes the form  \cite{Belin:2013uta}

\begin{equation}
\left\langle J_a \sigma_n(\mu) \right\rangle = \frac{i k_n(\mu)}{2 \pi} \frac{\tau_{a}}{y^{d-1}}\, 
\end{equation}
where $\tau_a$ is again a fixed structure determined by the geometry of the setup. The coefficient $k_n(\mu)$ is the magnetic response of the generalized twist operators, and for a spherical entangling surface it can be computed as
\begin{equation}
k_n(\mu) = 2\pi n R^{d-1} \rho(n, \mu)\, ,
\label{eq:kndef}
\end{equation}
where $\rho(n, \mu)$ is the charge density of the theory on $\mathbb{S}^1\times \mathbb{H}^{d-1}(R)$ at temperature $T=T_0/n$ and with chemical potential $\mu$. 

It is then interesting to consider the expansion of $h_n(\mu)$ and $k_n(\mu)$ around $n=1$ and $\mu=0$,

\begin{equation}
\begin{aligned}
h_n(\mu)=&\sum_{l=0}^{\infty}\sum_{m=0}^{\infty}\frac{1}{l!m!}h_{lm}(n-1)^l\mu^m\, ,\\
k_n(\mu)=&\sum_{l=0}^{\infty}\sum_{m=0}^{\infty}\frac{1}{l!m!}k_{lm}(n-1)^l\mu^m\, ,
\end{aligned}
\end{equation}
where

\begin{equation}
h_{lm}=(\partial_n)^l (\partial_{\mu})^m h_{n}(\mu)\Big|_{n=1,\mu=0}\, ,\qquad k_{lm}=(\partial_n)^l (\partial_{\mu})^m k_{n}(\mu)\Big|_{n=1,\mu=0}\, .
\end{equation}
As shown by Ref.~\cite{Belin:2013uta} (and by Refs.~\cite{Hung:2011nu,Hung:2014npa,Chu:2016tps} in the case of $h_n$ for $\mu=0$), these coefficients involve integrated correlators of the form $\braket{T\ldots T J\ldots J}$. In particular, the few first coefficients are related to two or three-point functions of $T$ and $J$, and therefore have a universal form for any CFT. These relations were derived in \cite{Hung:2011nu,Belin:2013uta,Hung:2014npa} from first principles, but here we will see that they can be equivalently derived by using holography with higher-derivative terms.

\subsubsection{Conformal dimension of generalized twist operators}
Let us start by studying the conformal dimension of the generalized twist operators, given by Eq.~\req{eq:hndef}. Holographically, the energy density $\E$ is nothing but the mass of a hyperbolic black hole  over the volume of the hyperbolic boundary, \textit{i.e.},

\begin{equation}
\E (T, \mu) = \frac{M (T, \mu)}{V_{-1, d-1} R^{d-1}}\, .
\label{eq:energydensity}
\end{equation}
This can be obtained from Eq.~\eqref{eq:massrplus} by setting $k=-1$. Observing that $M(T_0,\mu=0)=0$,  by virtue of \req{eq:hndef} we have 

\begin{equation}
\begin{aligned}\label{eq:hnvalue}
h_{n}(\mu)=&-\frac{n L^{d-1}}{8(d-1) \sqrt{f_{\infty}} G} \Bigg[(d-1)\Big(-x^{d-2}+x^d+\lambda x^{d-4}\Big)\\
&+\frac{2p^2 x^d}{(d-2)}\left(1-\frac{(d-2)}{x^2}(3(d-1)\alpha_1+\alpha_2)\right)-\frac{\beta  p^4 x^d}{(3d-4)}\Bigg]\, ,
\end{aligned}
\end{equation}
where as usual we have introduced $x=r_{+}/L$, $p=Q L/r_{+}^{d-1}$, which depend on $n$ and $\mu$ through the relations \req{eq:nrel} and \req{eq:prel}. For $n=1$ and $\mu=0$, those equations are solved by $x=1/\sqrt{f_{\infty}}$ and $p=0$. We can then perform an expansion around those values to find

\begin{equation}\label{eq:nmuexpansion}
\begin{aligned}
x=&\frac{1}{\sqrt{f_{\infty}}}-\frac{(n-1)}{(d-1) \sqrt{f_{\infty}}}+\left(\frac{\mu \ell_{*} R}{L}\right)^2\frac{(d-2)^2 f_{\infty}^{3/2} \left(1-\alpha_1  (3d+2) (d-1) f_{\infty}-\alpha_2 d f_{\infty}\right)}{(d-1)^2 (2-f_{\infty}) \left(\alpha_{\rm eff}^{\rm EQG}\right)^2}+\ldots\, ,\\
p=&\left(\frac{\mu \ell_{*} R}{L}\right)\left[\frac{(d-2)f_{\infty}}{\alpha_{\rm eff}^{\rm EQG}}\right.\\
&\left.+(n-1)\frac{(d-2) f_{\infty} \left(1+\alpha_1 \left(d-2\right)(d-1) f_{\infty}+\alpha_2 (d-2) f_{\infty}\right)}{(d-1)  \left(\alpha_{\rm eff}^{\rm EQG}\right)^2}+\ldots\right]+\ldots\, ,
\end{aligned}
\end{equation}
where we only show the terms that we will need. From these expressions it is straightforward to obtain the expansion of $h_n$ in \req{eq:hnvalue} and to read off the values of the derivatives. 
In the first place, we find

\begin{equation}
h_{10}=\frac{1 - 2\lambda f_\infty}{4 (d-1)  G} \left( \frac{L}{\sqrt{f_\infty}} \right)^{d-1}\, ,
\end{equation}
and comparing with the value of the central charge $C_T$ four our theory, given by  Eq.~\req{eq:CT}, we realize that this relation can be written as

\begin{equation}\label{h10}
h_{10}= 2 \pi^{d/2 + 1} \frac{\Gamma(d / 2)}{\Gamma(d + 2)} C_T\, .
\end{equation}
This is precisely the relation found in Ref.~\cite{Hung:2011nu}. In a similar way, the second derivative of $h_n$ at vanishing $\mu$, that is, $h_{20}$, is completely determined in terms of $C_T$ and the 3-point function coefficients $t_2$ and $t_4$ \cite{Hung:2014npa,Chu:2016tps}. Those relations have been shown to be identically satisfied for holographic higher-curvature gravities \cite{Chu:2016tps,Bueno:2018xqc,Bueno:2020odt}.
Thus, let us turn our attention to the derivatives of $h_n$ with respect to $\mu$, which, to the best of our knowledge, have not been studied in detail for higher-derivative theories.  

From \req{eq:nmuexpansion} and \req{eq:hnvalue} we find 
\begin{equation}
\begin{aligned}
h_{02}& =- \frac{(d-2)  \ell_{*}^2 R^2}{2 (d-1)^2 G } \left( \frac{L}{\sqrt{f_\infty}} \right)^{d-3} \frac{2d-3 - (d-2) f_\infty \left( (6d-1) (d-1) \alpha_1 + (2d-1) \alpha_2 \right)}{\left(\alpha_{\rm eff}^{\rm EQG}\right)^2} \, .
\end{aligned}
\end{equation}
Now, looking at Eqs.~\req{eq:CJgen}, \req{eq:alphaeffEQG} and \req{eq:a2EQG}, we see that this expression can be written in terms of the central charge $C_{J}$ and the flux parameter $a_2$ as

\begin{equation}
h_{02} =-(2\pi R)^2\frac{C_{J} \pi ^{\frac{d}{2}-1} \Gamma \left(\frac{d}{2}\right) }{(d-1)^3 \Gamma (d+1)} \left((d-1) d (2 d-3)+a_2 (d-2)^2\right)\, .
\end{equation}
Finally, we can write it in terms of the $\braket{TJJ}$ coefficients $\hat c$ and $\hat e$ using the relations \req{eq:CJed} and \req{eq:a2ed}, and we get\footnote{We could of course derive this relation directly from the values of $\hat{c}$ and  $\hat{e}$ for EQG, given by Eq.~\req{eq:TJJEQG}, but we find it interesting to show the intermediate expression in terms of $C_J$ and $a_2$.}

\begin{equation}\label{h02}
h_{02}= - (2 \pi R)^2 \frac{4 \pi^{d-1}}{\Gamma(d+1)} \left( \frac{2}{d} \hat{c} + \hat{e} \right)\, ,
\end{equation}
which is precisely the result in Eq.~(2.45) of \cite{Belin:2013uta} and which applies to any CFT.\footnote{Note that we have an additional $(2\pi R)^2$ factor with respect to \cite{Belin:2013uta}, which comes from the fact that they normalize the chemical potential with a factor of $1/(2\pi R)$, that we do not introduce.}

\subsubsection{Magnetic response of generalized twist operators}
Let us now take a look at the magnetic response $k_{n}(\mu)$, which we can compute using the relation \req{eq:kndef}. The charge density in the boundary theory is simply

\begin{equation}
\rho(n, \mu) = \frac{\ell_*q }{R^{d-1}V_{-1,d-1}}\, ,
\end{equation}
where $q$ is given by Eq.~\req{eq:physcharge}. Therefore we get

\begin{equation}
k_n(\mu) = \frac{n \ell_*Q }{2 G}=\frac{ \ell_*  L^{d-2}}{2 G} n p x^{d-1}\, .
\end{equation}
By using \req{eq:nmuexpansion} we can easily expand this quantity near $n=1$ and $\mu=0$ and read off its derivatives. For the first derivative with respect to $\mu$ we obtain

\begin{equation}\label{k01}
k_{01}= \frac{(d-2) \ell_{*}^2 R}{2 G\alpha_{\rm eff}^{\rm EQG}} \left( \frac{L}{\sqrt{f_\infty}} \right)^{d-3}=8 \pi^{d/2+1} R \frac{\Gamma \left( \frac{d + 2}{2} \right)}{\Gamma(d+1)} C_J\, , 
\end{equation}
where in the second equality we used \req{eq:CJgen}. Again, up to a factor of $(2\pi R)$ that arises from different normalization conventions for $\mu$, this coincides with Eq.~(2.57) of \cite{Belin:2013uta}. 
We can also compute the mixed partial derivative $k_{11}$, which yields

\begin{equation}
 k_{11}= \frac{(d-2) \ell_{*}^2 R \left[ 1 + (d-2) f_\infty \left( (d-1) \alpha_1 + \alpha_2 \right) \right] }{4 (d-1) G \left(\alpha_{\rm eff}^{\rm EQG} \right)^2 } \left( \frac{L}{\sqrt{f_\infty}} \right)^{d-3}\, .
\end{equation}
This can be express in terms of $C_J$ \req{eq:CJgen} and $a_2$  \req{eq:a2EQG} as 

\begin{equation}
 k_{11}= \frac{4 R C_{J} \pi ^{\frac{d}{2}+1}  \Gamma \left(\frac{d}{2}\right)}{(d-1)^2 \Gamma (d+1)} \left(d(d-1)-a_2 (d-2)^2\right)\, ,
\end{equation}
or in terms of the $\braket{TJJ}$ coefficients \req{eq:TJJEQG} as

\begin{equation}\label{k11}
 k_{11}=\frac{16 \pi^{d+1} R}{d \Gamma(d+1)} \left( 2 \hat{c} - d (d-3) \hat{e} \right)\, .
\end{equation}
Thus, we reproduce in this case Eq.~(2.56) of \cite{Belin:2013uta}. Let us remark that Ref.~\cite{Belin:2013uta} checked that these relations held for holographic Einstein-Maxwell theory, but that case is somewhat restricted as the dual theory has $a_2=0$. To the best of our knowledge, this is the first holographic derivation of these universal relationships in a theory with a general $\braket{TJJ}$ three-point function.

\section{Conclusions}
\label{sec:conclusions}

We have carried out a general analysis of the holographic aspects of the Electromagnetic Quasitopological gravity theory given by \req{eq:EQTfour}. This is a theory containing a $(d-2)$-form, but as we have seen it can be dualized into a theory with a vector field, and we use this formulation to make contact with holography. One of the most interesting aspects of this theory is that it contains non-minimal couplings, which affect the central charge of the two-point function $\braket{JJ}$, and more importantly, give rise to a non-vanishing parameter $a_2$ (see \req{eq:a2EQG}) that controls the angular distribution of the energy one-point function \req{eq:flux}. This in turn means that the boundary theory has a general $\braket{TJJ}$ correlator. Therefore, we can probe holographic CFTs beyond the universality class defined by Einstein-Maxwell theory. In addition, the special properties of the EQG theories allow us to carry out a fully analytic and exact study of many of their holographic aspects, so we do not have to restrict to the perturbative regime.  

One of the main questions we tried to answer is that of how the physics of the CFT can change while satisfying physically reasonable conditions. Thus, we have constrained the couplings of our bulk theory by demanding that the boundary CFT respects unitarity. This means that the central charges $C_T$ and $C_J$, as well as the energy fluxes  $\langle\mathcal E\left(\vec n\right)\rangle_{J}$ and $\langle\mathcal E\left(\vec n\right)\rangle_{T}$ (see resp. \req{eq:flux} and \req{eq:fluxT}), have to remain positive. We also studied the constraints coming from demanding causality in the bulk in the background of a planar neutral black hole. In the case of gravitational fluctuations, it is known that these causality constraints imply the positivity of the energy flux $\langle\mathcal E\left(\vec n\right)\rangle_{T}$ \cite{Camanho:2009vw,Buchel:2009tt,Camanho:2009hu,Camanho:2013pda}. Here we have shown that demanding that the electromagnetic waves do not propagate faster than light is equivalent to the constraints obtained from the positivity of $\langle\mathcal E\left(\vec n\right)\rangle_{J}$. These causality bounds follow from looking at the phase velocity of electromagnetic waves close to the boundary of AdS. We have not observed additional causality constraints when extending these conditions deeper into the bulk, but our analysis in this regard is limited due to the number of parameters involved, so it would be interesting to do a more thorough search of causality constraints in the bulk interior, as in Ref.~\cite{Camanho:2010ru}. Likewise, it would be desirable to extend these bounds to the case of charged black hole backgrounds. 

One of the novelties in our analysis was the inclusion of constraints from the weak gravity conjecture. As proposed by Ref.~\cite{Cheung:2018cwt} and recently explored by Ref.~\cite{Cremonini:2019wdk} in the case of AdS, the so-called mild form of the WGC demands that the corrections to the entropy of thermally stable black holes be positive in the microcanonical ensemble. This implies in particular that the charge-to-mass ratio of extremal black holes is corrected positively \cite{Goon:2019faz}, which is the most familiar form of the WGC for asymptotically flat black holes \cite{Kats:2006xp,Hamada:2018dde}. However, demanding the entropy corrections to be positive is a more general condition than that and is amenable to the AdS case. When applied to spherical (and planar) black holes, we obtain constraints on the signs of (certain combinations of) the couplings, and these become very powerful when combined with unitarity/causality bounds. In fact, we obtain that the couplings $\alpha_1$, $\alpha_2$ and $\lambda$ of \req{eq:EQTfour} can only lie in a very small compact set of $\mathbb{R}^3$ in $d=3,4,5$. The only parameter that can be unbounded is $\beta$, which is simply required to satisfy $\beta\ge 0$ by the WGC. However, we suspect that additional causality/unitarity conditions should provide an upper limit for $\beta$. 

Now, when the positivity-of-entropy bounds are implemented instead for hyperbolic black holes we find something quite remarkable that was not noticed in Ref.~\cite{Cremonini:2019wdk}: some of the constraints become incompatible with those coming from spherical black holes. For instance, demanding that the corrections to the entropy for large spherical black holes and for hyperbolic black holes (both of which are stable) be non-negative can only be achieved if the GB coupling is vanishing, $\lambda=0$. This looks like an unreasonably strong constraint, even more taking into account that a positive GB coupling (which is the sign imposed by $\delta S>0$ in spherical black holes) is explicitly realized in string theory effective actions \cite{Metsaev:1987zx,Bergshoeff:1989de}, which should be compatible with the WGC. This calls into question the validity of the WGC bounds for hyperbolic black holes, and thus we decided to trust only the conditions imposed by the spherical case. As we see later, this leads to quite reasonable physics even when hyperbolic black holes are concerned, \textit{e.g.}, for R\'enyi entropies --- we comment on this below. However, it is also worth mentioning that Ref.~\cite{Bobev:2021qxx} recently provided examples of string theory realizations in which $\lambda<0$. If the sign of $\lambda$ can be arbitrary within string theory, this would indeed contradict the positive-entropy bounds of \cite{Cheung:2018cwt} as well as the results of \cite{Cheung:2016wjt}. Hence, one would conclude that these bounds are too strong, and perhaps only a weaker version of them holds true --- for instance, one could think of applying these bounds only to spherical near-extremal black holes. Clearly, all of this deserves further attention.

Next, we studied the thermodynamics of charged plasmas for these holographic models. For Einstein-Maxwell theory, a single phase exists for any value of the temperature and chemical potential, so we focused on the question of whether new phases could appear in the higher-derivative theory. We have seen that several branches of solutions may indeed appear, and for large enough couplings we even find zeroth-order phase transitions from the usual Einstein-Maxwell-like branch to a new exotic phase.  However, this situation always comes accompanied of the quite unphysical phenomenon of not having a large $\mu$ regime --- \textit{i.e.}, no black hole solutions exist if $\mu$ is too large. An interesting question is whether this scenario can be avoided when the physical constraints are implemented. As we have seen, the coupling $\beta$ has a somewhat different status in our action, as it is poorly constrained, so we looked first at the case of $\beta=0$. We have seen that the exotic absence-of-solutions behavior (as well as the existence of phase transitions) is completely ruled out by the constraints in $d=3,4$ dimensions, for which the WGC bounds prove to be essential. In fact, for $d=4$ the constraint we get on $\alpha_1$ is exactly the one that marks the existence of the phase transition.
When $\beta\neq 0$ this kind of behavior in the thermodynamic phase space can take place, but as we have argued, this coupling is not properly constrained. Overall, we tend to think that these exotic phase transitions are unphysical and it would be interesting to investigate if additional unitarity/causality conditions could rule all of them out.

We further studied the shear viscosity of these plasmas, focusing on the physical case where a phase exists for arbitrary values of the chemical potential.  Our result for the shear viscosity to entropy density ratio $\eta/s$ provides a non-perturbative and $d$-dimensional generalization of the computation of Ref.~\cite{Myers:2009ij}. 
In general, this ratio --- see \req{eq:etas2} --- is a function of the chemical potential, and the departure from the $1/4\pi$ value is controlled by two terms, one of them proportional to the GB coupling $\lambda$, and another one proportional to the $a_2$ parameter. Taking into account unitarity and WGC bounds, we find that the behavior is quite different depending on the sign of $a_2$. A negative value $a_2\le 0$ leads to very reasonable results: the shear viscosity to entropy density ratio is always a growing function of $\mu$ and it has absolute minimum and maximum values, given by Eq.~\req{eq:etasmaxmin}, so that it never departs much from the $1/4\pi$ prediction. It is worth mentioning that QCD belongs to this class, $a^{\rm QCD}_2<0$, so it would be interesting to study if QCD quark-gluon plasma shares any of the qualitative properties observed on our holographic models.  
On the other hand, we have seen that $a_2>0$ can lead to $\eta=0$ without violating any of the physical conditions, as long as the chemical potential is large enough. In fact this can be achieved in a theory as simple as the one with only $\alpha_2$ active. It would be very interesting to understand, even in that simple model, if other mechanisms could prevent the vanishing of the shear viscosity. A typical argument to rule out large corrections to $\eta/s$ in holographic higher-order gravities is that of Ref.~\cite{Camanho:2014apa}, which, in essence, implies that only perturbative couplings are consistent in order to respect causality in theories with a non-Einsteinian three-point structure for the graviton. However, this reasoning certainly cannot be applied to the terms with $\alpha_2$ in \req{eq:EQTfour} (since they do not affect the three-point function of the graviton), so we do not have a concrete argument by which $\alpha_2$ should be perturbatively small. 
Still, it would be convenient to investigate other constraints, \textit{e.g.}, causality in charged backgrounds in the bulk interior or the existence of plasma instabilities \cite{Ge:2008ni,Ge:2009ac,Ge:2009eh}, which could avoid reaching too small values of $\eta$. Otherwise, this example could indicate that in certain CFTs it would possible to reach an arbitrarily low viscosity by turning on a chemical potential. 

In the final part of the paper we studied holographic charged R\'enyi entropies and their associated generalized twist operators, both of which are related to the thermodynamic properties of black holes with hyperbolic horizons. We observed that, providing that the dual CFT respects unitarity, the chemical potential always increases R\'enyi entropies with $n\ge 1$. Furthermore, standard R\'enyi entropies are known to satisfy a number of inequalities as a function of the index $n$ --- see \req{eq:inerenyi} --- so we wondered if these held in our higher-derivative theories as well. As it turns out, these seem to be always satisfied if one assumes all of the constraints we have studied. However, if one gives up the WGC bounds, it is found that the RE can violate some of these inequalities, and they could even become negative --- see Fig.~\ref{fig:renyiwgc}. It is quite remarkable that the WGC avoids this behavior, which points in the direction that the WGC bounds in AdS are necessary in order to give rise to a sensible theory in the boundary.  

We finally computed the scaling dimension $h_n(\mu)$ and magnetic response $k_n(\mu)$ of generalized twist operators, as introduced by Ref.~\cite{Belin:2013uta}. By using the entries for the holographic dictionary of the EQG \req{eq:EQTfour}, we have obtained a series of relationships between the derivatives of $h_n(\mu)$ and $k_n(\mu)$ at $n=1$, $\mu=0$ and $C_T$, $C_J$ and the coefficients of $\braket{TJJ}$ (see Eqs.~\req{h10}, \req{h02}, \req{k01} and \req{k11}). These are actually universal relations that hold for any CFT, and they were first derived from first principles in Refs.~\cite{Hung:2011nu,Belin:2013uta}. The fact that one can independently derive these results by using holographic higher-derivative theories is a proof of the power of this approach to learn about universality in CFTs. It is remarkable how everything comes together taking into account the number of computations involved in obtaining these formulas from two completely different approaches. 

Let us close by commenting on additional future directions. On general grounds, it is necessary to get a better understanding of the WGC in AdS space; in particular, to understand what are the implications of this conjecture for the dual CFT. Also, the issue about the positivity-of-entropy bounds in hyperbolic black holes, and in general, the regime of validity of these bounds, should be clarified. 
Regarding the Electromagnetic Quasitopological gravities we presented, there are some holographic aspects of these theories that we did not address here, but that could be worth pursuing.  This includes studying the thermodynamic phase space of CFTs in a sphere or carrying out a more general hydrodynamic analysis, including for instance the study of conductivities.  These theories could also be interesting in the context of holographic superconductors \cite{Gregory:2009fj,Jing:2010zp,Cai:2010cv,Edelstein:2022xlb}. The higher-order Electromagnetic Quasitopological gravities introduced in Sec.~\ref{subs:eqgsallorders} will also provide interesting holographic models for future endeavors.

\section*{Acknowledgments}
We would like to thank Brian McPeak for discussions on the WGC in AdS and Nikolay Bobev, Pablo Bueno and Robie Hennigar for useful comments.  
The work of P.A.C. is supported by a postdoctoral fellowship from the Research Foundation - Flanders
(FWO grant 12ZH121N). The work of  \'A. J. M. is funded by the Spanish FPU Grant No. FPU17/04964. \'A. J. M. was further supported by the MCIU/AEI/FEDER UE grant PGC2018-095205-B-I00 and by the Spanish Research Agency (Agencia Estatal de Investigación) through the Grant IFT Centro de Excelencia Severo
Ochoa No CEX2020-001007-S, funded by MCIN/AEI/10.13039/501100011033. A.R.S. is supported by the Spanish MECD grant FPU18/03719. The work of A.R.S. is further funded by AEI-Spain (under project PID2020-114157GB-I00 and Unidad de Excelencia Mar\'\i a de Maeztu MDM-2016-0692), by Xunta de Galicia-Conseller\'\i a de Educaci\' on (Centro singular de investigaci\' on de Galicia accreditation 2019-2022, and project ED431C-2021/14), and by the European Union FEDER. 
A.R.S. is pleased to thank KU Leuven, where a large part of this work was done, for their warm hospitality. The work of X.Z. is supported by the KU Leuven C1 grant ZKD1118 C16/16/005.

\appendix

\bibliographystyle{JHEP}
\bibliography{Gravities.bib}

\end{document}